\documentclass[a4paper, traditabstract, longauth]{aa} 
%\documentclass[referee, traditabstract, longauth]{aa} 

% for the abstract without structuration 
%                   (traditional abstract)                      
%
\usepackage{graphicx}
\usepackage{subcaption}
\usepackage{color}
\usepackage[hyphens]{url}
\usepackage[colorlinks, citecolor=blue]{hyperref}  \usepackage{breakurl}
\usepackage{natbib}
\usepackage{longtable,lscape}
\usepackage{txfonts}
\usepackage{amsfonts}
\usepackage{amssymb}
\usepackage{epstopdf}
\usepackage{newlfont}
\usepackage{textcomp}
\usepackage{times}
\usepackage{units}
\usepackage{ifthen}
\usepackage[switch,displaymath, mathlines,]{lineno}

\def\setsymbol#1#2{\expandafter\def\csname #1\endcsname{#2}}
\def\getsymbol#1{\csname #1\endcsname}

%-----------------------------------------------------------------------
% Planck
%-----------------------------------------------------------------------
\def\Planck{{\it Planck\/}}

%-----------------------------------------------------------------------
% The Planck Helium-4 JT cooler
%-----------------------------------------------------------------------

%-----------------------------------------------------------------------
% To include all Planck Early Results papers in the reference lists
%-----------------------------------------------------------------------

%-----------------------------------------------------------------------
% Tables
%-----------------------------------------------------------------------
\newbox\tablebox    \newdimen\tablewidth
\def\leaderfil{\leaders\hbox to 5pt{\hss.\hss}\hfil}
%
% use the following definition of \endPlancktable for ApJ style notes to tables, set to the %         width of the table
%\def\endPlancktable{\tablewidth=\wd\tablebox 
%
% use the following definition of \endPlancktable instead for A&A style notes set to full 
%         column width
\def\endPlancktable{\tablewidth=\columnwidth 
    $$\hss\copy\tablebox\hss$$
    \vskip-\lastskip\vskip -2pt}

\def\tablenote#1 #2\par{\begingroup \parindent=0.8em
    \abovedisplayshortskip=0pt\belowdisplayshortskip=0pt
    \noindent
    $$\hss\vbox{\hsize\tablewidth \hangindent=\parindent \hangafter=1 \noindent
    \hbox to \parindent{\sup{\rm #1}\hss}\strut#2\strut\par}\hss$$
    \endgroup}
\def\doubleline{\vskip 3pt\hrule \vskip 1.5pt \hrule \vskip 5pt}

%-----------------------------------------------------------------------
% useful macros
%-----------------------------------------------------------------------
%
\def\L2{\ifmmode L_2\else $L_2$\fi}

\def\DeltaT{\ifmmode \Delta T\else $\Delta T$\fi}
\def\deltat{\ifmmode \Delta t\else $\Delta t$\fi}
\def\fknee{\ifmmode f_{\rm knee}\else $f_{\rm knee}$\fi}
\def\Fmax{\ifmmode F_{\rm max}\else $F_{\rm max}$\fi}
\def\solar{\ifmmode{\rm M}_{\mathord\odot}\else${\rm M}_{\mathord\odot}$\fi}

\def\inv{\ifmmode^{-1}\else$^{-1}$\fi}
\def\mo{\ifmmode^{-1}\else$^{-1}$\fi}
\def\sup#1{\ifmmode ^{\rm #1}\else $^{\rm #1}$\fi}
\def\expo#1{\ifmmode \times 10^{#1}\else $\times 10^{#1}$\fi}
\def\,{\thinspace}
\def\lsim{\mathrel{\raise .4ex\hbox{\rlap{$<$}\lower 1.2ex\hbox{$\sim$}}}}
\def\gsim{\mathrel{\raise .4ex\hbox{\rlap{$>$}\lower 1.2ex\hbox{$\sim$}}}}

\def\simprop{\mathrel{\raise .4ex\hbox{\rlap{$\propto$}\lower 1.2ex\hbox{$\sim$}}}}
\def\deg{\ifmmode^\circ\else$^\circ$\fi}
\def\pdeg{\ifmmode $\setbox0=\hbox{$^{\circ}$}\rlap{\hskip.11\wd0 .}$^{\circ}
          \else \setbox0=\hbox{$^{\circ}$}\rlap{\hskip.11\wd0 .}$^{\circ}$\fi}
\def\arcs{\ifmmode {^{\scriptstyle\prime\prime}}
          \else $^{\scriptstyle\prime\prime}$\fi}
\def\arcm{\ifmmode {^{\scriptstyle\prime}}
          \else $^{\scriptstyle\prime}$\fi}
\newdimen\sa  \newdimen\sb
\def\parcs{\sa=.07em \sb=.03em
     \ifmmode \hbox{\rlap{.}}^{\scriptstyle\prime\kern -\sb\prime}\hbox{\kern -\sa}
     \else \rlap{.}$^{\scriptstyle\prime\kern -\sb\prime}$\kern -\sa\fi}
\def\parcm{\sa=.08em \sb=.03em
     \ifmmode \hbox{\rlap{.}\kern\sa}^{\scriptstyle\prime}\hbox{\kern-\sb}
     \else \rlap{.}\kern\sa$^{\scriptstyle\prime}$\kern-\sb\fi}
\def\ra[#1 #2 #3.#4]{#1\sup{h}#2\sup{m}#3\sup{s}\llap.#4}
\def\dec[#1 #2 #3.#4]{#1\deg#2\arcm#3\arcs\llap.#4}
\def\deco[#1 #2 #3]{#1\deg#2\arcm#3\arcs}
\def\rra[#1 #2]{#1\sup{h}#2\sup{m}}

\def\dots{\relax\ifmmode \ldots\else $\ldots$\fi}
%
%-----------------------------------------------------------------------
% units
%-----------------------------------------------------------------------
%
\def\WHzsr{\ifmmode $W\,Hz\mo\,sr\mo$\else W\,Hz\mo\,sr\mo\fi}
\def\mHz{\ifmmode $\,mHz$\else \,mHz\fi}
\def\GHz{\ifmmode $\,GHz$\else \,GHz\fi}
\def\mKs{\ifmmode $\,mK\,s$^{1/2}\else \,mK\,s$^{1/2}$\fi}
\def\muKs{\ifmmode \,\mu$K\,s$^{1/2}\else \,$\mu$K\,s$^{1/2}$\fi}
\def\muKRJs{\ifmmode \,\mu$K$_{\rm RJ}$\,s$^{1/2}\else \,$\mu$K$_{\rm RJ}$\,s$^{1/2}$\fi}
\def\muKHz{\ifmmode \,\mu$K\,Hz$^{-1/2}\else \,$\mu$K\,Hz$^{-1/2}$\fi}
\def\MJysr{\ifmmode \,$MJy\,sr\mo$\else \,MJy\,sr\mo\fi}
\def\MJysrmK{\ifmmode \,$MJy\,sr\mo$\,mK$_{\rm CMB}\mo\else \,MJy\,sr\mo\,mK$_{\rm CMB}\mo$\fi}
\def\microns{\ifmmode \,\mu$m$\else \,$\mu$m\fi}

\def\muK{\ifmmode \,\mu$K$\else \,$\mu$\hbox{K}\fi}
\def\microK{\ifmmode \,\mu$K$\else \,$\mu$\hbox{K}\fi}
\def\muW{\ifmmode \,\mu$W$\else \,$\mu$\hbox{W}\fi}
\def\kms{\ifmmode $\,km\,s$^{-1}\else \,km\,s$^{-1}$\fi}
\def\kmsMpc{\ifmmode $\,\kms\,Mpc\mo$\else \,\kms\,Mpc\mo\fi}
%
%
%----------------------------------------------------------------------

% LFI Center Frequency

\setsymbol{LFI:center:frequency:70GHz:units}{70.3\,GHz}
\setsymbol{LFI:center:frequency:44GHz:units}{44.1\,GHz}
\setsymbol{LFI:center:frequency:30GHz:units}{28.5\,GHz}

\setsymbol{LFI:center:frequency:70GHz}{70.3}
\setsymbol{LFI:center:frequency:44GHz}{44.1}
\setsymbol{LFI:center:frequency:30GHz}{28.5}

\setsymbol{LFI:center:frequency:LFI18:Rad:M:units}{71.7\GHz}
\setsymbol{LFI:center:frequency:LFI19:Rad:M:units}{67.5\GHz}
\setsymbol{LFI:center:frequency:LFI20:Rad:M:units}{69.2\GHz}
\setsymbol{LFI:center:frequency:LFI21:Rad:M:units}{70.4\GHz}
\setsymbol{LFI:center:frequency:LFI22:Rad:M:units}{71.5\GHz}
\setsymbol{LFI:center:frequency:LFI23:Rad:M:units}{70.8\GHz}
\setsymbol{LFI:center:frequency:LFI24:Rad:M:units}{44.4\GHz}
\setsymbol{LFI:center:frequency:LFI25:Rad:M:units}{44.0\GHz}
\setsymbol{LFI:center:frequency:LFI26:Rad:M:units}{43.9\GHz}
\setsymbol{LFI:center:frequency:LFI27:Rad:M:units}{28.3\GHz}
\setsymbol{LFI:center:frequency:LFI28:Rad:M:units}{28.8\GHz}
\setsymbol{LFI:center:frequency:LFI18:Rad:S:units}{70.1\GHz}
\setsymbol{LFI:center:frequency:LFI19:Rad:S:units}{69.6\GHz}
\setsymbol{LFI:center:frequency:LFI20:Rad:S:units}{69.5\GHz}
\setsymbol{LFI:center:frequency:LFI21:Rad:S:units}{69.5\GHz}
\setsymbol{LFI:center:frequency:LFI22:Rad:S:units}{72.8\GHz}
\setsymbol{LFI:center:frequency:LFI23:Rad:S:units}{71.3\GHz}
\setsymbol{LFI:center:frequency:LFI24:Rad:S:units}{44.1\GHz}
\setsymbol{LFI:center:frequency:LFI25:Rad:S:units}{44.1\GHz}
\setsymbol{LFI:center:frequency:LFI26:Rad:S:units}{44.1\GHz}
\setsymbol{LFI:center:frequency:LFI27:Rad:S:units}{28.5\GHz}
\setsymbol{LFI:center:frequency:LFI28:Rad:S:units}{28.2\GHz}

\setsymbol{LFI:center:frequency:LFI18:Rad:M}{71.7}
\setsymbol{LFI:center:frequency:LFI19:Rad:M}{67.5}
\setsymbol{LFI:center:frequency:LFI20:Rad:M}{69.2}
\setsymbol{LFI:center:frequency:LFI21:Rad:M}{70.4}
\setsymbol{LFI:center:frequency:LFI22:Rad:M}{71.5}
\setsymbol{LFI:center:frequency:LFI23:Rad:M}{70.8}
\setsymbol{LFI:center:frequency:LFI24:Rad:M}{44.4}
\setsymbol{LFI:center:frequency:LFI25:Rad:M}{44.0}
\setsymbol{LFI:center:frequency:LFI26:Rad:M}{43.9}
\setsymbol{LFI:center:frequency:LFI27:Rad:M}{28.3}
\setsymbol{LFI:center:frequency:LFI28:Rad:M}{28.8}
\setsymbol{LFI:center:frequency:LFI18:Rad:S}{70.1}
\setsymbol{LFI:center:frequency:LFI19:Rad:S}{69.6}
\setsymbol{LFI:center:frequency:LFI20:Rad:S}{69.5}
\setsymbol{LFI:center:frequency:LFI21:Rad:S}{69.5}
\setsymbol{LFI:center:frequency:LFI22:Rad:S}{72.8}
\setsymbol{LFI:center:frequency:LFI23:Rad:S}{71.3}
\setsymbol{LFI:center:frequency:LFI24:Rad:S}{44.1}
\setsymbol{LFI:center:frequency:LFI25:Rad:S}{44.1}
\setsymbol{LFI:center:frequency:LFI26:Rad:S}{44.1}
\setsymbol{LFI:center:frequency:LFI27:Rad:S}{28.5}
\setsymbol{LFI:center:frequency:LFI28:Rad:S}{28.2}

% LFI White Noise Sensitivity, \delta T_{\rm RJ}

\setsymbol{LFI:white:noise:sensitivity:70GHz:units}{134.7\muKs}
\setsymbol{LFI:white:noise:sensitivity:44GHz:units}{164.7\muKs}
\setsymbol{LFI:white:noise:sensitivity:30GHz:units}{143.4\muKs}

\setsymbol{LFI:white:noise:sensitivity:70GHz}{134.7}
\setsymbol{LFI:white:noise:sensitivity:44GHz}{164.7}
\setsymbol{LFI:white:noise:sensitivity:30GHz}{143.4}

%***************the following are in \delta T_{\rm CMB} units!****************
%***************this segment should be revised********************************

\setsymbol{LFI:white:noise:sensitivity:LFI18:Rad:M:units}{512.0\muKs}
\setsymbol{LFI:white:noise:sensitivity:LFI19:Rad:M:units}{581.4\muKs}
\setsymbol{LFI:white:noise:sensitivity:LFI20:Rad:M:units}{590.8\muKs}
\setsymbol{LFI:white:noise:sensitivity:LFI21:Rad:M:units}{455.2\muKs}
\setsymbol{LFI:white:noise:sensitivity:LFI22:Rad:M:units}{492.0\muKs}
\setsymbol{LFI:white:noise:sensitivity:LFI23:Rad:M:units}{507.7\muKs}
\setsymbol{LFI:white:noise:sensitivity:LFI24:Rad:M:units}{462.2\muKs}
\setsymbol{LFI:white:noise:sensitivity:LFI25:Rad:M:units}{413.6\muKs}
\setsymbol{LFI:white:noise:sensitivity:LFI26:Rad:M:units}{478.6\muKs}
\setsymbol{LFI:white:noise:sensitivity:LFI27:Rad:M:units}{277.7\muKs}
\setsymbol{LFI:white:noise:sensitivity:LFI28:Rad:M:units}{312.3\muKs}
\setsymbol{LFI:white:noise:sensitivity:LFI18:Rad:S:units}{465.7\muKs}
\setsymbol{LFI:white:noise:sensitivity:LFI19:Rad:S:units}{555.6\muKs}
\setsymbol{LFI:white:noise:sensitivity:LFI20:Rad:S:units}{623.2\muKs}
\setsymbol{LFI:white:noise:sensitivity:LFI21:Rad:S:units}{564.1\muKs}
\setsymbol{LFI:white:noise:sensitivity:LFI22:Rad:S:units}{534.4\muKs}
\setsymbol{LFI:white:noise:sensitivity:LFI23:Rad:S:units}{542.4\muKs}
\setsymbol{LFI:white:noise:sensitivity:LFI24:Rad:S:units}{399.2\muKs}
\setsymbol{LFI:white:noise:sensitivity:LFI25:Rad:S:units}{392.6\muKs}
\setsymbol{LFI:white:noise:sensitivity:LFI26:Rad:S:units}{418.6\muKs}
\setsymbol{LFI:white:noise:sensitivity:LFI27:Rad:S:units}{302.9\muKs}
\setsymbol{LFI:white:noise:sensitivity:LFI28:Rad:S:units}{285.3\muKs}

\setsymbol{LFI:white:noise:sensitivity:LFI18:Rad:M}{512.0}
\setsymbol{LFI:white:noise:sensitivity:LFI19:Rad:M}{581.4}
\setsymbol{LFI:white:noise:sensitivity:LFI20:Rad:M}{590.8}
\setsymbol{LFI:white:noise:sensitivity:LFI21:Rad:M}{455.2}
\setsymbol{LFI:white:noise:sensitivity:LFI22:Rad:M}{492.0}
\setsymbol{LFI:white:noise:sensitivity:LFI23:Rad:M}{507.7}
\setsymbol{LFI:white:noise:sensitivity:LFI24:Rad:M}{462.2}
\setsymbol{LFI:white:noise:sensitivity:LFI25:Rad:M}{413.6}
\setsymbol{LFI:white:noise:sensitivity:LFI26:Rad:M}{478.6}
\setsymbol{LFI:white:noise:sensitivity:LFI27:Rad:M}{277.7}
\setsymbol{LFI:white:noise:sensitivity:LFI28:Rad:M}{312.3}
\setsymbol{LFI:white:noise:sensitivity:LFI18:Rad:S}{465.7}
\setsymbol{LFI:white:noise:sensitivity:LFI19:Rad:S}{555.6}
\setsymbol{LFI:white:noise:sensitivity:LFI20:Rad:S}{623.2}
\setsymbol{LFI:white:noise:sensitivity:LFI21:Rad:S}{564.1}
\setsymbol{LFI:white:noise:sensitivity:LFI22:Rad:S}{534.4}
\setsymbol{LFI:white:noise:sensitivity:LFI23:Rad:S}{542.4}
\setsymbol{LFI:white:noise:sensitivity:LFI24:Rad:S}{399.2}
\setsymbol{LFI:white:noise:sensitivity:LFI25:Rad:S}{392.6}
\setsymbol{LFI:white:noise:sensitivity:LFI26:Rad:S}{418.6}
\setsymbol{LFI:white:noise:sensitivity:LFI27:Rad:S}{302.9}
\setsymbol{LFI:white:noise:sensitivity:LFI28:Rad:S}{285.3}

% LFI Knee Frequency

\setsymbol{LFI:knee:frequency:70GHz:units}{29.5\mHz}
\setsymbol{LFI:knee:frequency:44GHz:units}{56.2\mHz}
\setsymbol{LFI:knee:frequency:30GHz:units}{113.7\mHz}

\setsymbol{LFI:knee:frequency:70GHz}{29.5}
\setsymbol{LFI:knee:frequency:44GHz}{56.2}
\setsymbol{LFI:knee:frequency:30GHz}{113.7}

\setsymbol{LFI:knee:frequency:LFI18:Rad:M:units}{16.3\mHz}
\setsymbol{LFI:knee:frequency:LFI19:Rad:M:units}{15.1\mHz}
\setsymbol{LFI:knee:frequency:LFI20:Rad:M:units}{18.7\mHz}
\setsymbol{LFI:knee:frequency:LFI21:Rad:M:units}{37.2\mHz}
\setsymbol{LFI:knee:frequency:LFI22:Rad:M:units}{12.7\mHz}
\setsymbol{LFI:knee:frequency:LFI23:Rad:M:units}{34.6\mHz}
\setsymbol{LFI:knee:frequency:LFI24:Rad:M:units}{46.2\mHz}
\setsymbol{LFI:knee:frequency:LFI25:Rad:M:units}{24.9\mHz}
\setsymbol{LFI:knee:frequency:LFI26:Rad:M:units}{67.6\mHz}
\setsymbol{LFI:knee:frequency:LFI27:Rad:M:units}{187.4\mHz}
\setsymbol{LFI:knee:frequency:LFI28:Rad:M:units}{122.2\mHz}
\setsymbol{LFI:knee:frequency:LFI18:Rad:S:units}{17.7\mHz}
\setsymbol{LFI:knee:frequency:LFI19:Rad:S:units}{22.0\mHz}
\setsymbol{LFI:knee:frequency:LFI20:Rad:S:units}{8.7\mHz}
\setsymbol{LFI:knee:frequency:LFI21:Rad:S:units}{25.9\mHz}
\setsymbol{LFI:knee:frequency:LFI22:Rad:S:units}{15.8\mHz}
\setsymbol{LFI:knee:frequency:LFI23:Rad:S:units}{129.8\mHz}
\setsymbol{LFI:knee:frequency:LFI24:Rad:S:units}{100.9\mHz}
\setsymbol{LFI:knee:frequency:LFI25:Rad:S:units}{38.9\mHz}
\setsymbol{LFI:knee:frequency:LFI26:Rad:S:units}{58.9\mHz}
\setsymbol{LFI:knee:frequency:LFI27:Rad:S:units}{104.4\mHz}
\setsymbol{LFI:knee:frequency:LFI28:Rad:S:units}{40.7\mHz}

\setsymbol{LFI:knee:frequency:LFI18:Rad:M}{16.3}
\setsymbol{LFI:knee:frequency:LFI19:Rad:M}{15.1}
\setsymbol{LFI:knee:frequency:LFI20:Rad:M}{18.7}
\setsymbol{LFI:knee:frequency:LFI21:Rad:M}{37.2}
\setsymbol{LFI:knee:frequency:LFI22:Rad:M}{12.7}
\setsymbol{LFI:knee:frequency:LFI23:Rad:M}{34.6}
\setsymbol{LFI:knee:frequency:LFI24:Rad:M}{46.2}
\setsymbol{LFI:knee:frequency:LFI25:Rad:M}{24.9}
\setsymbol{LFI:knee:frequency:LFI26:Rad:M}{67.6}
\setsymbol{LFI:knee:frequency:LFI27:Rad:M}{187.4}
\setsymbol{LFI:knee:frequency:LFI28:Rad:M}{122.2}
\setsymbol{LFI:knee:frequency:LFI18:Rad:S}{17.7}
\setsymbol{LFI:knee:frequency:LFI19:Rad:S}{22.0}
\setsymbol{LFI:knee:frequency:LFI20:Rad:S}{8.7}
\setsymbol{LFI:knee:frequency:LFI21:Rad:S}{25.9}
\setsymbol{LFI:knee:frequency:LFI22:Rad:S}{15.8}
\setsymbol{LFI:knee:frequency:LFI23:Rad:S}{129.8}
\setsymbol{LFI:knee:frequency:LFI24:Rad:S}{100.9}
\setsymbol{LFI:knee:frequency:LFI25:Rad:S}{38.9}
\setsymbol{LFI:knee:frequency:LFI26:Rad:S}{58.9}
\setsymbol{LFI:knee:frequency:LFI27:Rad:S}{104.4}
\setsymbol{LFI:knee:frequency:LFI28:Rad:S}{40.7}

% LFI low frequency noise slope

\setsymbol{LFI:slope:70GHz:units}{$-1.03$\mHz}
\setsymbol{LFI:slope:44GHz:units}{$-0.89$\mHz}
\setsymbol{LFI:slope:30GHz:units}{$-0.87$\mHz}

\setsymbol{LFI:slope:70GHz}{$-1.03$}
\setsymbol{LFI:slope:44GHz}{$-0.89$}
\setsymbol{LFI:slope:30GHz}{$-0.87$}

\setsymbol{LFI:slope:LFI18:Rad:M:units}{$-1.04$\mHz}
\setsymbol{LFI:slope:LFI19:Rad:M:units}{$-1.09$\mHz}
\setsymbol{LFI:slope:LFI20:Rad:M:units}{$-0.69$\mHz}
\setsymbol{LFI:slope:LFI21:Rad:M:units}{$-1.56$\mHz}
\setsymbol{LFI:slope:LFI22:Rad:M:units}{$-1.01$\mHz}
\setsymbol{LFI:slope:LFI23:Rad:M:units}{$-0.96$\mHz}
\setsymbol{LFI:slope:LFI24:Rad:M:units}{$-0.83$\mHz}
\setsymbol{LFI:slope:LFI25:Rad:M:units}{$-0.91$\mHz}
\setsymbol{LFI:slope:LFI26:Rad:M:units}{$-0.95$\mHz}
\setsymbol{LFI:slope:LFI27:Rad:M:units}{$-0.87$\mHz}
\setsymbol{LFI:slope:LFI28:Rad:M:units}{$-0.88$\mHz}
\setsymbol{LFI:slope:LFI18:Rad:S:units}{$-1.15$\mHz}
\setsymbol{LFI:slope:LFI19:Rad:S:units}{$-1.00$\mHz}
\setsymbol{LFI:slope:LFI20:Rad:S:units}{$-0.95$\mHz}
\setsymbol{LFI:slope:LFI21:Rad:S:units}{$-0.92$\mHz}
\setsymbol{LFI:slope:LFI22:Rad:S:units}{$-1.01$\mHz}
\setsymbol{LFI:slope:LFI23:Rad:S:units}{$-0.95$\mHz}
\setsymbol{LFI:slope:LFI24:Rad:S:units}{$-0.73$\mHz}
\setsymbol{LFI:slope:LFI25:Rad:S:units}{$-1.16$\mHz}
\setsymbol{LFI:slope:LFI26:Rad:S:units}{$-0.79$\mHz}
\setsymbol{LFI:slope:LFI27:Rad:S:units}{$-0.82$\mHz}
\setsymbol{LFI:slope:LFI28:Rad:S:units}{$-0.91$\mHz}

\setsymbol{LFI:slope:LFI18:Rad:M}{$-1.04$}
\setsymbol{LFI:slope:LFI19:Rad:M}{$-1.09$}
\setsymbol{LFI:slope:LFI20:Rad:M}{$-0.69$}
\setsymbol{LFI:slope:LFI21:Rad:M}{$-1.56$}
\setsymbol{LFI:slope:LFI22:Rad:M}{$-1.01$}
\setsymbol{LFI:slope:LFI23:Rad:M}{$-0.96$}
\setsymbol{LFI:slope:LFI24:Rad:M}{$-0.83$}
\setsymbol{LFI:slope:LFI25:Rad:M}{$-0.91$}
\setsymbol{LFI:slope:LFI26:Rad:M}{$-0.95$}
\setsymbol{LFI:slope:LFI27:Rad:M}{$-0.87$}
\setsymbol{LFI:slope:LFI28:Rad:M}{$-0.88$}
\setsymbol{LFI:slope:LFI18:Rad:S}{$-1.15$}
\setsymbol{LFI:slope:LFI19:Rad:S}{$-1.00$}
\setsymbol{LFI:slope:LFI20:Rad:S}{$-0.95$}
\setsymbol{LFI:slope:LFI21:Rad:S}{$-0.92$}
\setsymbol{LFI:slope:LFI22:Rad:S}{$-1.01$}
\setsymbol{LFI:slope:LFI23:Rad:S}{$-0.95$}
\setsymbol{LFI:slope:LFI24:Rad:S}{$-0.73$}
\setsymbol{LFI:slope:LFI25:Rad:S}{$-1.16$}
\setsymbol{LFI:slope:LFI26:Rad:S}{$-0.79$}
\setsymbol{LFI:slope:LFI27:Rad:S}{$-0.82$}
\setsymbol{LFI:slope:LFI28:Rad:S}{$-0.91$}

% LFI Beam FWHM

\setsymbol{LFI:FWHM:70GHz:units}{13\parcm01}
\setsymbol{LFI:FWHM:44GHz:units}{27\parcm92}
\setsymbol{LFI:FWHM:30GHz:units}{32\parcm65}

\setsymbol{LFI:FWHM:70GHz}{13.01}
\setsymbol{LFI:FWHM:44GHz}{27.92}
\setsymbol{LFI:FWHM:30GHz}{32.65}

\setsymbol{LFI:FWHM:LFI18:units}{13\parcm39}
\setsymbol{LFI:FWHM:LFI19:units}{13\parcm01}
\setsymbol{LFI:FWHM:LFI20:units}{12\parcm75}
\setsymbol{LFI:FWHM:LFI21:units}{12\parcm74}
\setsymbol{LFI:FWHM:LFI22:units}{12\parcm87}
\setsymbol{LFI:FWHM:LFI23:units}{13\parcm27}
\setsymbol{LFI:FWHM:LFI24:units}{22\parcm98}
\setsymbol{LFI:FWHM:LFI25:units}{30\parcm46}
\setsymbol{LFI:FWHM:LFI26:units}{30\parcm31}
\setsymbol{LFI:FWHM:LFI27:units}{32\parcm65}
\setsymbol{LFI:FWHM:LFI28:units}{32\parcm66}

\setsymbol{LFI:FWHM:LFI18}{13.39}
\setsymbol{LFI:FWHM:LFI19}{13.01}
\setsymbol{LFI:FWHM:LFI20}{12.75}
\setsymbol{LFI:FWHM:LFI21}{12.74}
\setsymbol{LFI:FWHM:LFI22}{12.87}
\setsymbol{LFI:FWHM:LFI23}{13.27}
\setsymbol{LFI:FWHM:LFI24}{22.98}
\setsymbol{LFI:FWHM:LFI25}{30.46}
\setsymbol{LFI:FWHM:LFI26}{30.31}
\setsymbol{LFI:FWHM:LFI27}{32.65}
\setsymbol{LFI:FWHM:LFI28}{32.66}

% LFI Beam FWHM Uncertainty
% When uncertainties are routinely available for all quantities, we'll likely change the format to build them into 
% the \setsymbol command.  For now, this will be a bit easier.

%\setsymbol{LFI:FWHM:uncertainty:70GHz}{TBD\arcm}
%\setsymbol{LFI:FWHM:uncertainty:44GHz}{TBD\arcm}
%\setsymbol{LFI:FWHM:uncertainty:30GHz}{TBD\arcm}

\setsymbol{LFI:FWHM:uncertainty:LFI18:units}{0.170\arcm}
\setsymbol{LFI:FWHM:uncertainty:LFI19:units}{0.174\arcm}
\setsymbol{LFI:FWHM:uncertainty:LFI20:units}{0.170\arcm}
\setsymbol{LFI:FWHM:uncertainty:LFI21:units}{0.156\arcm}
\setsymbol{LFI:FWHM:uncertainty:LFI22:units}{0.164\arcm}
\setsymbol{LFI:FWHM:uncertainty:LFI23:units}{0.171\arcm}
\setsymbol{LFI:FWHM:uncertainty:LFI24:units}{0.652\arcm}
\setsymbol{LFI:FWHM:uncertainty:LFI25:units}{1.075\arcm}
\setsymbol{LFI:FWHM:uncertainty:LFI26:units}{1.131\arcm}
\setsymbol{LFI:FWHM:uncertainty:LFI27:units}{1.266\arcm}
\setsymbol{LFI:FWHM:uncertainty:LFI28:units}{1.287\arcm}

\setsymbol{LFI:FWHM:uncertainty:LFI18}{0.170}
\setsymbol{LFI:FWHM:uncertainty:LFI19}{0.174}
\setsymbol{LFI:FWHM:uncertainty:LFI20}{0.170}
\setsymbol{LFI:FWHM:uncertainty:LFI21}{0.156}
\setsymbol{LFI:FWHM:uncertainty:LFI22}{0.164}
\setsymbol{LFI:FWHM:uncertainty:LFI23}{0.171}
\setsymbol{LFI:FWHM:uncertainty:LFI24}{0.652}
\setsymbol{LFI:FWHM:uncertainty:LFI25}{1.075}
\setsymbol{LFI:FWHM:uncertainty:LFI26}{1.131}
\setsymbol{LFI:FWHM:uncertainty:LFI27}{1.266}
\setsymbol{LFI:FWHM:uncertainty:LFI28}{1.287}

% HFI Center Frequency

\setsymbol{HFI:center:frequency:100GHz:units}{100\,GHz}
\setsymbol{HFI:center:frequency:143GHz:units}{143\,GHz}
\setsymbol{HFI:center:frequency:217GHz:units}{217\,GHz}
\setsymbol{HFI:center:frequency:353GHz:units}{353\,GHz}
\setsymbol{HFI:center:frequency:545GHz:units}{545\,GHz}
\setsymbol{HFI:center:frequency:857GHz:units}{857\,GHz}

\setsymbol{HFI:center:frequency:100GHz}{100}
\setsymbol{HFI:center:frequency:143GHz}{143}
\setsymbol{HFI:center:frequency:217GHz}{217}
\setsymbol{HFI:center:frequency:353GHz}{353}
\setsymbol{HFI:center:frequency:545GHz}{545}
\setsymbol{HFI:center:frequency:857GHz}{857}

% HFI Number of Detectors

\setsymbol{HFI:Ndetectors:100GHz}{8}
\setsymbol{HFI:Ndetectors:143GHz}{11}
\setsymbol{HFI:Ndetectors:217GHz}{12}
\setsymbol{HFI:Ndetectors:353GHz}{12}
\setsymbol{HFI:Ndetectors:545GHz}{3}
\setsymbol{HFI:Ndetectors:857GHz}{4}

% HFI FWHM on maps

\setsymbol{HFI:FWHM:Maps:100GHz:units}{9\parcm88}
\setsymbol{HFI:FWHM:Maps:143GHz:units}{7\parcm18}
\setsymbol{HFI:FWHM:Maps:217GHz:units}{4\parcm87}
\setsymbol{HFI:FWHM:Maps:353GHz:units}{4\parcm65}
\setsymbol{HFI:FWHM:Maps:545GHz:units}{4\parcm72}
\setsymbol{HFI:FWHM:Maps:857GHz:units}{4\parcm39}
\setsymbol{HFI:FWHM:Maps:100GHz}{9.88}
\setsymbol{HFI:FWHM:Maps:143GHz}{7.18}
\setsymbol{HFI:FWHM:Maps:217GHz}{4.87}
\setsymbol{HFI:FWHM:Maps:353GHz}{4.65}
\setsymbol{HFI:FWHM:Maps:545GHz}{4.72}
\setsymbol{HFI:FWHM:Maps:857GHz}{4.39}

% HFI Beam Ellipticity on maps

\setsymbol{HFI:beam:ellipticity:Maps:100GHz}{1.15}
\setsymbol{HFI:beam:ellipticity:Maps:143GHz}{1.01}
\setsymbol{HFI:beam:ellipticity:Maps:217GHz}{1.06}
\setsymbol{HFI:beam:ellipticity:Maps:353GHz}{1.05}
\setsymbol{HFI:beam:ellipticity:Maps:545GHz}{1.14}
\setsymbol{HFI:beam:ellipticity:Maps:857GHz}{1.19}

% HFI optical Beam FWHM from Mars; time response deconvolved: frequency  average of values in table 4 in HFI instrument paper

\setsymbol{HFI:FWHM:Mars:100GHz:units}{9\parcm37}
\setsymbol{HFI:FWHM:Mars:143GHz:units}{7\parcm04}
\setsymbol{HFI:FWHM:Mars:217GHz:units}{4\parcm68}
\setsymbol{HFI:FWHM:Mars:353GHz:units}{4\parcm43}
\setsymbol{HFI:FWHM:Mars:545GHz:units}{3\parcm80}
\setsymbol{HFI:FWHM:Mars:857GHz:units}{3\parcm67}

\setsymbol{HFI:FWHM:Mars:100GHz}{9.37}
\setsymbol{HFI:FWHM:Mars:143GHz}{7.04}
\setsymbol{HFI:FWHM:Mars:217GHz}{4.68}
\setsymbol{HFI:FWHM:Mars:353GHz}{4.43}
\setsymbol{HFI:FWHM:Mars:545GHz}{3.80}
\setsymbol{HFI:FWHM:Mars:857GHz}{3.67}

% HFI optical Beam Ellipticity from Mars; time response deconvolved: frequency average of values in table 4 in HFI instrument paper

\setsymbol{HFI:beam:ellipticity:Mars:100GHz}{1.18}
\setsymbol{HFI:beam:ellipticity:Mars:143GHz}{1.03}
\setsymbol{HFI:beam:ellipticity:Mars:217GHz}{1.14}
\setsymbol{HFI:beam:ellipticity:Mars:353GHz}{1.09}
\setsymbol{HFI:beam:ellipticity:Mars:545GHz}{1.25}
\setsymbol{HFI:beam:ellipticity:Mars:857GHz}{1.03}

% HFI CMB relative calibration accuracy

\setsymbol{HFI:CMB:relative:calibration:100GHz}{$\lsim 1\%$}
\setsymbol{HFI:CMB:relative:calibration:143GHz}{$\lsim 1\%$}
\setsymbol{HFI:CMB:relative:calibration:217GHz}{$\lsim 1\%$}
\setsymbol{HFI:CMB:relative:calibration:353GHz}{$\lsim 1\%$}
\setsymbol{HFI:CMB:relative:calibration:545GHz}{}
\setsymbol{HFI:CMB:relative:calibration:857GHz}{}

% HFI CMB absolute calibration accuracy

\setsymbol{HFI:CMB:absolute:calibration:100GHz}{$\lsim 2\%$}
\setsymbol{HFI:CMB:absolute:calibration:143GHz}{$\lsim 2\%$}
\setsymbol{HFI:CMB:absolute:calibration:217GHz}{$\lsim 2\%$}
\setsymbol{HFI:CMB:absolute:calibration:353GHz}{$\lsim 2\%$}
\setsymbol{HFI:CMB:absolute:calibration:545GHz}{}
\setsymbol{HFI:CMB:absolute:calibration:857GHz}{}

% HFI FIRAS gain calibration accuracy: statistical

\setsymbol{HFI:FIRAS:gain:calibration:accuracy:statistical:100GHz}{}
\setsymbol{HFI:FIRAS:gain:calibration:accuracy:statistical:143GHz}{}
\setsymbol{HFI:FIRAS:gain:calibration:accuracy:statistical:217GHz}{}
\setsymbol{HFI:FIRAS:gain:calibration:accuracy:statistical:353GHz}{2.5\%}
\setsymbol{HFI:FIRAS:gain:calibration:accuracy:statistical:545GHz}{1\%}
\setsymbol{HFI:FIRAS:gain:calibration:accuracy:statistical:857GHz}{0.5\%}

% HFI FIRAS gain calibration accuracy: systematic

\setsymbol{HFI:FIRAS:gain:calibration:accuracy:systematic:100GHz}{}
\setsymbol{HFI:FIRAS:gain:calibration:accuracy:systematic:143GHz}{}
\setsymbol{HFI:FIRAS:gain:calibration:accuracy:systematic:217GHz}{}
\setsymbol{HFI:FIRAS:gain:calibration:accuracy:systematic:353GHz}{}
\setsymbol{HFI:FIRAS:gain:calibration:accuracy:systematic:545GHz}{7\%}
\setsymbol{HFI:FIRAS:gain:calibration:accuracy:systematic:857GHz}{7\%}

% HFI FIRAS zero point accuracy:

\setsymbol{HFI:FIRAS:zero:point:accuracy:100GHz:units}{0.8\MJysr}
\setsymbol{HFI:FIRAS:zero:point:accuracy:143GHz:units}{}
\setsymbol{HFI:FIRAS:zero:point:accuracy:217GHz:units}{}
\setsymbol{HFI:FIRAS:zero:point:accuracy:353GHz:units}{1.4\MJysr}
\setsymbol{HFI:FIRAS:zero:point:accuracy:545GHz:units}{2.2\MJysr}
\setsymbol{HFI:FIRAS:zero:point:accuracy:857GHz:units}{1.7\MJysr}

\setsymbol{HFI:FIRAS:zero:point:accuracy:100GHz}{0.8}
\setsymbol{HFI:FIRAS:zero:point:accuracy:143GHz}{}
\setsymbol{HFI:FIRAS:zero:point:accuracy:217GHz}{}
\setsymbol{HFI:FIRAS:zero:point:accuracy:353GHz}{1.4}
\setsymbol{HFI:FIRAS:zero:point:accuracy:545GHz}{2.2}
\setsymbol{HFI:FIRAS:zero:point:accuracy:857GHz}{1.7}

% HFI diffuse source sensitivity unit conversion

\setsymbol{HFI:unit:conversion:100GHz:units}{0.2415\MJysrmK}
\setsymbol{HFI:unit:conversion:143GHz:units}{0.3694\MJysrmK}
\setsymbol{HFI:unit:conversion:217GHz:units}{0.4811\MJysrmK}
\setsymbol{HFI:unit:conversion:353GHz:units}{0.2883\MJysrmK}
\setsymbol{HFI:unit:conversion:545GHz:units}{0.05826\MJysrmK}
\setsymbol{HFI:unit:conversion:857GHz:units}{0.002238\MJysrmK}

\setsymbol{HFI:unit:conversion:100GHz}{0.2415}
\setsymbol{HFI:unit:conversion:143GHz}{0.3694}
\setsymbol{HFI:unit:conversion:217GHz}{0.4811}
\setsymbol{HFI:unit:conversion:353GHz}{0.2883}
\setsymbol{HFI:unit:conversion:545GHz}{0.05826}
\setsymbol{HFI:unit:conversion:857GHz}{0.002238}

% HFI Colour Correction for \alpha = -2, for V1.01 of the spectral bands

\setsymbol{HFI:colour:correction:alpha=-2:V1.01:100GHz}{0.9893}
\setsymbol{HFI:colour:correction:alpha=-2:V1.01:143GHz}{0.9759}
\setsymbol{HFI:colour:correction:alpha=-2:V1.01:217GHz}{1.0007}
\setsymbol{HFI:colour:correction:alpha=-2:V1.01:353GHz}{1.0028}
\setsymbol{HFI:colour:correction:alpha=-2:V1.01:545GHz}{1.0019}
\setsymbol{HFI:colour:correction:alpha=-2:V1.01:857GHz}{0.9889}

% HFI Colour Correction for \alpha = 0, for V1.01 of the spectral bands

\setsymbol{HFI:colour:correction:alpha=0:V1.01:100GHz}{1.0008}
\setsymbol{HFI:colour:correction:alpha=0:V1.01:143GHz}{1.0148}
\setsymbol{HFI:colour:correction:alpha=0:V1.01:217GHz}{0.9909}
\setsymbol{HFI:colour:correction:alpha=0:V1.01:353GHz}{0.9888}
\setsymbol{HFI:colour:correction:alpha=0:V1.01:545GHz}{0.9878}
\setsymbol{HFI:colour:correction:alpha=0:V1.01:857GHz}{1.0014}

\def\reff@jnl#1{{\rm#1\/}}
\def\apj{\reff@jnl{ApJ}}       % Astrophysical Journal dfdfsfsd
\def\apjs{\reff@jnl{ApJS}}     % Astrophysical Journal, Supplement
\def\aaps{\reff@jnl{A\&AS}}    % Astronomy and Astrophysics, Supplement
\def\mnras{\reff@jnl{MNRAS}}   % Monthly Notices of the RAS
\def\prd{\reff@jnl{Phys.\ Rev.\ D}}    % Physical Review D

 % N_side
   % N_pix
   % N_tau: one sided length of kernel

 % transpose
\newcommand{\beq}{\begin{equation}}
\newcommand{\eeq}{\end{equation}}

\newcommand{\be}{\begin{equation}}
\newcommand{\ee}{\end{equation}}
\newcommand{\bea}{\begin{eq}}
\newcommand{\eea}{\end{equation}}

{\begin{enumerate}\setlength{\itemsep}{0mm}}
{\end{enumerate}}
{\begin{enumerate}\setlength{\itemsep}{0mm}}%
{\end{enumerate}}

\newcommand{\bc}{\begin{center}}
\newcommand{\ec}{\end{center}}
\newcommand{\bi}{\begin{itemize}}
\newcommand{\ei}{\end{itemize}}
\newcommand{\ben}{\begin{enumerate}}
\newcommand{\een}{\end{enumerate}}

\usepackage{mathrsfs}
\usepackage{mathbbol}

\newfont{\gwpfont}{cmssq8 scaled 1000}

%%%%%%%%%%%%%%%%%%%%%%%%%%%%%%%%%%%%%%%%%%%%%%%%%%%%%%%%%%%%

\begin{document}

\title{\Planck\ 2015 results XXII.  A map of the thermal Sunyaev-Zeldovich effect}

%\author{N. Aghanim, M.~Arnaud, B.~Comis, M.~Douspis, J.M.~Diego, D.~Herranz, G.~Hurier, F.~Lacasa, M.~Lopez-Caniego, J.B. Melin, J.F. ~Macias-Perez, E.~Pointecouteau, G.~Pratt, M.~Remazeilles} 
%
\providecommand{\sorthelp}[1]{}

 %\input{AuthorList_P05b_SZ_PowerSpectrum_Bispectrum_Py_authors_and_institutes.tex}
%This author list corresponds to \title{Author list for A28\_ySZ\_map}
%Prepared by M. Lopez-Caniego (Marcos.Lopez.Caniego@sciops.esa.int), ESAC/ESA
%This version is from Tue Feb  3 16:19:52 2015 CET
%\subtitle{There are 201 co-authors in this list}
\author{\small
Planck Collaboration: N.~Aghanim\inst{59}
\and
M.~Arnaud\inst{72}
\and
M.~Ashdown\inst{68, 6}
\and
J.~Aumont\inst{59}
\and
C.~Baccigalupi\inst{82}
\and
A.~J.~Banday\inst{91, 10}
\and
R.~B.~Barreiro\inst{64}
\and
J.~G.~Bartlett\inst{1, 66}
\and
N.~Bartolo\inst{29, 65}
\and
E.~Battaner\inst{93, 94}
% ...
}
\institute{\small
APC, AstroParticule et Cosmologie, Universit\'{e} Paris Diderot, CNRS/IN2P3, CEA/lrfu, Observatoire de Paris, Sorbonne Paris Cit\'{e}, 10, rue Alice Domon et L\'{e}onie Duquet, 75205 Paris Cedex 13, France\goodbreak
\and
Aalto University Mets\"{a}hovi Radio Observatory and Dept of Radio Science and Engineering, P.O. Box 13000, FI-00076 AALTO, Finland\goodbreak
\and
African Institute for Mathematical Sciences, 6-8 Melrose Road, Muizenberg, Cape Town, South Africa\goodbreak
\and
Agenzia Spaziale Italiana Science Data Center, Via del Politecnico snc, 00133, Roma, Italy\goodbreak
\and
Aix Marseille Universit\'{e}, CNRS, LAM (Laboratoire d'Astrophysique de Marseille) UMR 7326, 13388, Marseille, France\goodbreak
\and
Astrophysics Group, Cavendish Laboratory, University of Cambridge, J J Thomson Avenue, Cambridge CB3 0HE, U.K.\goodbreak
\and
Astrophysics \& Cosmology Research Unit, School of Mathematics, Statistics \& Computer Science, University of KwaZulu-Natal, Westville Campus, Private Bag X54001, Durban 4000, South Africa\goodbreak
\and
Atacama Large Millimeter/submillimeter Array, ALMA Santiago Central Offices, Alonso de Cordova 3107, Vitacura, Casilla 763 0355, Santiago, Chile\goodbreak
\and
CITA, University of Toronto, 60 St. George St., Toronto, ON M5S 3H8, Canada\goodbreak
\and
CNRS, IRAP, 9 Av. colonel Roche, BP 44346, F-31028 Toulouse cedex 4, France\goodbreak
\and
CRANN, Trinity College, Dublin, Ireland\goodbreak
\and
California Institute of Technology, Pasadena, California, U.S.A.\goodbreak
\and
Centre for Theoretical Cosmology, DAMTP, University of Cambridge, Wilberforce Road, Cambridge CB3 0WA, U.K.\goodbreak
\and
Centro de Estudios de F\'{i}sica del Cosmos de Arag\'{o}n (CEFCA), Plaza San Juan, 1, planta 2, E-44001, Teruel, Spain\goodbreak
\and
Computational Cosmology Center, Lawrence Berkeley National Laboratory, Berkeley, California, U.S.A.\goodbreak
\and
DSM/Irfu/SPP, CEA-Saclay, F-91191 Gif-sur-Yvette Cedex, France\goodbreak
\and
DTU Space, National Space Institute, Technical University of Denmark, Elektrovej 327, DK-2800 Kgs. Lyngby, Denmark\goodbreak
\and
D\'{e}partement de Physique Th\'{e}orique, Universit\'{e} de Gen\`{e}ve, 24, Quai E. Ansermet,1211 Gen\`{e}ve 4, Switzerland\goodbreak
\and
Departamento de F\'{\i}sica, Universidad de Oviedo, Avda. Calvo Sotelo s/n, Oviedo, Spain\goodbreak
\and
Department of Astrophysics/IMAPP, Radboud University Nijmegen, P.O. Box 9010, 6500 GL Nijmegen, The Netherlands\goodbreak
\and
Department of Physics \& Astronomy, University of British Columbia, 6224 Agricultural Road, Vancouver, British Columbia, Canada\goodbreak
\and
Department of Physics and Astronomy, Dana and David Dornsife College of Letter, Arts and Sciences, University of Southern California, Los Angeles, CA 90089, U.S.A.\goodbreak
\and
Department of Physics and Astronomy, University College London, London WC1E 6BT, U.K.\goodbreak
\and
Department of Physics, Florida State University, Keen Physics Building, 77 Chieftan Way, Tallahassee, Florida, U.S.A.\goodbreak
\and
Department of Physics, Gustaf H\"{a}llstr\"{o}min katu 2a, University of Helsinki, Helsinki, Finland\goodbreak
\and
Department of Physics, Princeton University, Princeton, New Jersey, U.S.A.\goodbreak
\and
Department of Physics, University of California, Santa Barbara, California, U.S.A.\goodbreak
\and
Department of Physics, University of Illinois at Urbana-Champaign, 1110 West Green Street, Urbana, Illinois, U.S.A.\goodbreak
\and
Dipartimento di Fisica e Astronomia G. Galilei, Universit\`{a} degli Studi di Padova, via Marzolo 8, 35131 Padova, Italy\goodbreak
\and
Dipartimento di Fisica e Scienze della Terra, Universit\`{a} di Ferrara, Via Saragat 1, 44122 Ferrara, Italy\goodbreak
\and
Dipartimento di Fisica, Universit\`{a} La Sapienza, P. le A. Moro 2, Roma, Italy\goodbreak
\and
Dipartimento di Fisica, Universit\`{a} degli Studi di Milano, Via Celoria, 16, Milano, Italy\goodbreak
\and
Dipartimento di Fisica, Universit\`{a} degli Studi di Trieste, via A. Valerio 2, Trieste, Italy\goodbreak
\and
Dipartimento di Matematica, Universit\`{a} di Roma Tor Vergata, Via della Ricerca Scientifica, 1, Roma, Italy\goodbreak
\and
Discovery Center, Niels Bohr Institute, Blegdamsvej 17, Copenhagen, Denmark\goodbreak
\and
Dpto. Astrof\'{i}sica, Universidad de La Laguna (ULL), E-38206 La Laguna, Tenerife, Spain\goodbreak
\and
European Southern Observatory, ESO Vitacura, Alonso de Cordova 3107, Vitacura, Casilla 19001, Santiago, Chile\goodbreak
\and
European Space Agency, ESAC, Planck Science Office, Camino bajo del Castillo, s/n, Urbanizaci\'{o}n Villafranca del Castillo, Villanueva de la Ca\~{n}ada, Madrid, Spain\goodbreak
\and
European Space Agency, ESTEC, Keplerlaan 1, 2201 AZ Noordwijk, The Netherlands\goodbreak
\and
Facolt\`{a} di Ingegneria, Universit\`{a} degli Studi e-Campus, Via Isimbardi 10, Novedrate (CO), 22060, Italy\goodbreak
\and
Gran Sasso Science Institute, INFN, viale F. Crispi 7, 67100 L'Aquila, Italy\goodbreak
\and
HGSFP and University of Heidelberg, Theoretical Physics Department, Philosophenweg 16, 69120, Heidelberg, Germany\goodbreak
\and
Helsinki Institute of Physics, Gustaf H\"{a}llstr\"{o}min katu 2, University of Helsinki, Helsinki, Finland\goodbreak
\and
ICTP South American Institute for Fundamental Research, Instituto de F\'{\i}sica Te\'{o}rica, Universidade Estadual Paulista, S\~{a}o Paulo, Brazil\goodbreak
\and
INAF - Osservatorio Astronomico di Padova, Vicolo dell'Osservatorio 5, Padova, Italy\goodbreak
\and
INAF - Osservatorio Astronomico di Roma, via di Frascati 33, Monte Porzio Catone, Italy\goodbreak
\and
INAF - Osservatorio Astronomico di Trieste, Via G.B. Tiepolo 11, Trieste, Italy\goodbreak
\and
INAF/IASF Bologna, Via Gobetti 101, Bologna, Italy\goodbreak
\and
INAF/IASF Milano, Via E. Bassini 15, Milano, Italy\goodbreak
\and
INFN, Sezione di Bologna, Via Irnerio 46, I-40126, Bologna, Italy\goodbreak
\and
INFN, Sezione di Roma 1, Universit\`{a} di Roma Sapienza, Piazzale Aldo Moro 2, 00185, Roma, Italy\goodbreak
\and
INFN, Sezione di Roma 2, Universit\`{a} di Roma Tor Vergata, Via della Ricerca Scientifica, 1, Roma, Italy\goodbreak
\and
INFN/National Institute for Nuclear Physics, Via Valerio 2, I-34127 Trieste, Italy\goodbreak
\and
IPAG: Institut de Plan\'{e}tologie et d'Astrophysique de Grenoble, Universit\'{e} Grenoble Alpes, IPAG, F-38000 Grenoble, France, CNRS, IPAG, F-38000 Grenoble, France\goodbreak
\and
Imperial College London, Astrophysics group, Blackett Laboratory, Prince Consort Road, London, SW7 2AZ, U.K.\goodbreak
\and
Infrared Processing and Analysis Center, California Institute of Technology, Pasadena, CA 91125, U.S.A.\goodbreak
\and
Institut N\'{e}el, CNRS, Universit\'{e} Joseph Fourier Grenoble I, 25 rue des Martyrs, Grenoble, France\goodbreak
\and
Institut Universitaire de France, 103, bd Saint-Michel, 75005, Paris, France\goodbreak
\and
Institut d'Astrophysique Spatiale, CNRS (UMR8617) Universit\'{e} Paris-Sud 11, B\^{a}timent 121, Orsay, France\goodbreak
\and
Institut d'Astrophysique de Paris, CNRS (UMR7095), 98 bis Boulevard Arago, F-75014, Paris, France\goodbreak
\and
Institute of Astronomy, University of Cambridge, Madingley Road, Cambridge CB3 0HA, U.K.\goodbreak
\and
Institute of Theoretical Astrophysics, University of Oslo, Blindern, Oslo, Norway\goodbreak
\and
Instituto de Astrof\'{\i}sica de Canarias, C/V\'{\i}a L\'{a}ctea s/n, La Laguna, Tenerife, Spain\goodbreak
\and
Instituto de F\'{\i}sica de Cantabria (CSIC-Universidad de Cantabria), Avda. de los Castros s/n, Santander, Spain\goodbreak
\and
Istituto Nazionale di Fisica Nucleare, Sezione di Padova, via Marzolo 8, I-35131 Padova, Italy\goodbreak
\and
Jet Propulsion Laboratory, California Institute of Technology, 4800 Oak Grove Drive, Pasadena, California, U.S.A.\goodbreak
\and
Jodrell Bank Centre for Astrophysics, Alan Turing Building, School of Physics and Astronomy, The University of Manchester, Oxford Road, Manchester, M13 9PL, U.K.\goodbreak
\and
Kavli Institute for Cosmology Cambridge, Madingley Road, Cambridge, CB3 0HA, U.K.\goodbreak
\and
LAL, Universit\'{e} Paris-Sud, CNRS/IN2P3, Orsay, France\goodbreak
\and
LAPTh, Univ. de Savoie, CNRS, B.P.110, Annecy-le-Vieux F-74941, France\goodbreak
\and
LERMA, CNRS, Observatoire de Paris, 61 Avenue de l'Observatoire, Paris, France\goodbreak
\and
Laboratoire AIM, IRFU/Service d'Astrophysique - CEA/DSM - CNRS - Universit\'{e} Paris Diderot, B\^{a}t. 709, CEA-Saclay, F-91191 Gif-sur-Yvette Cedex, France\goodbreak
\and
Laboratoire Traitement et Communication de l'Information, CNRS (UMR 5141) and T\'{e}l\'{e}com ParisTech, 46 rue Barrault F-75634 Paris Cedex 13, France\goodbreak
\and
Laboratoire de Physique Subatomique et Cosmologie, Universit\'{e} Grenoble-Alpes, CNRS/IN2P3, 53, rue des Martyrs, 38026 Grenoble Cedex, France\goodbreak
\and
Laboratoire de Physique Th\'{e}orique, Universit\'{e} Paris-Sud 11 \& CNRS, B\^{a}timent 210, 91405 Orsay, France\goodbreak
\and
Lebedev Physical Institute of the Russian Academy of Sciences, Astro Space Centre, 84/32 Profsoyuznaya st., Moscow, GSP-7, 117997, Russia\goodbreak
\and
Max-Planck-Institut f\"{u}r Astrophysik, Karl-Schwarzschild-Str. 1, 85741 Garching, Germany\goodbreak
\and
National University of Ireland, Department of Experimental Physics, Maynooth, Co. Kildare, Ireland\goodbreak
\and
Niels Bohr Institute, Blegdamsvej 17, Copenhagen, Denmark\goodbreak
\and
Optical Science Laboratory, University College London, Gower Street, London, U.K.\goodbreak
\and
SB-ITP-LPPC, EPFL, CH-1015, Lausanne, Switzerland\goodbreak
\and
SISSA, Astrophysics Sector, via Bonomea 265, 34136, Trieste, Italy\goodbreak
\and
School of Physics and Astronomy, Cardiff University, Queens Buildings, The Parade, Cardiff, CF24 3AA, U.K.\goodbreak
\and
Sorbonne Universit\'{e}-UPMC, UMR7095, Institut d'Astrophysique de Paris, 98 bis Boulevard Arago, F-75014, Paris, France\goodbreak
\and
Space Research Institute (IKI), Russian Academy of Sciences, Profsoyuznaya Str, 84/32, Moscow, 117997, Russia\goodbreak
\and
Space Sciences Laboratory, University of California, Berkeley, California, U.S.A.\goodbreak
\and
Special Astrophysical Observatory, Russian Academy of Sciences, Nizhnij Arkhyz, Zelenchukskiy region, Karachai-Cherkessian Republic, 369167, Russia\goodbreak
\and
Sub-Department of Astrophysics, University of Oxford, Keble Road, Oxford OX1 3RH, U.K.\goodbreak
\and
Theory Division, PH-TH, CERN, CH-1211, Geneva 23, Switzerland\goodbreak
\and
UPMC Univ Paris 06, UMR7095, 98 bis Boulevard Arago, F-75014, Paris, France\goodbreak
\and
Universit\'{e} de Toulouse, UPS-OMP, IRAP, F-31028 Toulouse cedex 4, France\goodbreak
\and
University Observatory, Ludwig Maximilian University of Munich, Scheinerstrasse 1, 81679 Munich, Germany\goodbreak
\and
University of Granada, Departamento de F\'{\i}sica Te\'{o}rica y del Cosmos, Facultad de Ciencias, Granada, Spain\goodbreak
\and
University of Granada, Instituto Carlos I de F\'{\i}sica Te\'{o}rica y Computacional, Granada, Spain\goodbreak
% ...
}

 \abstract {
 We have constructed all-sky  Compton parameters maps ($y$\/-maps) of the thermal Sunyaev-Zeldovich
 (tSZ) effect by applying specifically tailored component separation
 algorithms to the 30 to 857\,GHz frequency channel maps from the \Planck\
 satellite survey. These reconstructed $y$\/-maps are delivered as part of the \Planck\ 2015 release. 
 The $y$\/-maps are characterized in terms of noise properties and
 residual foreground contamination, mainly thermal dust emission at large angular scales and
 CIB and extragalactic point sources at small angular scales. Specific masks are defined to minimize
 foreground residuals and systematics. Using these masks, we compute the $y$\/-map angular power spectrum and higher order statistics.
 From these we conclude that the $y$\/-map is dominated by tSZ signal in the multipole range, 
 $20 <  \ell  < 600$.  We compare the measured tSZ power spectrum and higher order statistics to various physically motivated
 models and discuss the implications of our results in terms of cluster physics and cosmology.
}


\authorrunning{Planck Collaboration}
\titlerunning{A map of the thermal Sunyaev-Zeldovich effect}

\maketitle
%\allearlypapers
%\alltwentyfifteenresultspapers
%\tableofcontents   
\clearpage
\section{Introduction}
\label{sec:introduction}

This paper, one of a set associated with the 2015 release of data from the
\Planck
\footnote{\Planck\ (\url{http://www.esa.int/Planck}) is a project of the European Space Agency  (ESA) with instruments provided by two scientific consortia funded by ESA member states and led by Principal Investigators from France and Italy, telescope reflectors provided through a collaboration between ESA and a scientific consortium led and funded by Denmark, and additional contributions from NASA (USA).} mission, describes the
\Planck\ Compton parameter ($y$) map, which is part of the \Planck\ 2015 data release.
%and the determination of
%its angular power spectrum and high-order statistics.

The thermal Sunyaev-Zeldovich (tSZ) effect \citep{SZ} is produced by the
inverse Compton scattering of cosmic microwave background (CMB)
photons by hot electrons along the line of sight and in particular in clusters of galaxies. 
The tSZ effect has proved to be a major tool to study the
physics of clusters of galaxies as well as structure formation in the
Universe.  Catalogues of clusters of galaxies selected via the tSZ
effect have become available in the last few years, including for example those from the
\Planck\ satellite \citep{planck2011-5.1a,planck2013-p05a}, the
Atacama Cosmology Telescope
\citep[ACT,][]{Hasselfield:2013p2303} and the
South Pole Telescope \citep[SPT,][]{Reichardt:2013p2252, Bleem_2014}.  These
catalogues and their associated sky surveys have been used to study
the physics of clusters of galaxies
\citep{planck2011-5.2c,planck2011-5.2b,planck2011-5.2a} and their cosmological implications
\citep{planck2013-p15,Benson:2013p2263,Das:2013p2167,Wilson:2012p2102,Mak:2012p2066}.

The study of cluster number counts and their evolution with
redshift using tSZ selected catalogues is
an important cosmological test \citep{car02,Dunkley:2013p2181,Benson:2013p2263,planck2013-p15,Hou_2014}.
This study is well complemented by the measurement of the tSZ effect power spectrum \citep{Komatsu:2002p1799}, for which 
no explicit measurement of cluster masses is required.   
Low mass clusters, thus fainter in tSZ, which can not be detected individually,
also contribute statistically to the measured signal \citep{Battaglia2010,Shaw2010}.
Another complementary approach \citep[as pointed out in][]{RubinoMartin:2003p1790}
is to study the higher order statistics of the tSZ signal, and in particular the skewness or, equivalently, the
bispectrum. The bispectrum of the tSZ
effect signal is dominated by massive clusters at intermediate
redshifts \citep{Bhattacharya:2012p2458}, for which high-precision X-ray observations exist. 
This contrasts with the power spectrum, for which most of the signal comes from
the lower mass and higher redshift groups and clusters
\citep[e.g.,][]{Trac:2011p1795}. Thus, theoretical uncertainties in the
tSZ bispectrum are expected to be significantly smaller than those
in the estimation of the tSZ power spectrum. Therefore, combined measurements of the power spectrum
and the bispectrum can be used to distinguish the contribution to
the power spectrum from different cluster masses and redshift
ranges. However, contamination from point
sources \citep{RubinoMartin:2003p1790,Taburet2010a} and other
foregrounds \citep{planck2013-p15} needs to be dealt with carefully.

As shown in the \Planck\ 2013 results \citep{planck2013-p15},
the all-sky coverage and unprecedented wide
frequency range of \Planck\ allowed us to
produce all-sky tSZ Compton parameter maps, also referred to as $y$\/-maps.
From this map an accurate measurement of the tSZ power spectrum at intermediate and large
angular scales, for which the tSZ fluctuations are almost insensitive to
the cluster core physics, can be obtained. Furthermore, the expected non-Gaussianity properties
of the tSZ signal can be studied using higher order statistical estimators, such as the skewness and the bispectrum. 
The $y$-map can be also used to extract the tSZ signal on regions centered at cluster positions and
in particular to perform stacking analysis. Finally,  this map may also be cross correlated with
other cosmological probes \citep[see for example][]{2014arXiv1404.4808M,2014PhRvD..89b3508V, Hill_2014} as a consistency test. \\

In this paper we construct revised tSZ all-sky maps from the individual \Planck\ frequency maps.
With respect to the \Planck\ Compton parameter map in the \Planck\ 2013 results
\citep{planck2013-p15} we have extended the analysis to the full mission data set and performed an in depth
characterization of its statistical properties. This extended analysis allows us to deliver this map as part of the \Planck\ 2015 data release. 
The paper is structured as
follows. Section~\ref{datasimu} describes the \Planck\ data used to
compute the tSZ all-sky maps and the simulations used to characterize
it.  Section~\ref{sec:allskymap} presents the reconstruction of the \Planck\ all-sky Compton
parameter map. In Sect.~\ref{sec:pixelanalysis} we discuss the validation of the reconstructed $y$\/-map in 
the pixel domain including signal and noise characterization. Section~\ref{sec:powerspec} describes the power
spectrum analysis.  Cross-checks using higher order statistics are
presented in Sect.~\ref{sec:higorderstat}. The cosmological
interpretation of the results is discussed in
Sect.~\ref{sec:cosmo}, and we finally present our conclusions in
Sect.~\ref{conclusions}.

\begin{figure}
  \begin{center}
   \includegraphics[width=0.9\columnwidth]{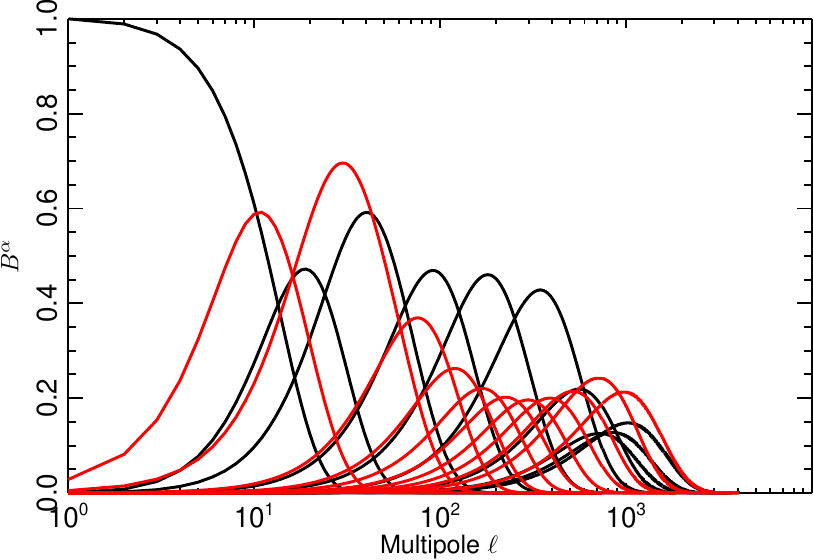}
  \end{center}
\caption{Window functions corresponding to the spectral localisation of the {\tt NILC} and {\tt MILCA} algorithms. For {\tt NILC}  (black) there are10 Gaussian window functions defining 10 needlet scales. {\tt MILCA} (red) uses 11 Gaussian overlapping windows.}
\label{Fig:bands}
\end{figure}

\section{Data and  simulations}
\label{datasimu}

\begin{table}[tmb]
\begingroup
\newdimen\tblskip \tblskip=5pt
\caption{\label{table:summary} Conversion factors for tSZ Compton parameter
$y$ to CMB temperature units and the FWHM of the beam of the
\Planck\ channel maps.}                          
\nointerlineskip
\vskip -4mm
\footnotesize
\setbox\tablebox=\vbox{
   \newdimen\digitwidth 
   \setbox0=\hbox{\rm 0} 
   \digitwidth=\wd0 
   \catcode`*=\active 
   \def*{\kern\digitwidth}
   \newdimen\dpwidth 
   \setbox0=\hbox{.} 
   \dpwidth=\wd0 
   \catcode`!=\active 
   \def!{\kern\dpwidth}
\halign{\hbox to 1.5cm{#\leaderfil}\tabskip 1em&
     \hfil#\hfil \tabskip 1em&
     \hfil#\hfil \tabskip 0em \cr
\noalign{\doubleline}
\omit  Frequency &  $ T_{\mathrm{CMB}}  \ g (\nu)$ & FWHM\cr
\omit\hfil$[$GHz$]$\hfil& $[$K$_{\mathrm{CMB}}]$ & [arcmin]\cr
\noalign{\vskip 3pt\hrule\vskip 5pt}
100& $-4.031$& 9.66\cr
143& $-2.785$& 7.27\cr
217& $*0.187$& 5.01\cr
353& $*6.205$& 4.86\cr
545& $14.455$& 4.84\cr
857& $26.335$& 4.63\cr
\noalign{\vskip 3pt\hrule\vskip 5pt}
}
}
\endPlancktable 
\endgroup
\end{table}

\begin{figure*}
\begin{center}
\includegraphics[trim=0.3cm 1cm 1.2cm 5cm, clip=true,width=1.71\columnwidth]{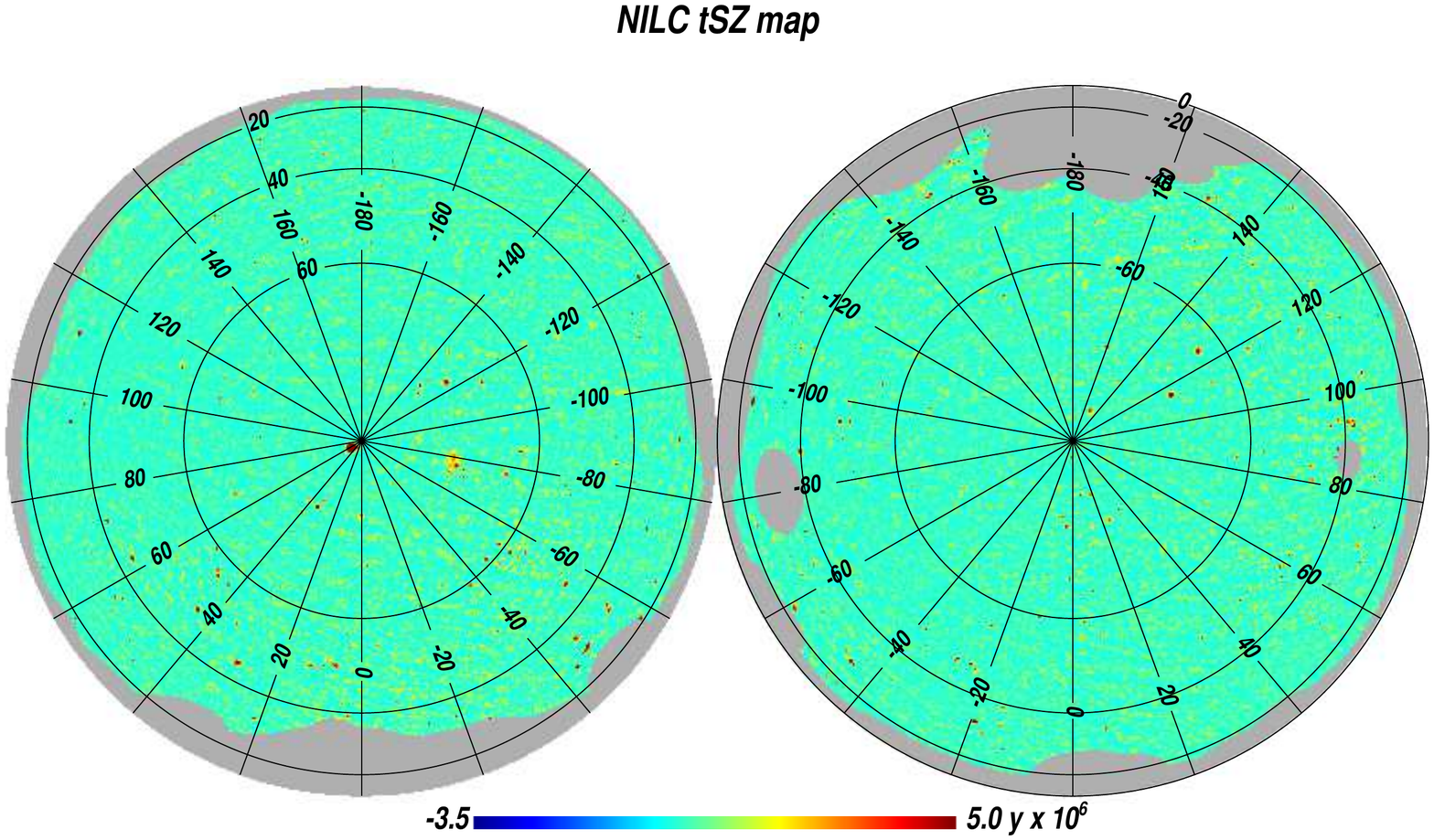}
\includegraphics[trim=0.3cm 1cm 1.2cm 5cm, clip=true,width=1.71\columnwidth]{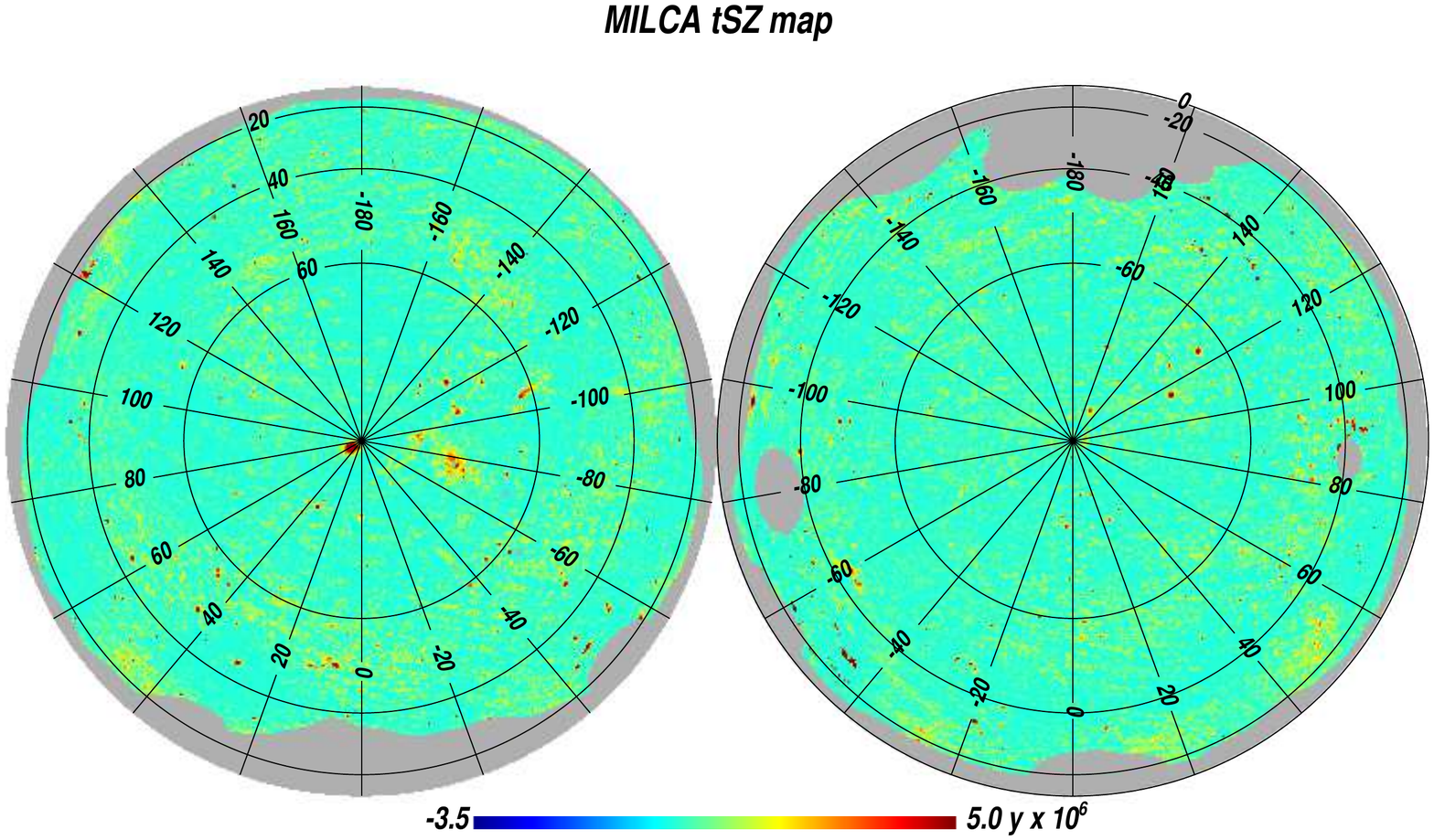}
\end{center}
\caption{Reconstructed \Planck\ all-sky Compton parameter maps for {\tt NILC}
(\textit{top}) and {\tt MILCA} (\textit{bottom}) in orthographic projections.
The apparent difference in contrast observed between the {\tt NILC} and
{\tt MILCA} maps comes from differences in the residual foreground contamination and from the differences in the
filtering applied for display purposes to the original Compton parameter maps. For the MILCA method filtering out low multipoles 
reduces significantly the level of foreground emission in the final $y$\/-map. The wavelet basis used in the NILC method was tailored for tSZ extraction. 
For details see \citet{planck2014-a28}.
\label{fig:planck_y_map}}
\end{figure*}

\begin{figure*}
\begin{center}
\includegraphics[trim=2cm 3.1cm 2cm 2cm, clip=true,width=0.87\columnwidth]{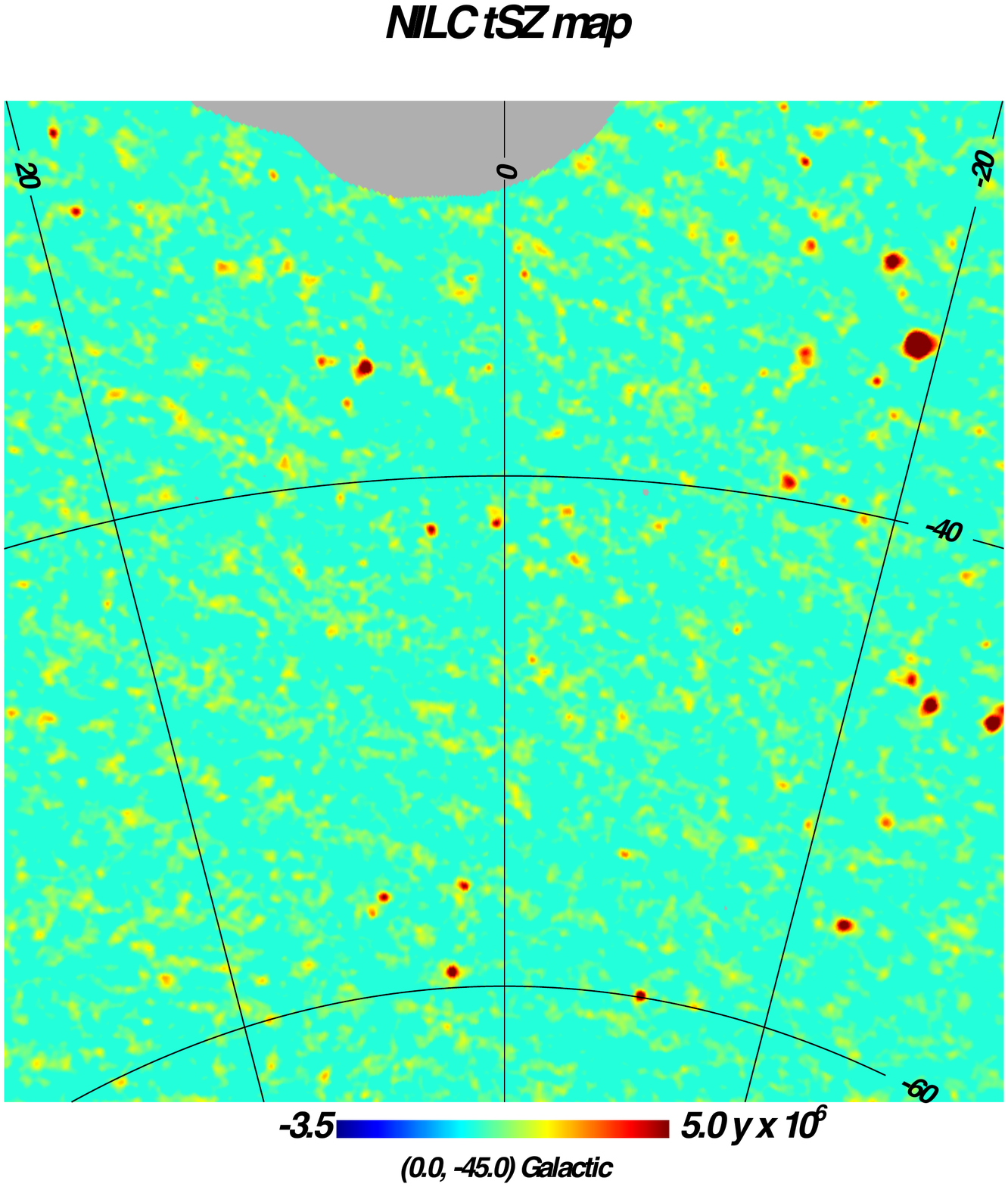}
\includegraphics[trim=2cm 3.1cm 2cm 2cm, clip=true,width=0.87\columnwidth]{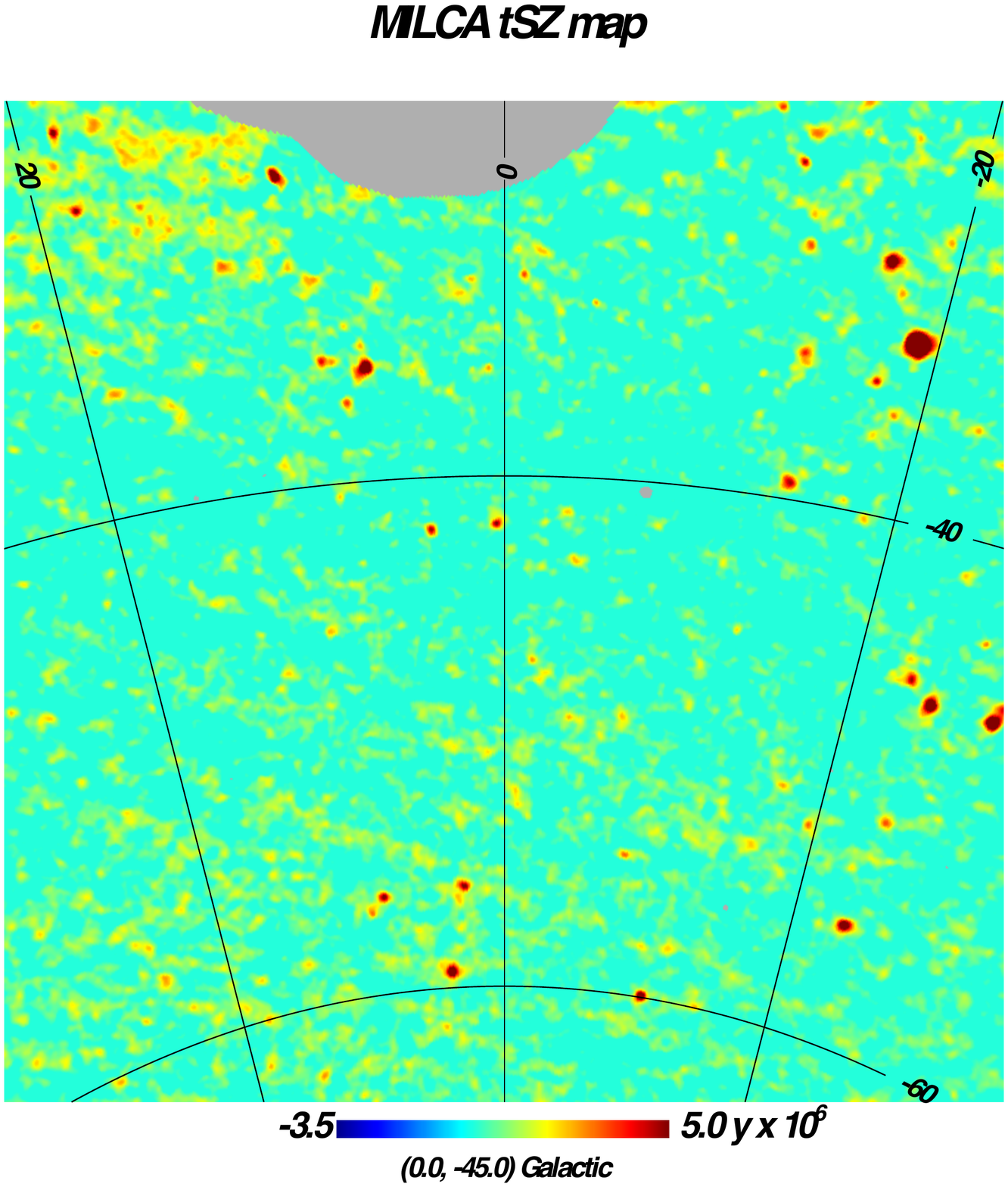}
\end{center}
\caption{A small region of the reconstructed \Planck\ all-sky Compton parameter maps for {\tt NILC} (left) and {\tt MILCA} (right) 
at intermediate Galactic latitudes in the southern sky centered at $(0,-45)$ degrees in Galactic coordinates.
\label{fig:planck_y_map_zoom}}
\end{figure*}

\subsection{The \Planck\ data}
\label{subsec:planckdata}

This paper is based on the \Planck's full mission,
corresponding to five and eight full-sky surveys for the High Frequency Instrument (HFI)
and Low Frequency Instrument (LFI) data respectively. 
\\
\\
The \Planck\ channel maps are provided in {\tt HEALPix} \citep{healpix}
pixelization scheme at $N_{\mathrm{side}}=2048$. A noise map is
associated with each channel map. This map is obtained from the half difference of maps 
made from the first and second half of each stable pointing period (also called ring).
In the half-difference maps the astrophysical emission cancels out, which makes them a good representation of the statistical instrumental noise. 
These half-difference maps are used to estimate the noise in the final Compton parameter map. In addition, survey maps, which are also available for each channel, will be used
to estimate possible residual systematic effects in the $y$-map.

For the purpose of this paper we approximate the \Planck\ effective beams by 
circular Gaussians, the FWHM estimates of which are given in Table~\ref{table:summary} for each frequency channel.
%Although tests have been performed using both LFI and HFI channel
%maps, the work presented here is based  on HFI data.

\subsection{Simulations}
\label{subsec:simulations}
We also use simulated \Planck\ frequency maps
obtained from the  Full Focal Plane (FFP) simulations \citep{planck2014-a14}, which are described in the \Planck\ Explanatory
Supplement \citep{planck2013-p28}.

These simulated maps include the most relevant sky components at microwave and
millimetre frequencies, based on foregrounds from the \Planck\ Sky Model
\citep[PSM,][]{Delabrouille2012}:
CMB, thermal SZ effect, diffuse Galactic
emissions (synchrotron, free-free, thermal and spinning dust and CO), radio
and infrared point sources, and the clustered CIB.
The tSZ signal was constructed using hydrodynamical simulations of clusters of galaxies
up to redshift 0.3. For higher redshifts pressure profile-based simulations
of individual clusters of galaxies randomly drawn on the sky have been added.  The
noise in the maps was obtained from realizations of Gaussian random 
noise in the time domain and therefore accounts for noise
inhomogeneities in the maps.

The simulation set also includes Monte Carlo noise-only realizations for each \Planck\ channel
map. These will also be used to estimate the noise properties in the final $y$-map.

%%%%%%%%%%%%%%%%%%%%%%%%%%%%%%%%%%%%%%%%%%%%%%%%%%%%%

%%%%%%%%%%%%%%%%%%%%%%%%%%%%%%%%%%%%%%%%%%%%%%%%%%%%%%

\section{Reconstruction of the all-sky tSZ maps}
\label{sec:allskymap}

The thermal SZ Compton parameter \citep{SZ} in a given direction, $\vec{n}$, can be expressed as
\begin{equation}
y (\vec{n}) = \int n_{\mathrm{e}} \frac{k_{\mathrm{B}}
 T_{\mathrm{e}}}{m_{\mathrm{e}} c^{2} } \sigma_{\mathrm{T}} \  \mathrm{d}s,
\end{equation}
where $k_{\mathrm{B}}$ is the Boltzmann contsant, $m_{\mathrm{e}}$ the electron
mass, $\sigma_{\mathrm{T}}$ the Thomson cross-section,
d$s$ the distance along the line of sight, $\vec{n}$, and
$n_{\mathrm{e}}$ and $T_{\mathrm{e}}$ are the electron number density and
temperature.

In CMB temperature units the tSZ effect contribution
to the \Planck\ maps for a given observational frequency $\nu$ is
given by
\begin{equation}
\frac{\Delta T}{T_{\mathrm{CMB}} }= g(\nu) \ y.
\end{equation}
Neglecting relativistic corrections, 
$g(\nu)~=~x \  \coth(x/2) - 4 $,
with $ x~=~h \nu/(k_{\mathrm{B}} T_{\mathrm{CMB}})$.  Table~\ref{table:summary}
presents the conversion factors for Compton parameter to CMB temperature,
K$_{\mathrm{CMB}}$, for each frequency channel, after integrating over
the instrumental bandpass \citep{planck2013-p03d}.

\begin{figure*}
\begin{center}
\setlength{\unitlength}{\columnwidth}
\begin{picture}(1.5,1.45)
\put(-0.2,0.7){\includegraphics[trim=1cm 2cm 0cm 10cm, clip=true,width=0.83\columnwidth]{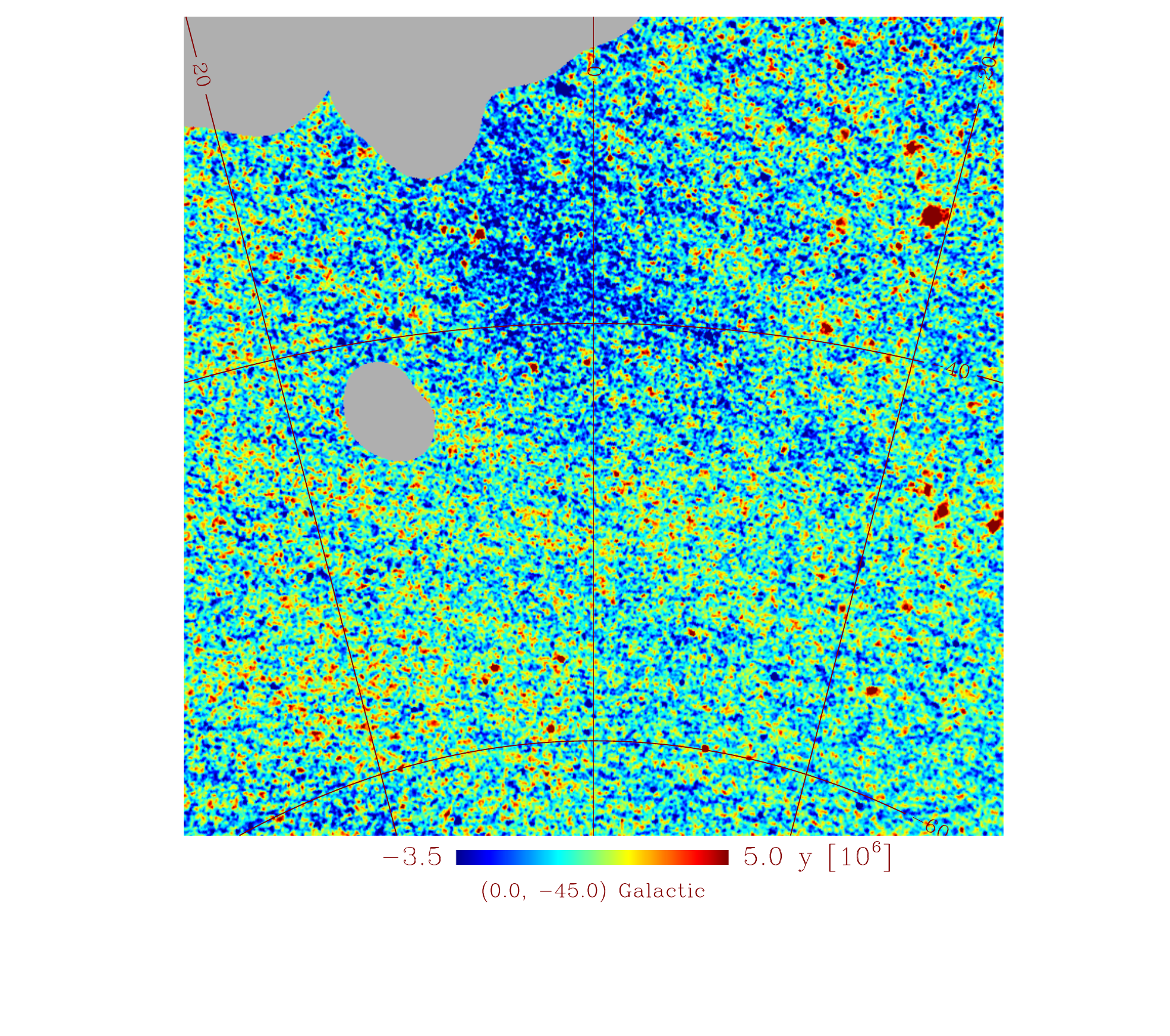}}
\put(0.4,0.7){\includegraphics[trim=1cm 2cm 0cm 10cm, clip=true,width=0.83\columnwidth]{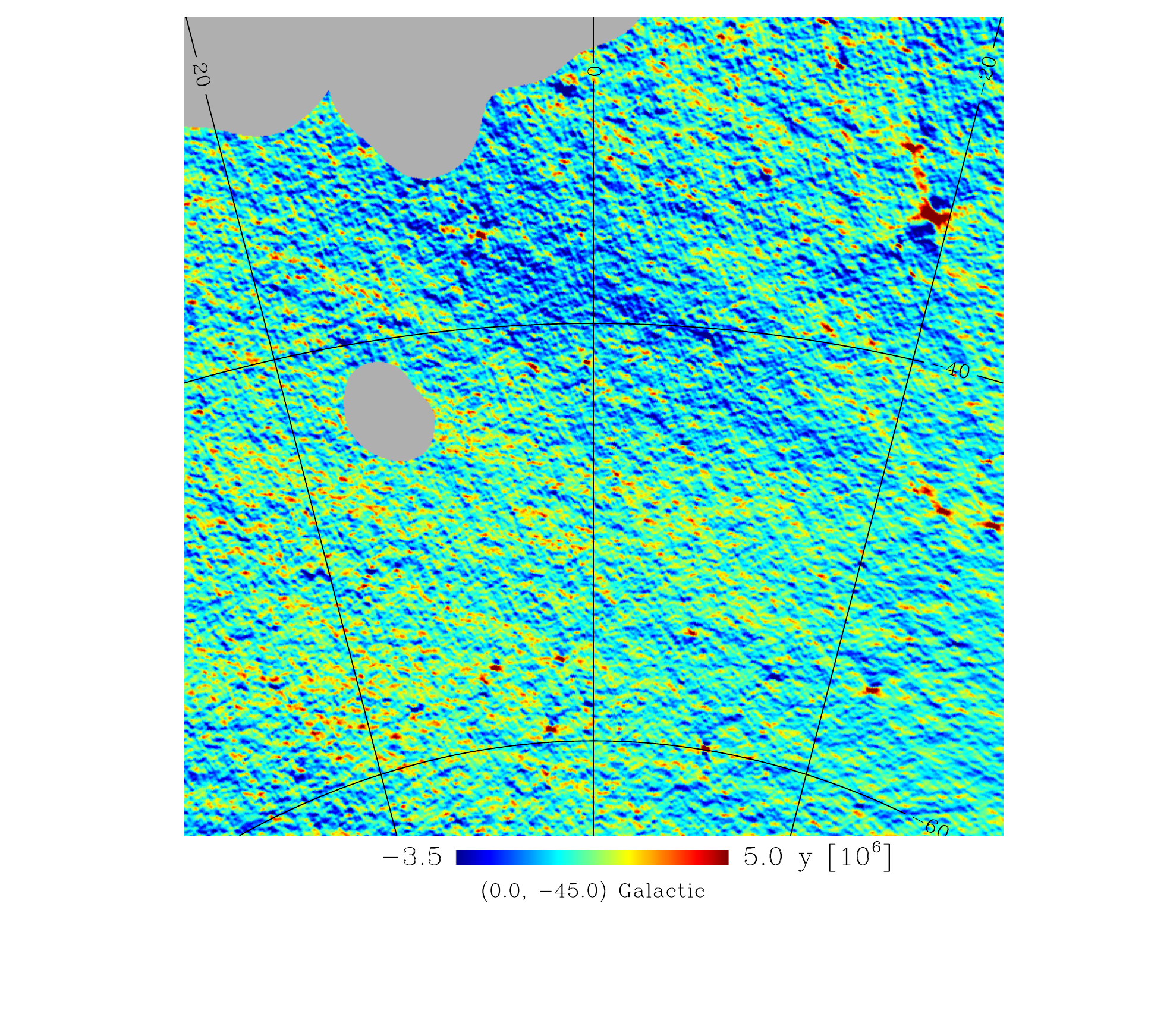}}
\put(1,0.7){\includegraphics[trim=1cm 2cm 0cm 10cm, clip=true,width=0.83\columnwidth]{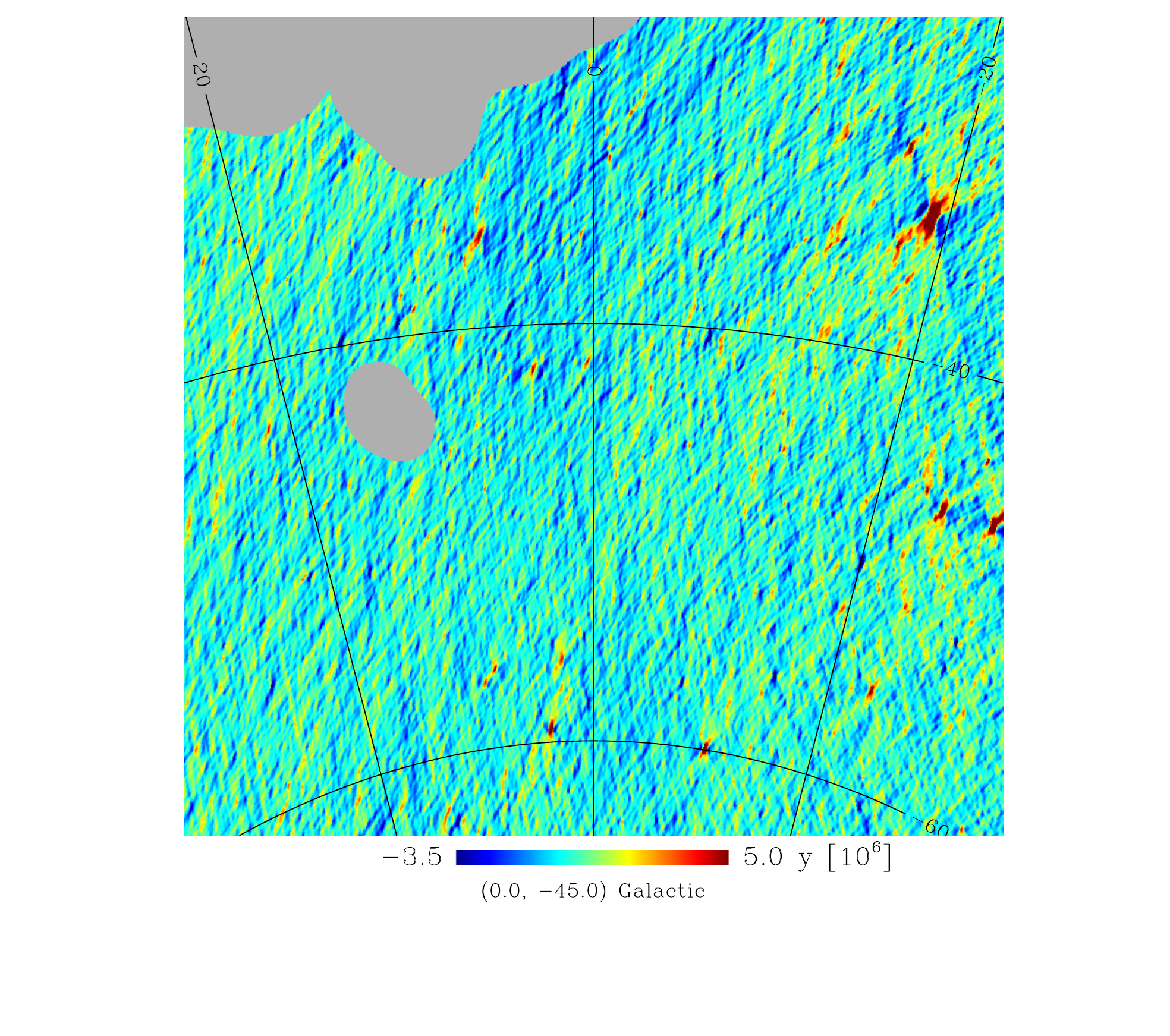}}
\put(-0.2,0){\includegraphics[trim=1cm 2cm 0cm 10cm, clip=true,width=0.83\columnwidth]{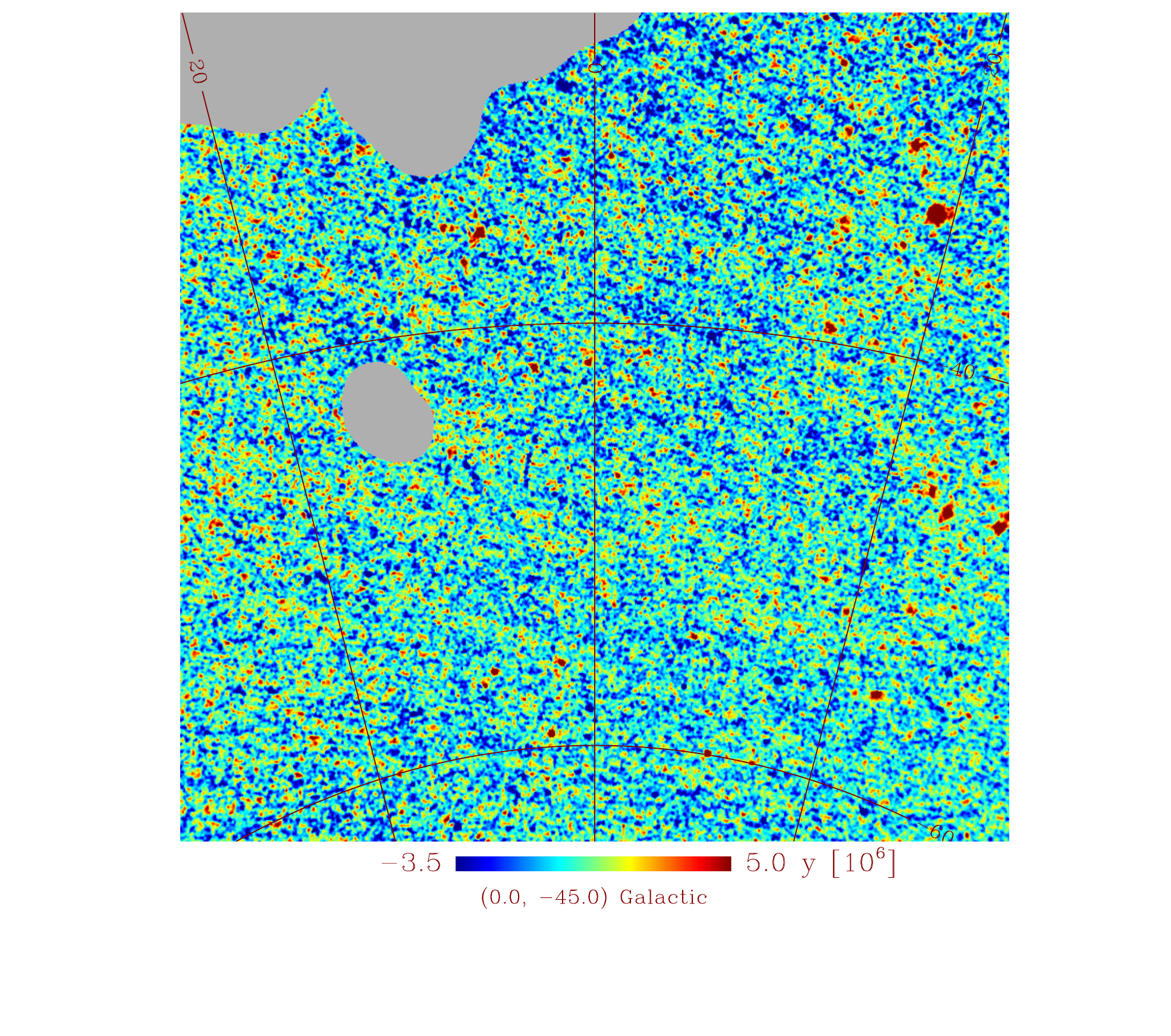}}
\put(0.4,0){\includegraphics[trim=1cm 2cm 0cm 10cm, clip=true,width=0.83\columnwidth]{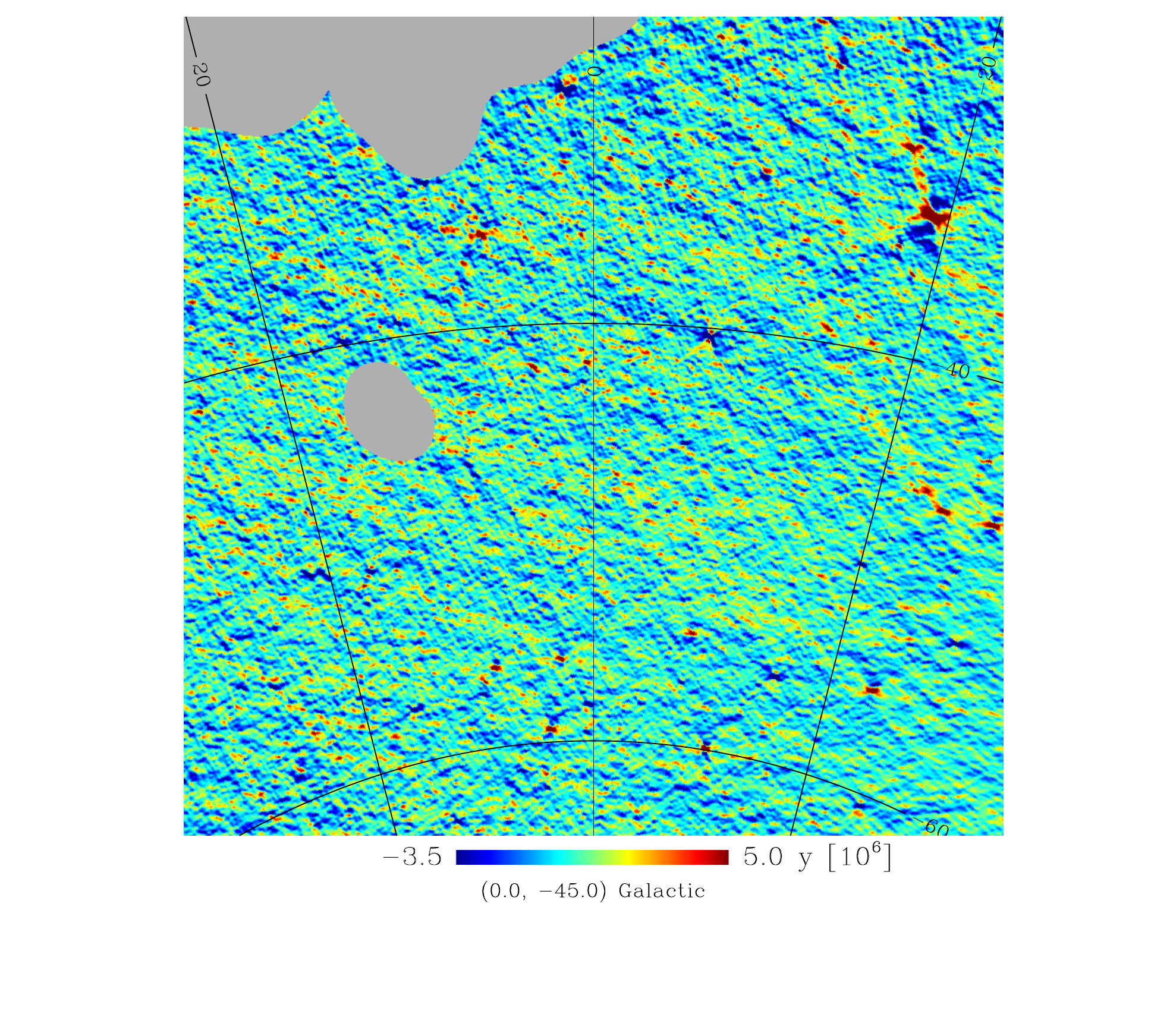}}
\put(1,0){\includegraphics[trim=1cm 2cm 0cm 10cm, clip=true,width=0.83\columnwidth]{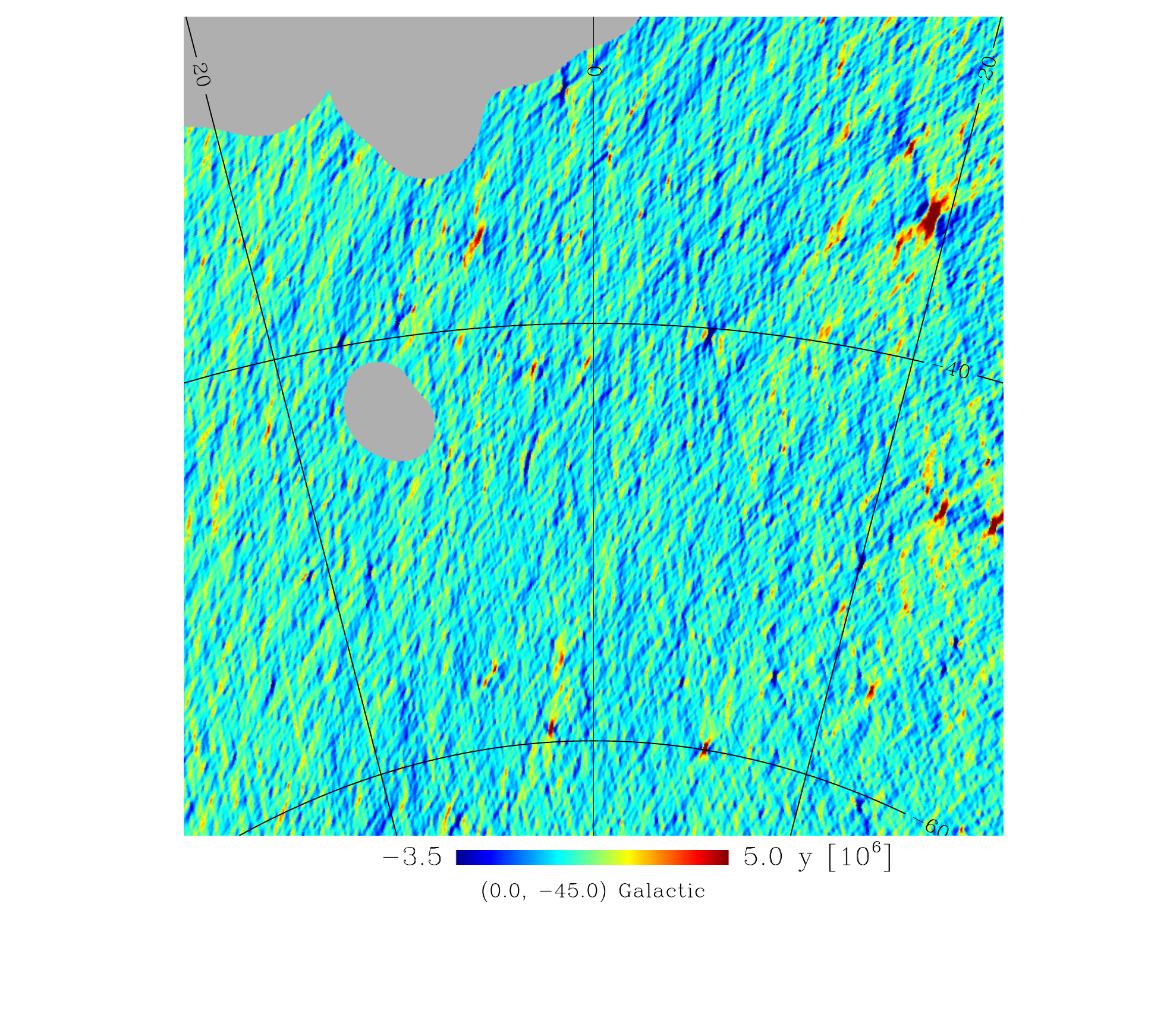}}
\end{picture}
\caption{In and cross scan contributions in the {\tt NILC} (top line)  and {\tt MILCA} (bottom line) $y$\/-maps. From left to right
we present the original $y$\/-maps, and their in and cross scan contributions for a small region at intermediate
Galactic latitudes in the southern sky centered at $(0,-45)$ degrees in Galactic coordinates. \label{fig:stripemaps}}
\end{center}
\end{figure*}
\subsection{Reconstruction methods}
\label{subsec:compsep}
The tSZ effect signal in the \Planck\ frequency maps is
subdominant with respect to the CMB and other foreground
emissions. By contrast to the CMB emission, the tSZ effect from galaxy clusters is spatially
localized and leads to a highly non-Gaussian signal. 
CMB-oriented component-separation methods                                     
\citep{planck2013-p06} are not adequate to recover the tSZ signal. We
therefore use specifically tailored component separation
algorithms to reconstruct the tSZ signal from the
\Planck\ frequency channel maps as in \citet{planck2013-p15}. 
These algorithms rely on the spatial localization of the different
astrophysical components and on their spectral diversity to separate
them. As in \citet{planck2013-p15} we consider here the {\tt
MILCA} \citep[Modified Internal Linear Combination Algorithm,][]{2013A&A...558A.118H} and
{\tt NILC} \citep[Needlet Independent Linear
Combination,][]{Remazeilles2011} methods. Both are based on the Internal
Linear Combination (ILC) approach that searches for the linear combination
of the input maps that minimizes the variance of the final
reconstructed map under the constraint of offering unit gain to the
component of interest (here the tSZ effect, whose frequency
dependence is known). Both algorithms have been extensively tested on
simulated \Planck\ data. 

For both methods, the \Planck\ HFI maps from 100 to 857\,GHz, convolved to a common resolution of
10\arcm, are used. In the case of {\tt NILC} we also use the LFI data at large angular scales ($\ell < 300$).
Similarly, for both methods the 857\,GHz map, which traces the thermal dust emission on large angular scales, is
only used for multipoles $\ell < 300$ to minimise residuals from IR point sources and clustered Cosmic Infrared
Bacground (CIB) emission  in the final $y$\/-maps.

\subsubsection{{\tt NILC}}

In the multi-component extensions of {\tt NILC}~\citep{Delabrouille:2009p2535,Remazeilles2011},
initially developed to extract the CMB, the weights for component separation
(i.e., covariances) are computed independently in domains of a needlet
decomposition (in the spherical wavelet frame). The needlet decomposition
provides localization of the ILC filters both in pixel and in multipole space,
allowing us to deal with local contamination conditions varying both in position
and in scale. We imposed constraints to remove the CMB contamination and
preserve the tSZ effect. To avoid strong foreground effects, part of the
Galactic plane (corresponding to about 2\% of the sky) was masked before applying {\tt NILC} to the \Planck\ frequency
maps.

The localisation in multipole space is achieved by using ten Gaussian window functions $\{h_j(\ell)\}_{1\leq j\leq 10}$ as bandpass filters\footnote{Note that in \citet{planck2013-p05b} cosine window functions were used.}, so-called needlet bands, allowing for smooth localisation in $\ell$ (see Fig. \ref{Fig:bands}). NILC performs a weighted linear combination of the bandpass filtered Planck maps for each needlet scale independently. 
%The Gaussian window functions are defined such that $\sum_{j=1}^{10} h_j(\ell)^2 = 1$, so the complete signal is conserved when co-adding the ten ILC maps from different needlet scales.
%\newline
The localisation in the spatial domain is achieved by defining scale-dependent zones over the sky on which the covariance matrices and ILC weights are computed.
More precisely, the pixel domain on which the covariance is computed, is defined from the smoothing of the product of the relevant needlet maps with a symmetric Gaussian window in pixel space whose FWHM depends on the needlet scale considered. This avoids artificial discontinuities at pixel edges.
%\newline
The localisation reduces the number of modes on which the statistics is computed, this may be responsible for an ``ILC bias'' due to chance correlations between SZ and foregrounds \citep{Delabrouille_2009}. At the coarsest scale in particular (first needlet band), the area of the spatial localisation must be large enough to counterbalance the lack of modes in multipole space due to spectral localisation. 
%In our current implementation of NILC, 
In practice, the zones for spatial localisation are not pre-defined but their area is automatically adjusted to the needlet scale considered . %in order to control the ILC bias.
The ILC bias $b_{\mathrm{ILC}}$ is related to the number of channels $N_{\mathrm{ch}}$ and to the number of modes $N_m$ \cite{Delabrouille_2009} as
\begin{equation}
b_{\mathrm{ILC}} = -\sigma^2_{\mathrm{SZ}}{N_{\mathrm{ ch}} -1 \over N_m}.
\end{equation}
This offers the possibility to adapt $N_m$ and $N_{\mathrm{ch}}$ in order to control the ILC bias to a set threshold, as discussed in \cite{Remazeilles_2013}: given both the number of channels and the number of spectral modes in $\ell$-space at the needlet scale considered, our NILC algorithm consistently computes the number of spatial modes (similarly, the zone area for spatial localisation) required to control the ILC bias.

\subsubsection{{\tt MILCA}}

{\tt MILCA}~\citep{2013A&A...558A.118H}, when applied to the
extraction of the tSZ signal, also uses two spectral constraints: preservation of
the tSZ signal (the tSZ spectral signature discussed above is assumed); and
removal of the CMB contamination in the final SZ map, making use
of the well known spectrum of the CMB. In addition, to compute the
weights of the linear combination, we have used the extra degrees of
freedom in the linear system to minimize residuals from other components
(two degrees of freedom) and from the noise (two additional degrees).
The noise covariance matrix was estimated from the half-difference maps
described in Section~\ref{subsec:planckdata}.  
To improve the efficiency of the {\tt MILCA} algorithm, weights are allowed to vary
as a function of multipole $\ell$, and are computed independently on
different sky regions. We have used 11 filters in $\ell$
space as shown in Figure~\ref{Fig:bands}.  These filters have an overall transmission of one, except for $\ell < 8$. 
For these large angular scales we have used a Gaussian filter to reduce
foreground contamination.  To ensure sufficient spatial localization
for each required resolution the size of the independent sky regions was
adapted to the multipole range. We used a minimum of 12 regions at low resolution and a maximum of 3072 regions at high resolution.

\subsection{Reconstructed Compton parameter map}
\label{subsec:ymaps}

Figure~\ref{fig:planck_y_map} shows the reconstructed
\Planck\ all-sky Compton parameter map for {\tt NILC} (top panel) and
{\tt MILCA} (bottom panel).  For display purposes, the
maps are filtered using the procedure described in the first paragraph of 
Sect.~\ref{methods:1D PDF}. Clusters appear as positive sources: the Coma
cluster and Virgo supercluster are clearly visible near the north Galactic
pole. The Galactic plane is masked in both maps, leaving 67\% of the sky.  
Residual Galactic contamination is also visible as
diffuse positive structures in the {\tt MILCA} $y$\/-map. We can also observe
a granular structure in the {\tt NILC} $y$\/-map that corresponds to an excess of
noise at large angular scales as discussed in Sect.~\ref{subsec:noise}.

Weaker and more compact clusters are visible in the zoomed region of the
Southern cap, shown in Fig.~\ref{fig:planck_y_map_zoom}.
Strong Galactic and extragalactic radio sources show up as negative bright
spots on the maps and were masked prior to any scientific analysis, as
discussed below in Sect.~\ref{subsec:pointsourcecont}. We can again observe
residual Galactic contamination around the edges of the masked area, which is more important for {\tt MILCA}. 
Finally, we note in the {\tt NILC} and {\tt MILCA} $y$\/-maps the presence of systematic residuals along the scanning direction that show
up as stripes. These are the consequence of stripes in the original \Planck\ maps \citep[see][for a detailed discussion]{planck2014-a09}.
We discuss the level of residual stripes in Sect.~\ref{subsec:systematics}.

In addition to the full Compton parameter maps, we also produce the
so-called first and last (F and L hereafter) Compton parameter maps from the first and
second halves of the survey rings (i.e., pointing periods). These maps are
used for  power spectrum estimation in~Sect.~\ref{sec:powerspec}
as well as to estimate the noise properties in the $y$\/-maps (see Sect.~\ref{subsec:noise}).

\subsection{Comparison to other Compton parameter maps in the literature}

 \cite{2014PhRvD..89b3508V} and  \cite{Hill_2014} have used the Planck nominal data to reconstruct a Compton parameter map over a large fraction of the sky. 
As for the present work, they use an ILC approach imposing spectral constraints to have unitary gain for the SZ component and null contribution from CMB, but
they do not use the spectral and spatial separation discussed above.
\cite{2014PhRvD..89b3508V} consider only the four HFI channels between 100 and 353 GHz and force a null contribution from dust emission. On their side \cite{Hill_2014} also include the 545 GHz map, while using the 857 GHz channel only as a template for dust emission: a flux cut is imposed in order to build a mask that keeps only the 30\% of the sky. They also apply a point source mask (radio and IR) before computing the ILC coefficients reducing the final sky fraction to about the 25\%.
Furthermore, \cite{2014PhRvD..89b3508V} and  \cite{Hill_2014} aim mainly at studying the cross-correlation with the gravitational lensing mass map from the Canada France Hawaii Telescope Lensing Survey (CFHTLenS) and the publicly-released Planck CMB lensing potential map, respectively. Their $y$\/-map reconstruction methods are tailored to fulfill these objectives and as a consequence they present larger overall foreground contamination, in the case of \cite{Hill_2014} also a significantly smaller sky fraction.

\begin{figure}
\begin{center}
\setlength{\unitlength}{\columnwidth}
\begin{picture}(1,2)
\put(0,1.4){\includegraphics[trim=0cm 0cm 0cm 0.5cm, clip=true,width=\columnwidth]{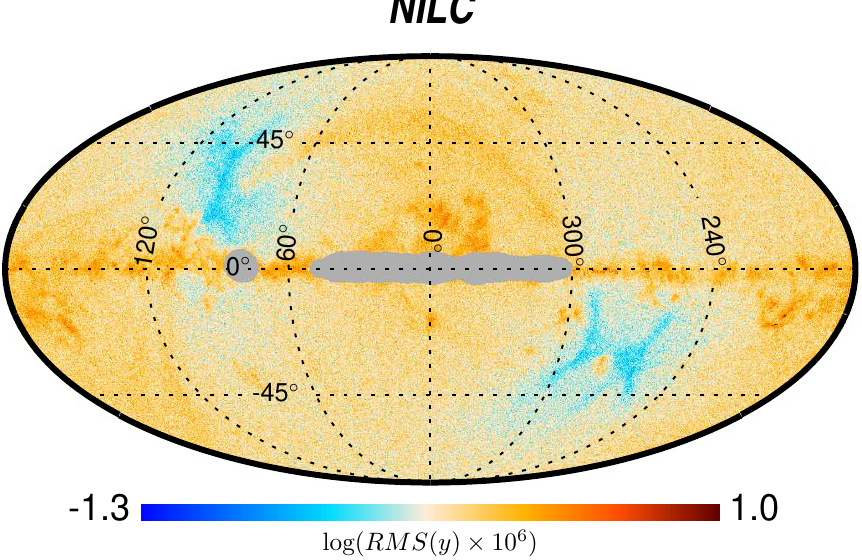}}
\put(0,0.7){\includegraphics[trim=0cm 0cm 0cm 0.5cm, clip=true,width=\columnwidth]{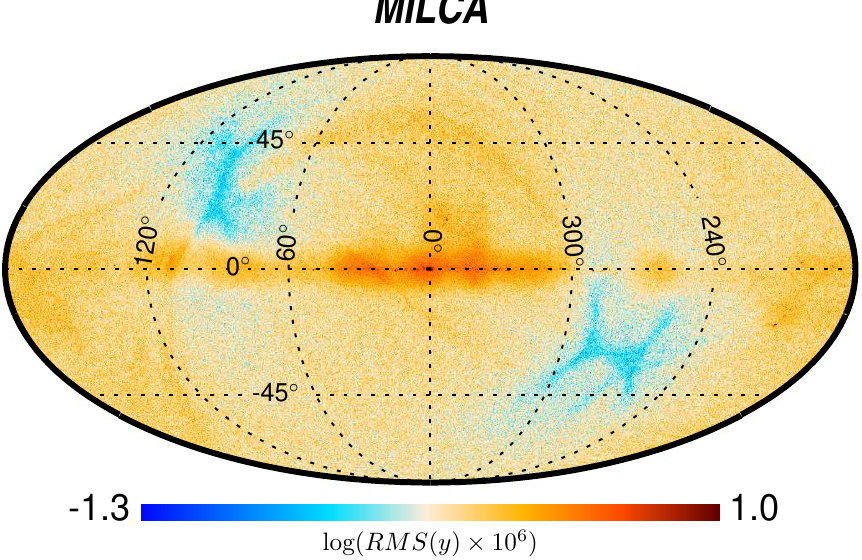}}
\put(0,0){\includegraphics[trim=0cm 0cm 0cm 0cm, clip=true,width=\columnwidth]{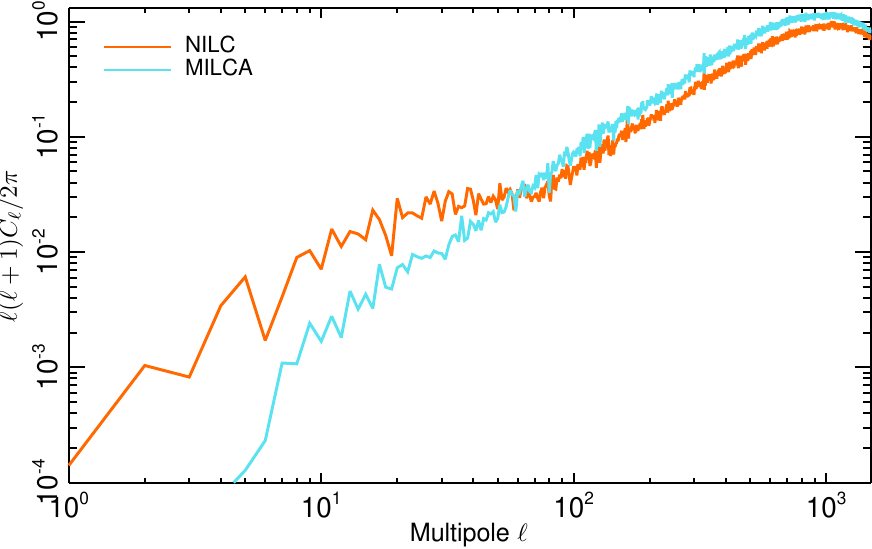}}
\put(0.45,2){\bf {\tt NILC}}
\put(0.45,1.3){\bf {\tt MILCA}}
\end{picture}
\caption{{\it Top}: Standard deviation maps for the {\tt NILC} (top) and {\tt MILCA} (middle)  $y$\/-maps corresponding
to the inhomogenous noise contribution computed from the half difference of the half-rings maps.
{\it Bottom}: Angular power spectrum of the homogenous noise contribution for the {\tt NILC} (orange) and {\tt MILCA} (blue) $y$\/-maps (see main text
for details). \label{fig:noiseflmaps}}
\end{center}
\end{figure}

\section{Pixel space analysis}
\label{sec:pixelanalysis}

\subsection{Stripes in the $y$\/-map}
\label{subsec:systematics}

As discussed in Sect.~\ref{subsec:ymaps}, residual stripes are visible in the {\tt NILC} and {\tt MILCA} $y$\/-maps. These stripes are mainly due to residuals in-scan direction systematics after subtraction of an offset for each \Planck\ stable pointing period \citep{planck2014-a09}. To study how these stripes contaminate the final $y$\/-maps we have decomposed them in their in-scan and cross-scan direction contributions. We first convert the $y$\/-maps from Galactic to Equatorial coordinates for which the scan direction corresponds mainly to a fixed longitude value. Secondly, we apply a Galactic mask to the Equatorial $y$\/-maps and decompose them in spherical harmonics. Third, we select the in-scan (cross-scan) direction components by nullifying the spherical harmonic coefficients, $a_{\ell,m}$,  for $\ell > m$ ($\ell < m$). Finally, we construct maps from those transformed coefficients and convert back to Galactic coordinates. We have chosen to mask the brightest 40\% of the sky in the 857 GHz \Planck\ map to keep Galactic ringing residuals negligible and a sufficiently large fraction of the sky for the analysis. 

Figure~\ref{fig:stripemaps} shows from left to the right the original, in-scan and cross-scan $y$\/-maps for {\tt NILC} (top) and {\tt MILCA} (bottom) in the southern sky region presented in Figure~\ref{fig:planck_y_map_zoom}. We note that the $y$\/-maps look noisier as they are not filtered using the procedure described in the first paragraph of Sect.~\ref{methods:1D PDF}. The stripes are apparent in both the original and in-scan $y$\/-maps as expected. We find that the ratio of the rms of the in and cross scan maps is 1.16  (1.17) for {\tt NILC} ({\tt MILCA}), consistent with residual stripe contamination (for Gaussian noise only maps we find a ratio of 1). Here we have measured an overall increase of the rms of the maps due to stripe contamination, a more detailed estimate of the effect in terms of tSZ power spectrum is presented later in Sect.~\ref{sec:lowell}.

The in-scan and cross-scan decomposition method modifies the cluster signal significantly due to ringing effects -- see positive and negative patterns around clusters in Figure~\ref{fig:stripemaps}. To explore this effect we have also applied the in-scan and cross-scan decomposition method to the simulated Compton parameter map for the detected and confirmed clusters of galaxies in the \Planck\ catalogue~\citep{planck2013-p05a}, which is presented in Section~\ref{subsec:resolvedclusters}. We find that those negative and positive patterns are also present in the decomposed maps and the ratio between the rms of the in-scan and cross-scan maps is 1 as expected. We finally stress that the in-scan and cross-scan decomposition of the $y$\/-maps does not preserve the positiveness of the tSZ signal. As a consequence we can not use them to estimate the contamination of the stripe systematic effect in the analysis presented in Section~\ref{sec:higorderstat}.

\begin{figure}
\includegraphics[width=\columnwidth]{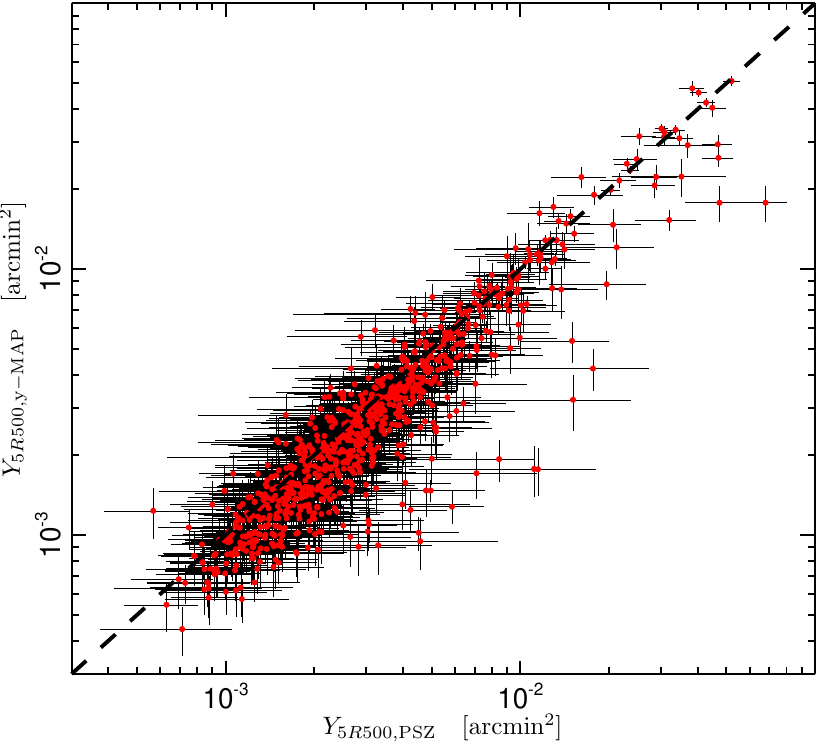}
\includegraphics[width=\columnwidth]{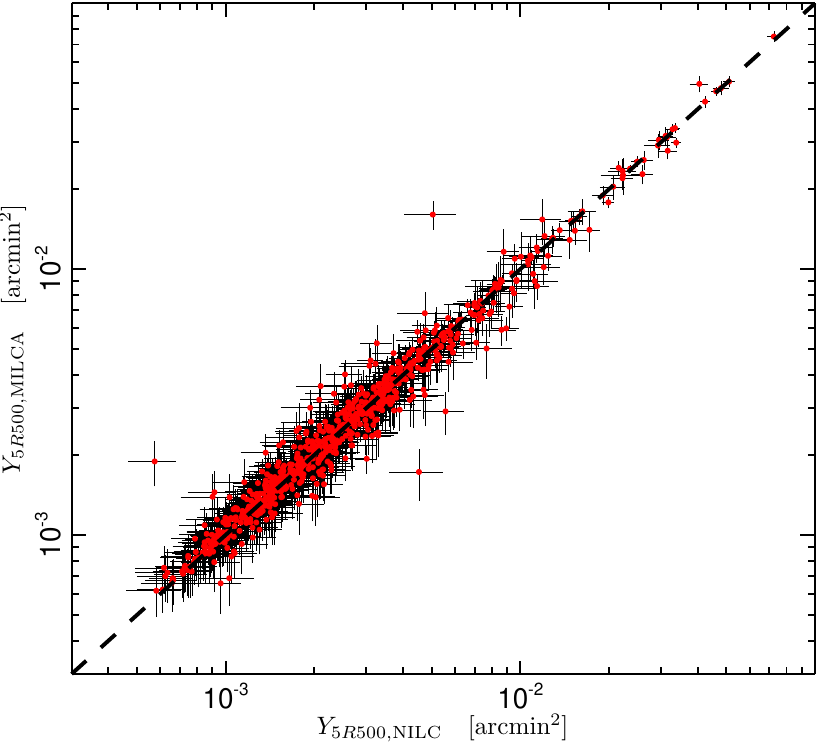}
\caption{{\it Top}: Comparison between the measured tSZ flux reported in the Planck cluster sample and that estimated directly on the  {\tt NILC} $y$\/-map for the blindly detected
common sources. {\it Bottom}:  As above but for the  {\tt NILC} and  {\tt MILCA} measured fluxes. \label{fig:photopsz2}}
\end{figure}

\subsection{Noise distribution on the $y$\/-map}
\label{subsec:noise}
The \Planck\ maps present a highly non-homogeneous structure of the noise that is mainly due to the inhomogeneous scanning
strategy. Such complicated structure is propagated into the $y$\/-map and needs to be considered for further analysis. 
Here, we have chosen to describe the noise structure in the $y$\/-map by  A) a variance map, which will capture the inhomogeneity of the
noise, and B) the angular power spectrum of an homogenous noise map, which it is obtained after correcting for inhomogeneity.
The variance map and homogeneous noise power spectrum have been obtained using two different methods:
\begin{enumerate}
\item {\it Half-difference}. The half difference of the half rings $y$\/-maps is used to obtain an estimate of the noise in the {\tt NILC} and {\tt MILCA} $y$\/-maps. From this half difference map we compute an estimate of the noise variance at lower resolution by squaring and downgrading it to {\tt Healpix} N$_{\mathrm{side}}$=128. The homogenous noise contribution is
obtained by dividing the half difference map by the variance map after upgrading it to {\tt Healpix} N$_{side}$=2048.
\item {\it Simulations-based}.100 noise only realisation from the simulations described in Sect.~\ref{subsec:simulations} are used to compute an estimate of the variance per pixel at full resolution, which is then averaged to {\tt Healpix} N$_{\mathrm{side}}$=128 resolution. The homogenous noise contribution map is then computed as above.
\end{enumerate}

The two methods give consistent results at high Galactic latitudes but differ at low Galactic latitudes where we expect larger foreground
residuals that show up in the {\it Half-difference} method. This is visible in the top panel of Figure~\ref{fig:noiseflmaps} where we display variance maps
obtained from the {\it Half-difference} method for {\tt NILC} (left) and {\tt MILCA} (right). For a wide range of analyses with the $y$\/-map  
(i.e computing the radial profile, surface density or total flux of a cluster, stacking of faint sources, detection of shocks, study of the ICM) an estimate of the overall uncertainties per pixel (including foreground contribution) is required. Thus, the half difference maps are released with the $y$\/-map. 
We also show in the bottom panel of Figure~\ref{fig:noiseflmaps} the power spectrum of the homogeneous noise contribution. 
We observe a significant large angular scale (low multipoles) component in the the homogeneous noise contribution.
This low multipole component is significantly larger for the {\tt NILC} $y$\/-map than for the {\tt MILCA} one. 
%(see bottom plot of Figure~\ref{fig:variance maps}) that needs to be accounted for further analysis.
%%%% SECTION

\subsection{Foreground contamination and masking}

\subsubsection{Galactic emission}
\label{subsec:galacticemission}

As discussed above, residual galactic foreground contribution is observed at low Galactic latitudes in both the {\tt NILC} and {\tt MILCA} $y$\/-maps,
while being more important for  the {\tt MILCA} $y$\/-map. This contribution is mainly induced by thermal dust emission, as discussed in details in~\citet{planck2013-p05a},
and is therefore mainly associated to the Galactic plane. When dealing with individual objects or with the stacking of faint objects we  recommend to account for it using the variance maps discussed above. For statistical analyses as those presented in Sect.~\ref{sec:powerspec} and \ref{sec:higorderstat}, specific masks need to be defined and they are fully discussed in those sections.

\subsection{tSZ signal from resolved sources}
\begin{figure*}
\includegraphics[width=0.53\columnwidth]{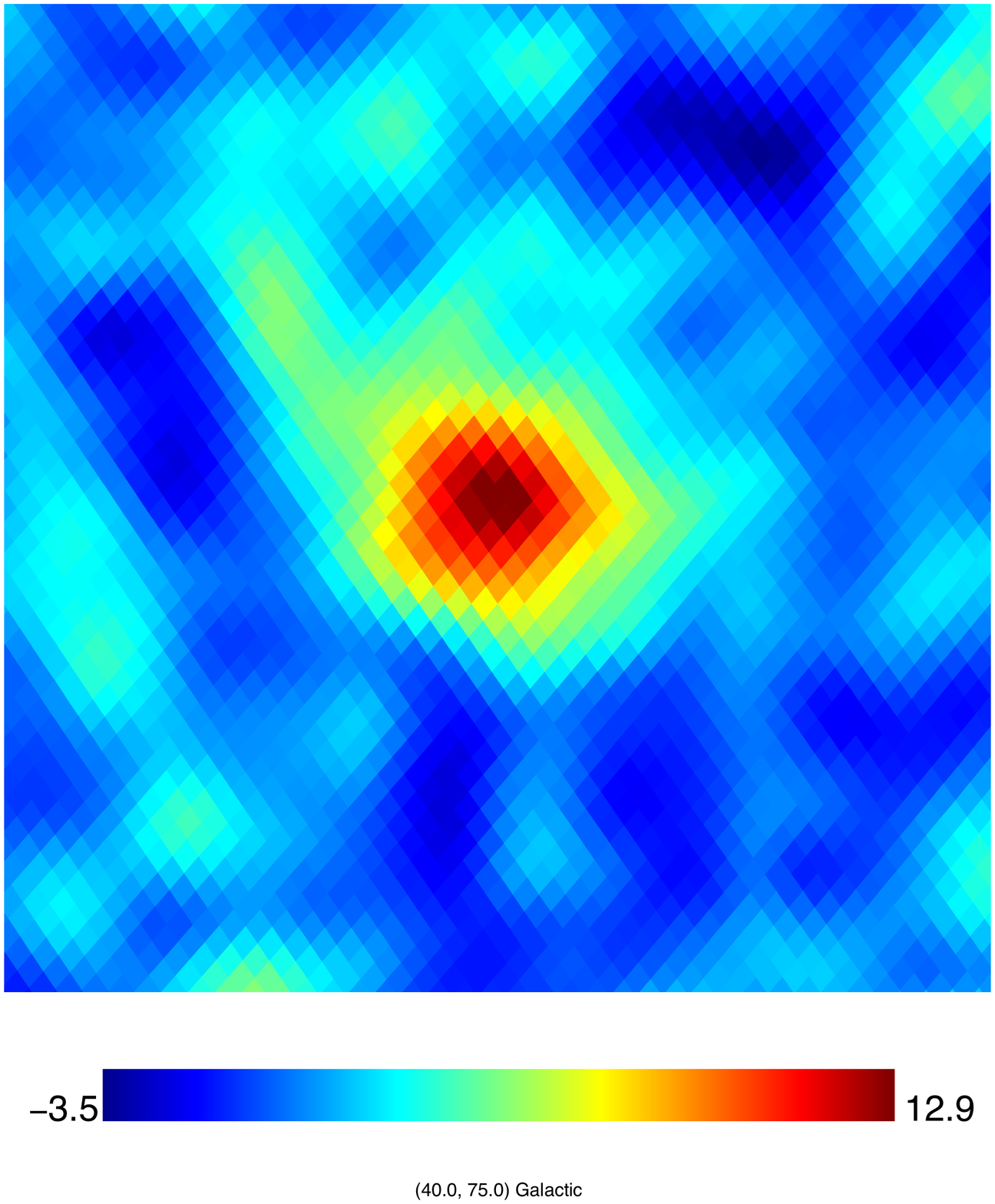}
\includegraphics[width=0.53\columnwidth]{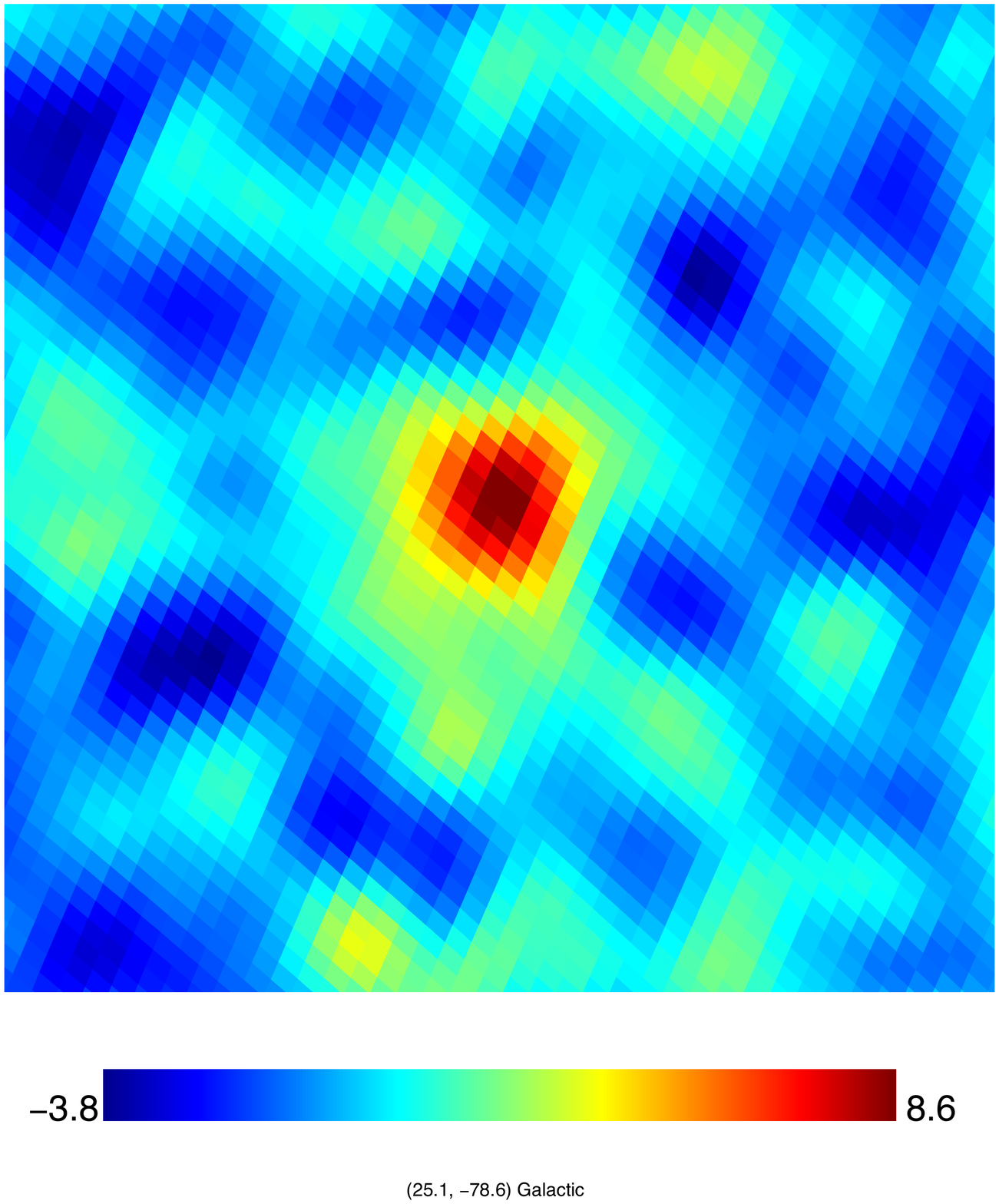}
\includegraphics[width=0.53\columnwidth]{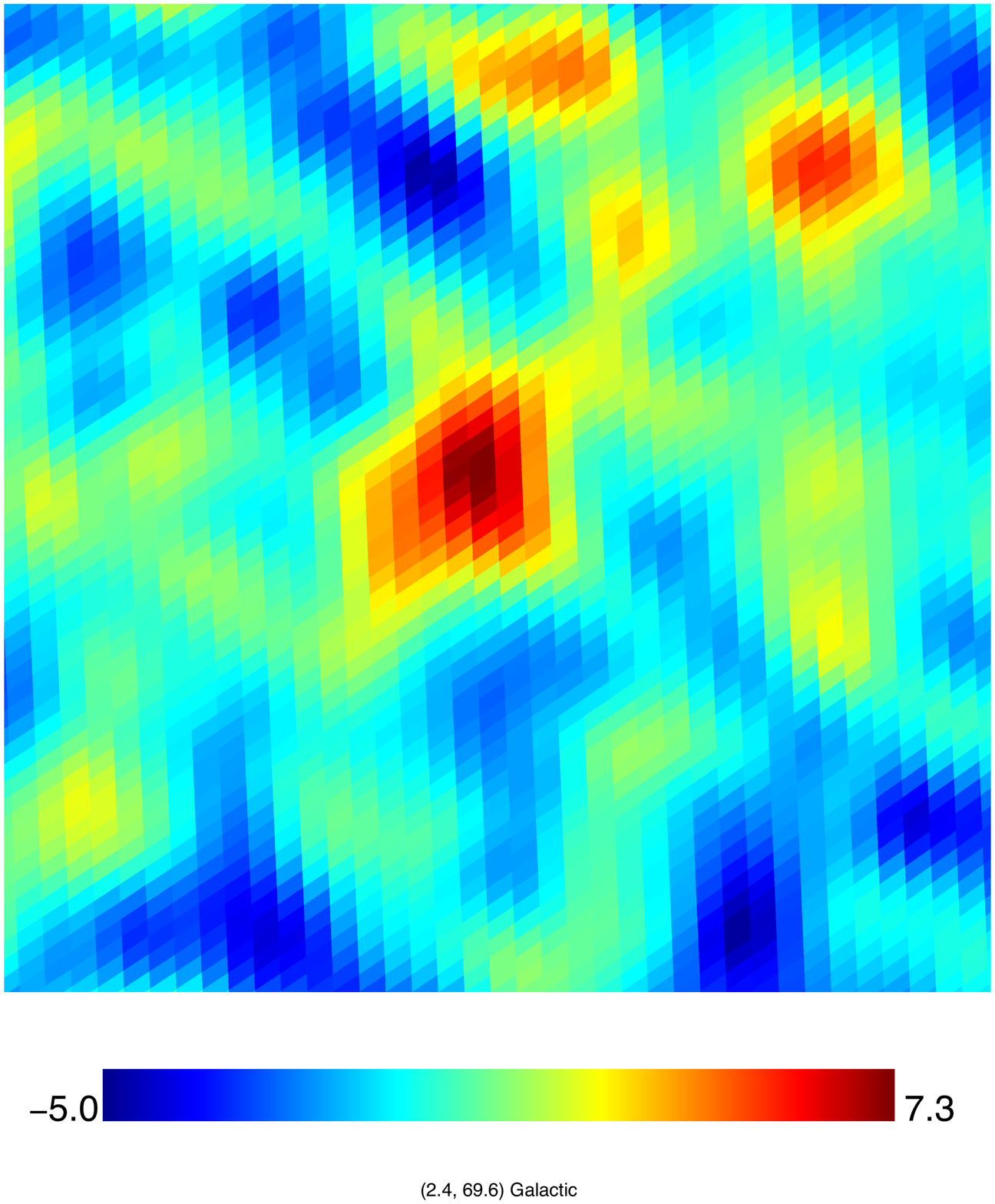}
\includegraphics[width=0.53\columnwidth]{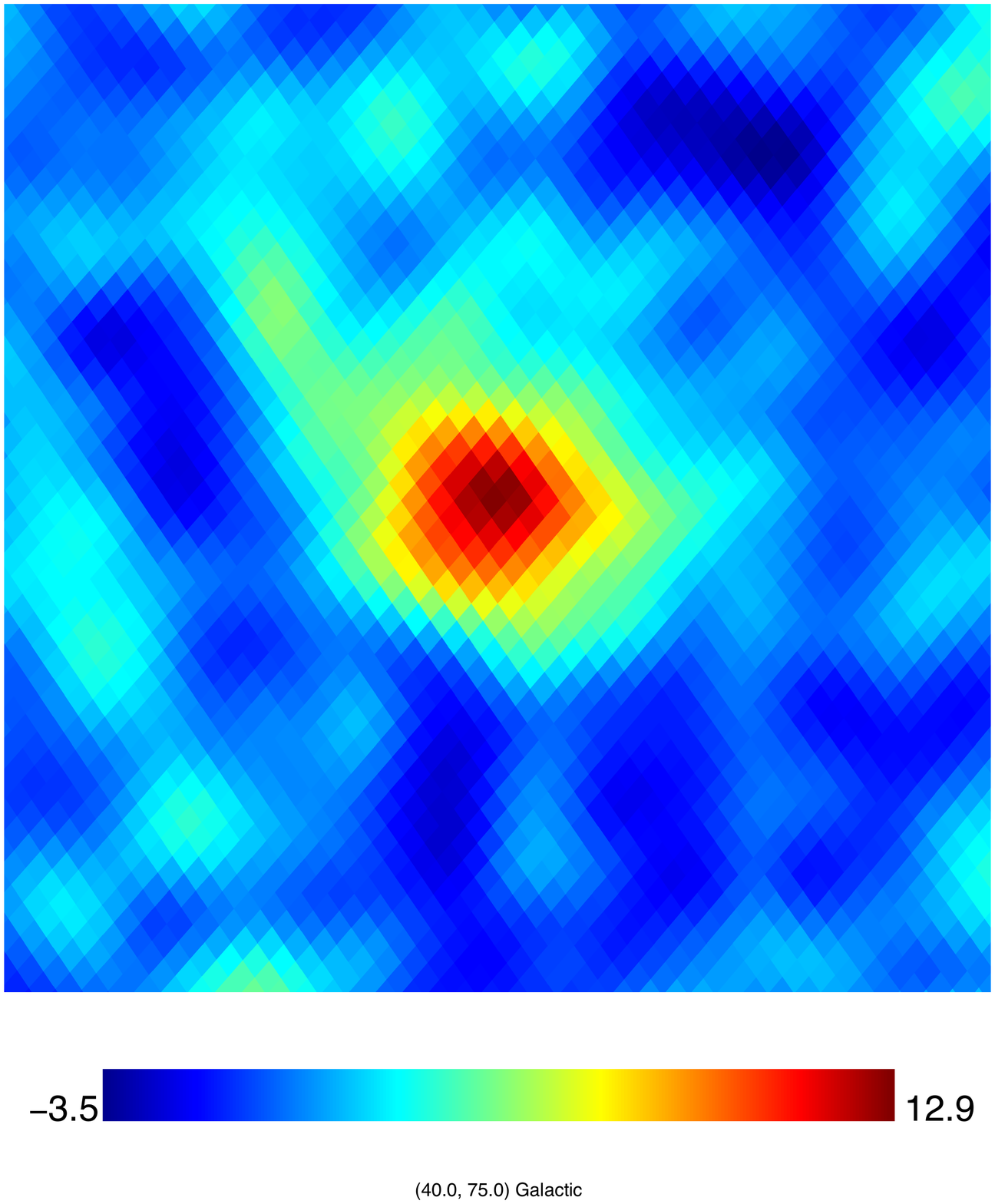}
\includegraphics[width=0.53\columnwidth]{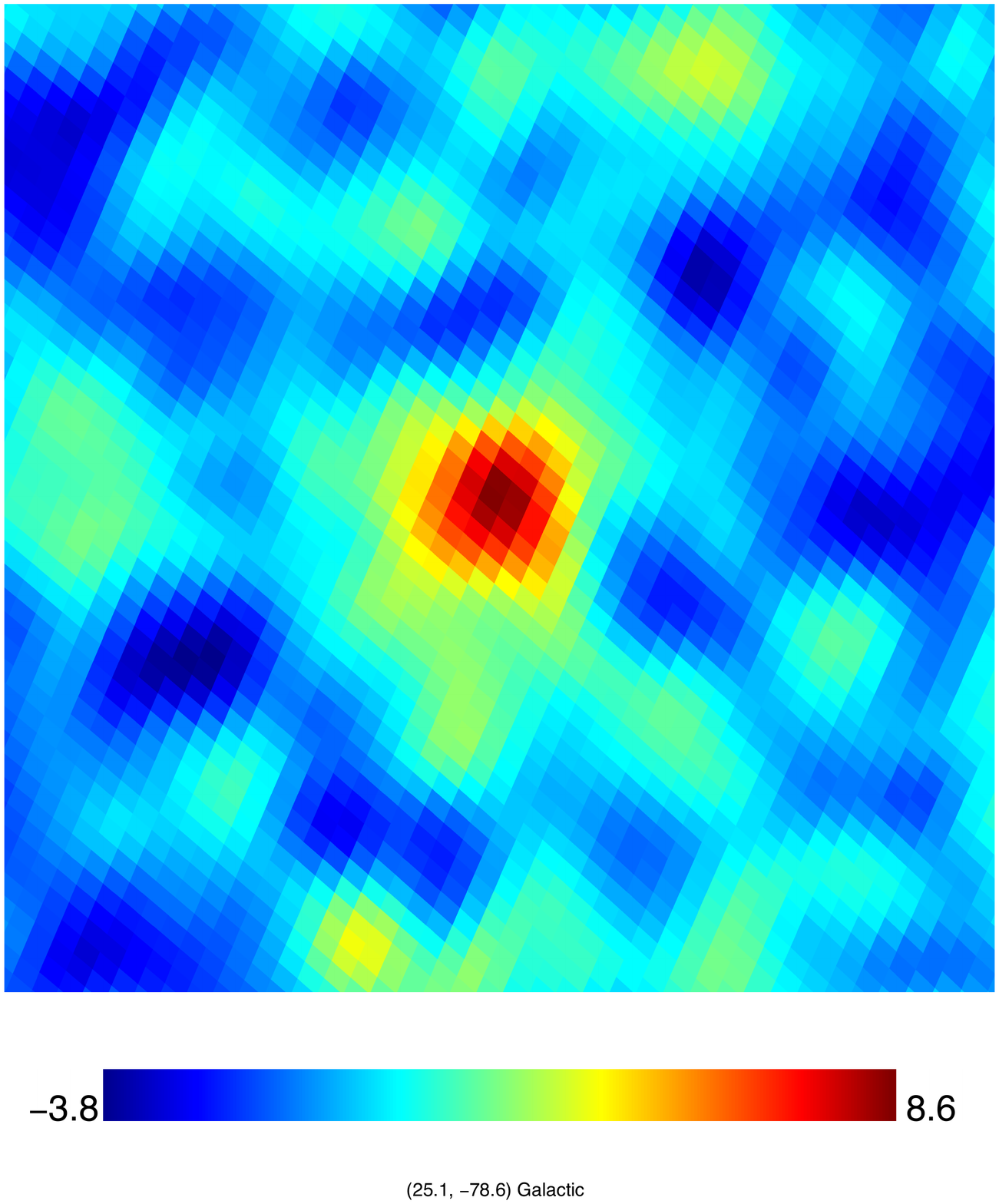}
\includegraphics[width=0.53\columnwidth]{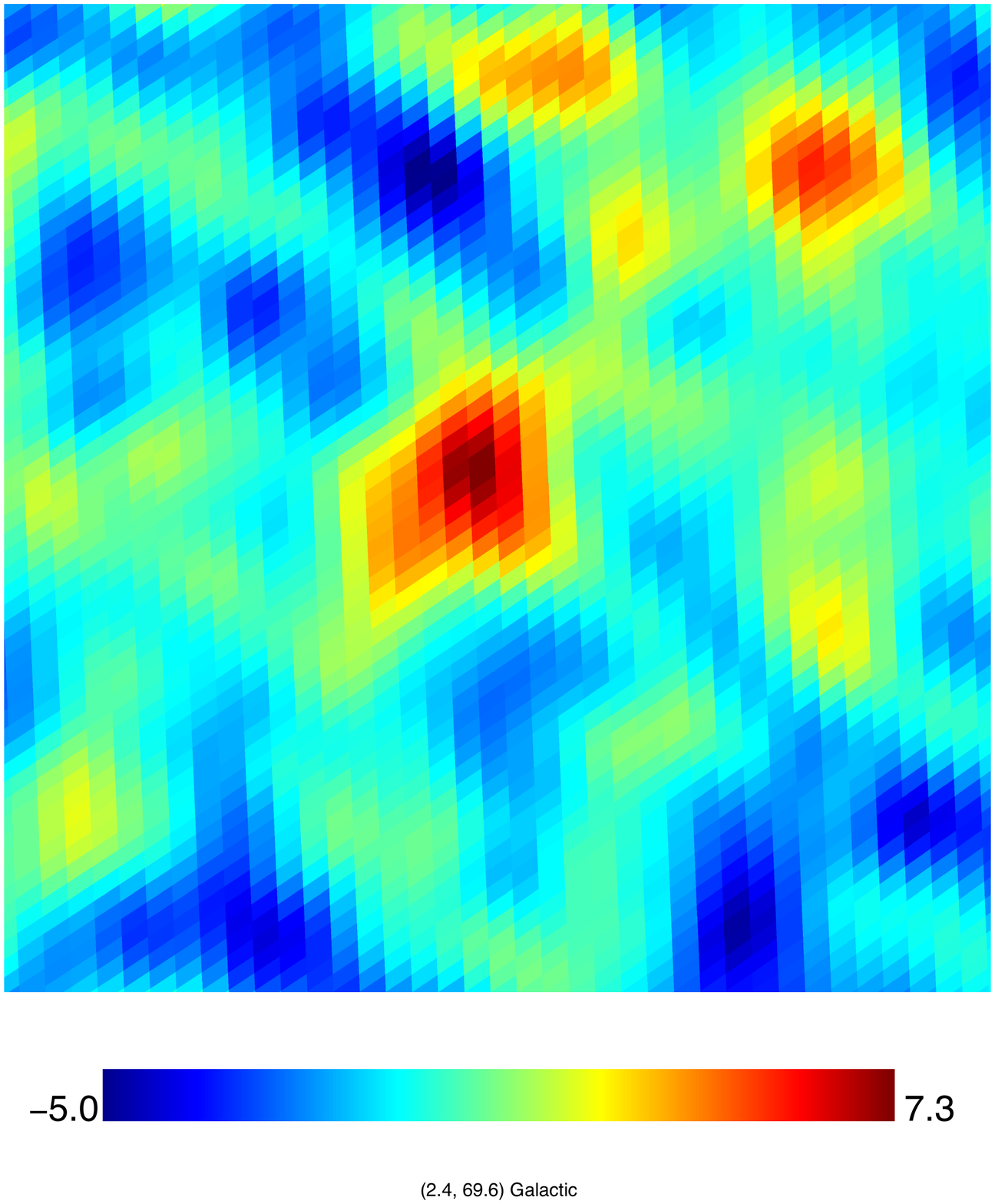}
\includegraphics[width=0.7\columnwidth]{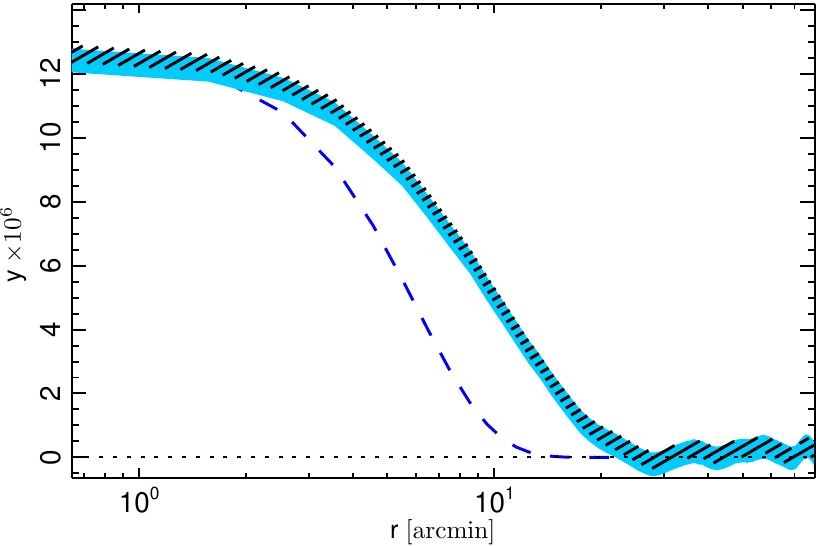}
\includegraphics[width=0.7\columnwidth]{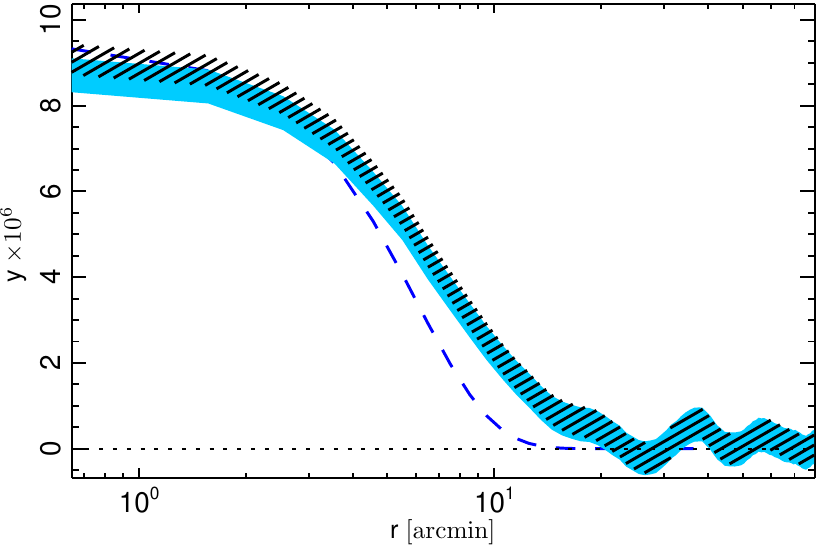}
\includegraphics[width=0.7\columnwidth]{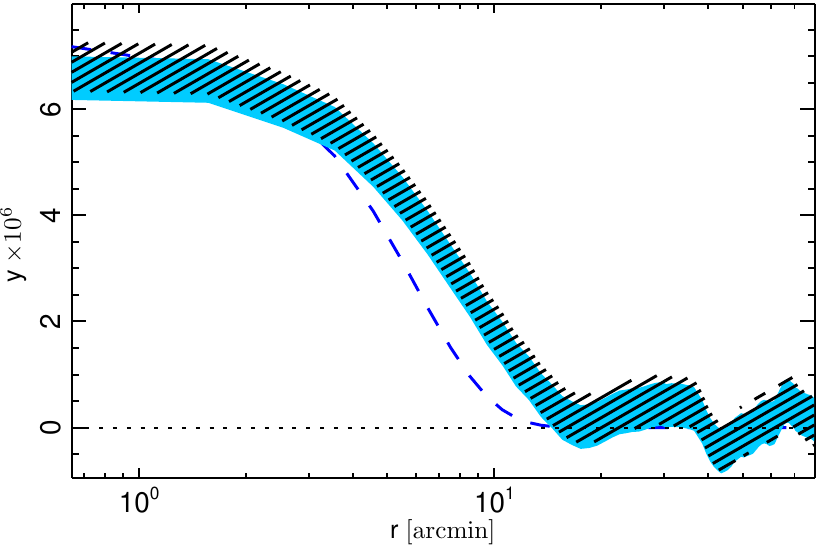}
\caption{Compton parameter {\tt MILCA} (top row) and {\tt NILC} (middle row) maps for a selection of PSZ2 cluster candidates with signal to noise
ratio of 9.3, 6.2 and 4.6 from left to right. The maps are centred at the positions of the clusters in galactic coordinates, which are given at the bottom of the
plot. Color scale is in units of  $y \times 10^{-6}$. The bottom row presents the cluster radial profiles for {\tt MILCA} (black) and {\tt NILC} (blue). The beam profile is shown as a blue dashed line. \label{fig:lowsnsources}}
\end{figure*}

\subsubsection{Point sources}
\label{subsec:pointsourcecont}

Point source contamination is an important issue for the cosmological
interpretation of the \Planck\ Compton parameter map. 

In the reconstructed tSZ maps radio sources will appear as negative peaks, while infrared sources will show up as positive
peaks, mimicking the cluster signal. To avoid contamination from these 
sources we introduce a point source mask (PSMASK, hereafter). This mask is the union of the
individual frequency point-source masks discussed in~\citet{planck2013-p05}. 
The reliability of this mask was verified by looking at the 1D PDF of the pixel signal (as it will be discussed in Section \ref{methods:1D PDF}). We found that a more robust mask can be obtained by enlarging the mask size around the strongest radio sources in order to minimize their contribution.

Alternatively, we have also performed a blind search for negative sources in the $y$\/-maps using the {\tt MHW2} algorithm \citep{LopezCaniego:2006p2546}.  We have detected 997 and 992 negative sources for {\tt NILC}  and {\tt MILCA} respectively.  We find that 67 and 42 (for {\tt NILC} and {\tt MILCA}, respectively) of those detected sources are not masked by the PSMASK. However, most of them (54 for for {\tt NILC} and 36 for {\tt MILCA}) are false detections made by the algorithm in regions surrounding very strong positive sources (i.e. galaxy clusters). for those detected as additional negative sources, the PSMASK has been updated to account for them, even though the strongest are already excluded when applying the 50\% Galactic mask used for the cosmological analysis below.

For infrared sources, estimating the efficiency of the masking is hampered by the tSZ signal itself. 
The residual contamination from point sources is discussed in
Sects.~\ref{subsec:forecont} and~\ref{sec:higorderstat}. It is
also important to note that the PSMASK may also exclude some
clusters of galaxies. This is particularly true in the case of clusters
with strong central radio sources, such as the Perseus
cluster~\citep[see][]{planck2013-p05a}.

\begin{figure*}
\setlength{\unitlength}{1.7\columnwidth}
\begin{picture}(1,1.5)
\put(0,1){\includegraphics[trim=1cm 2cm 3cm 4cm, clip=true,width=0.9\columnwidth]{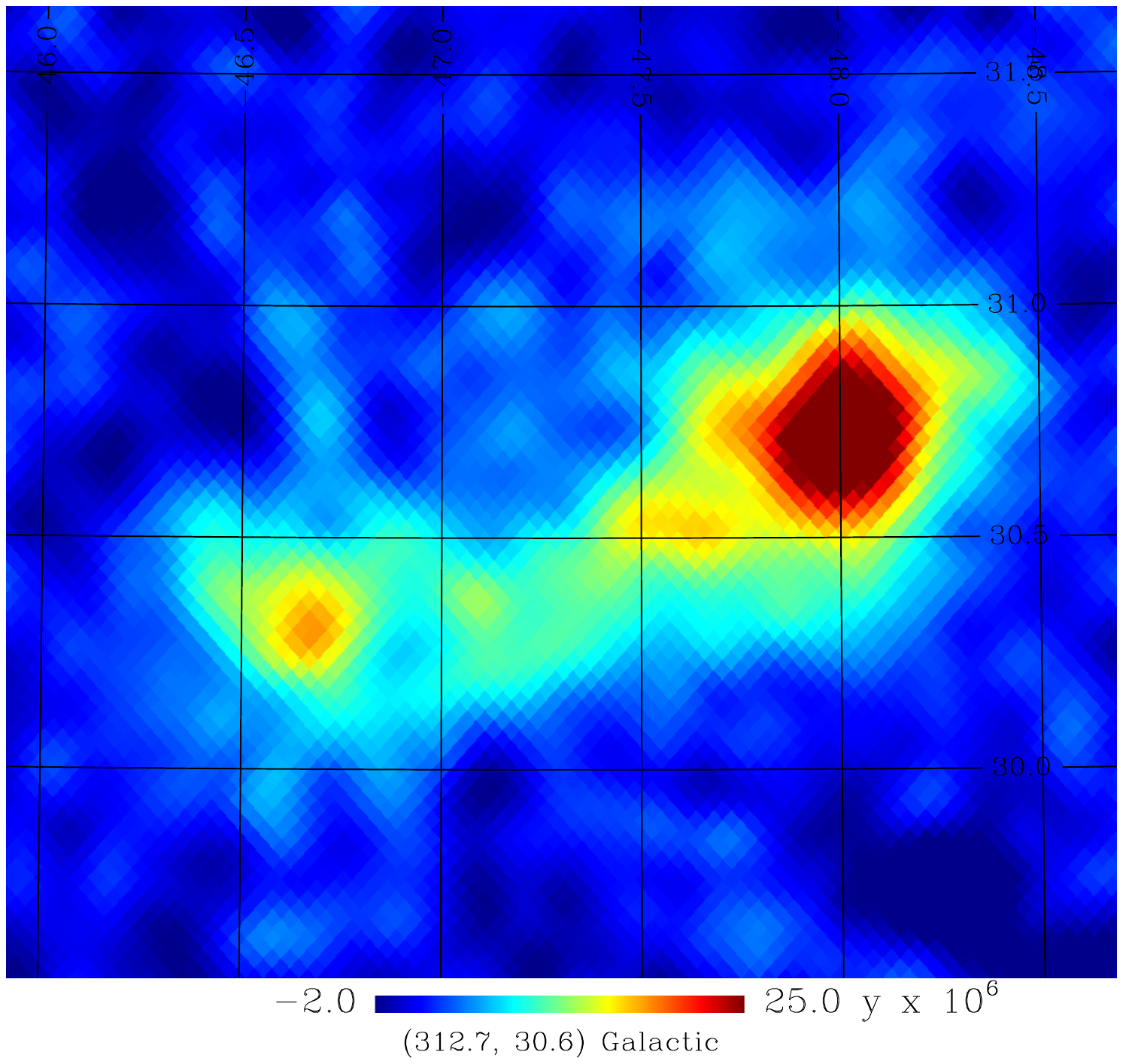}}
\put(0,0.5){\includegraphics[trim=1cm 2cm 3cm 4cm, clip=true,width=0.9\columnwidth]{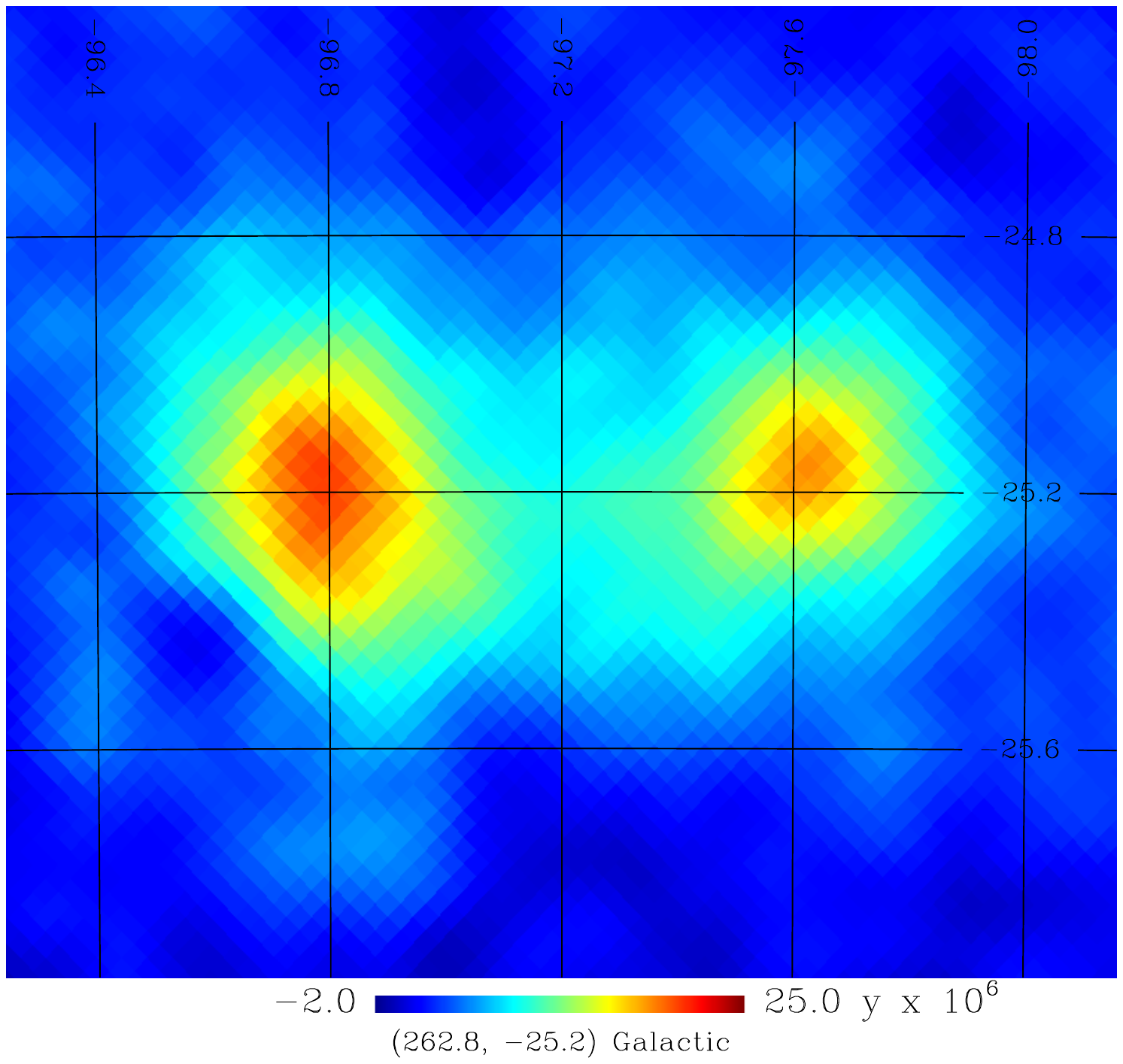}}
\put(0,0){\includegraphics[trim=1cm 2cm 3cm 4cm, clip=true,width=0.9\columnwidth]{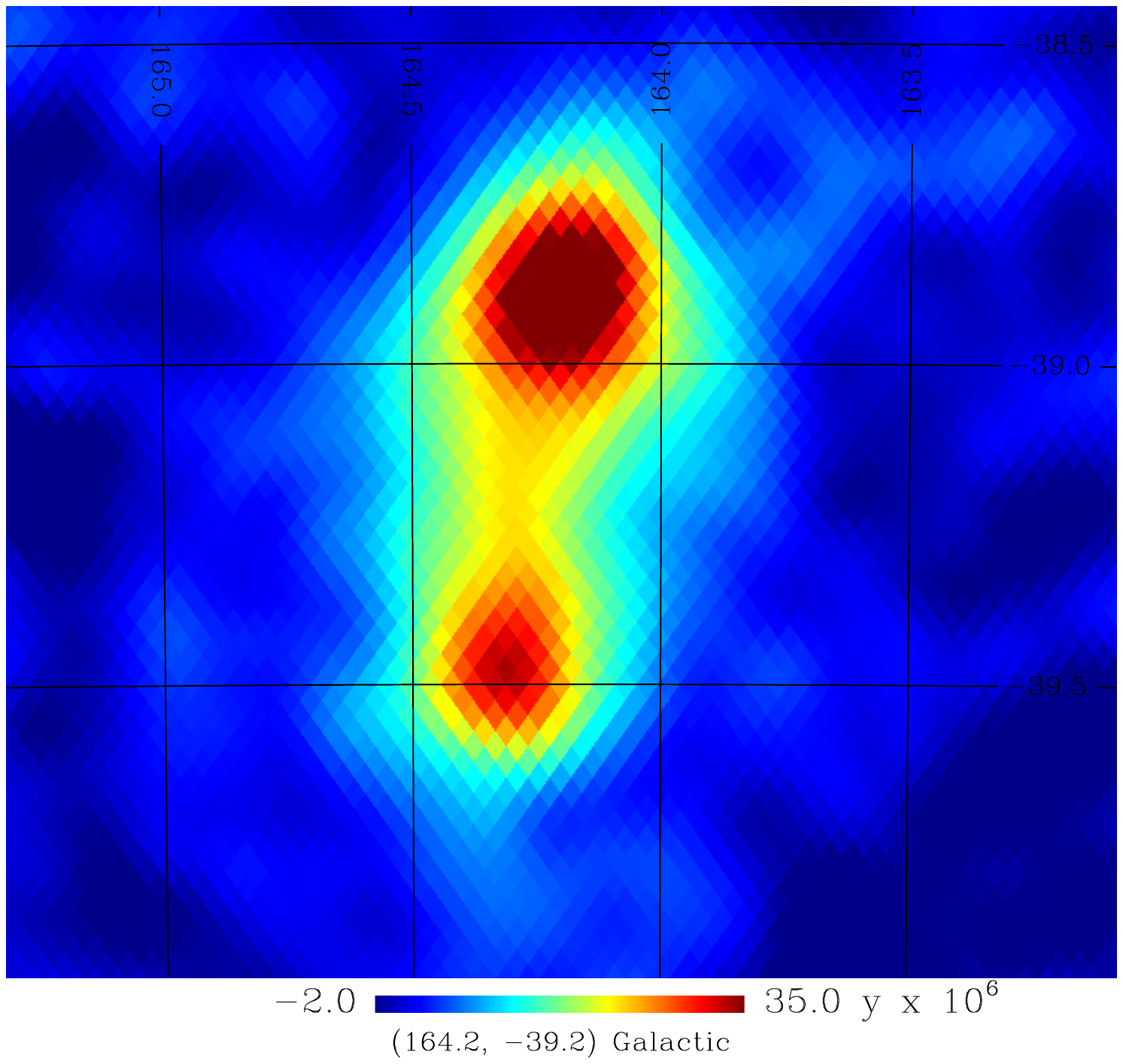}}
\put(0.5,1){\includegraphics[trim=1cm 2cm 3cm 4cm, clip=true,width=0.9\columnwidth]{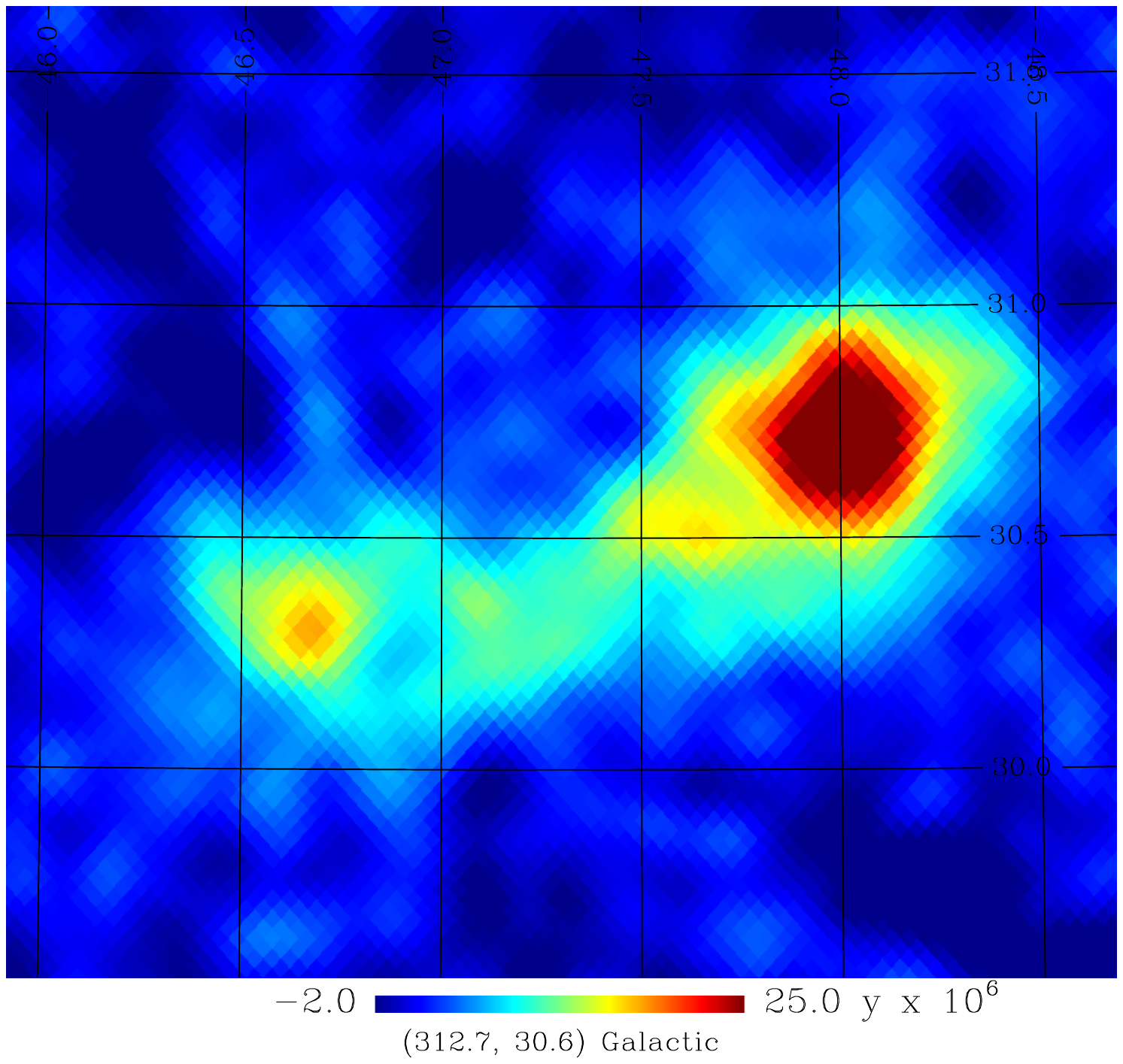}}
\put(0.5,0.5){\includegraphics[trim=1cm 2cm 3cm 4cm, clip=true,width=0.9\columnwidth]{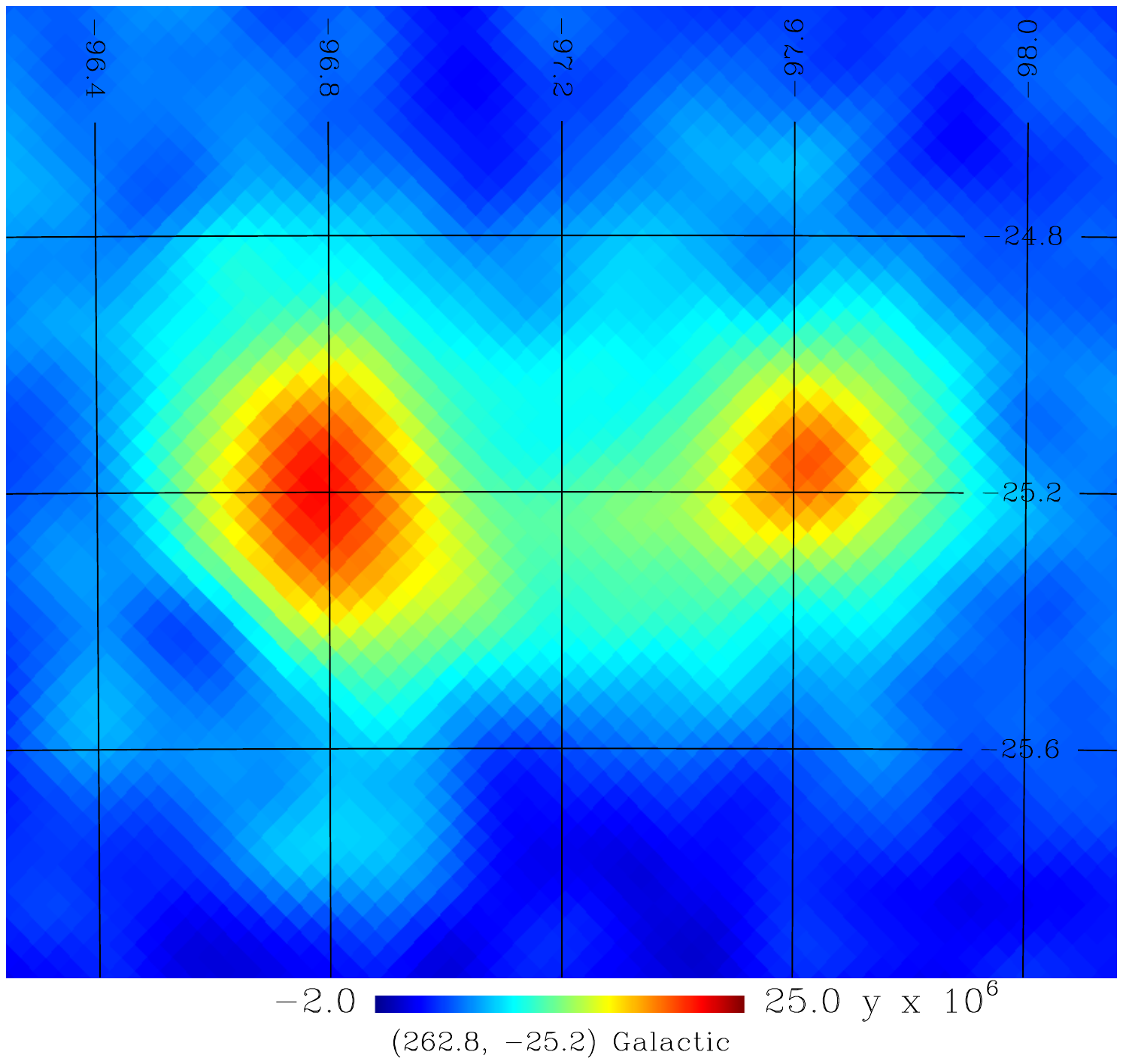}}
\put(0.5,0){\includegraphics[trim=1cm 2cm 3cm 4cm, clip=true,width=0.9\columnwidth]{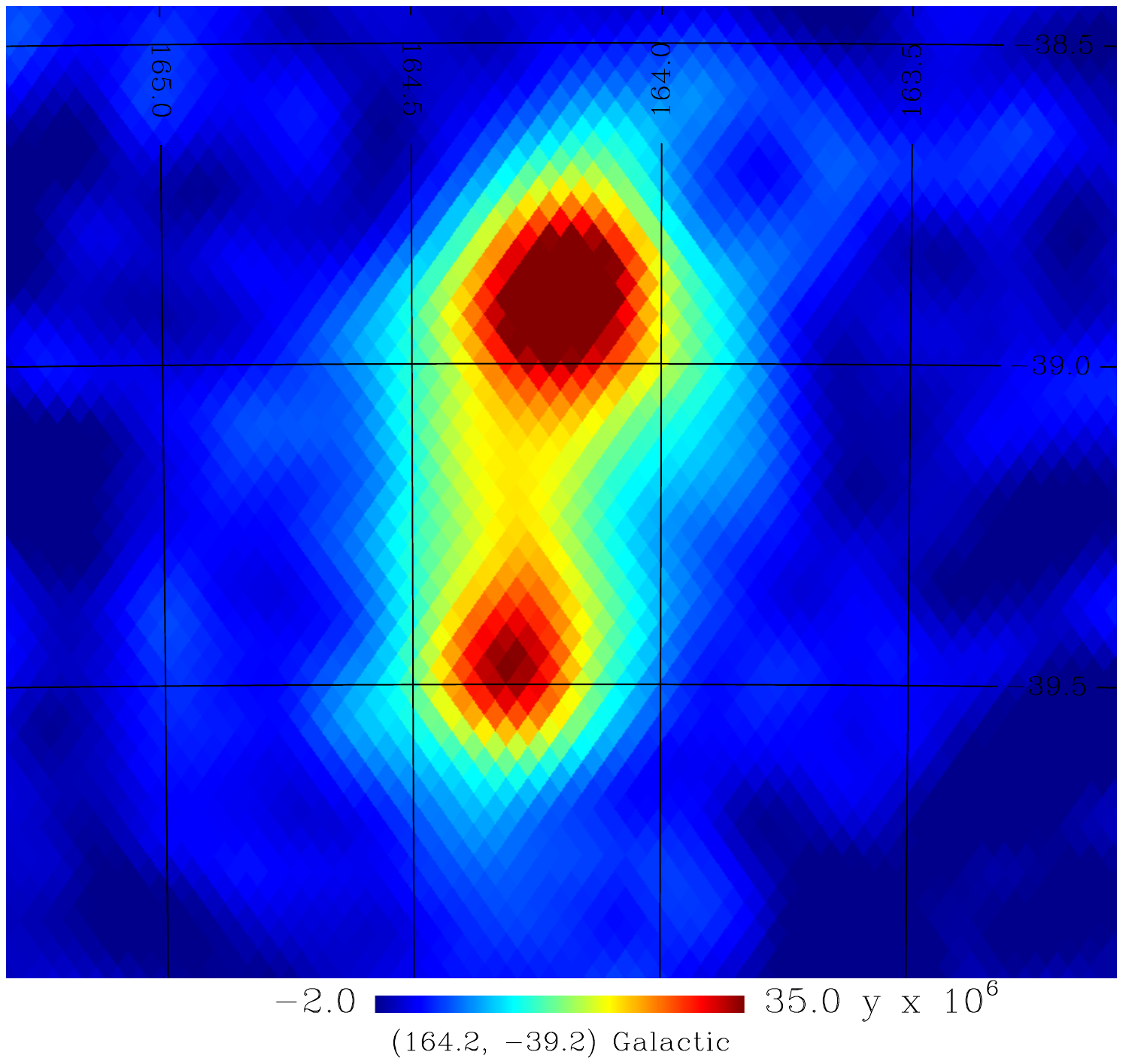}}
\put(0.05,1.49){(a) Shapley supercluster}
%\put(0.47,1.50){{\bf \tt MILCA}}
%\put(0.97,1.50){{\bf \tt NILC}}
\put(0.05,0.99){(b) A3395-A3391 merger system}
\put(0.05,0.49){(c) A339-A401 merger system}
\end{picture}
\caption{Compton parameter maps of well known merging systems for MILCA (left column) and for NILC (right column). \label{fig:mergers}}
\end{figure*}

\subsubsection{Blind search for clusters}
Following \citet{planck2013-p05b} a blind search for tSZ (positive) sources has been also performed on the all-sky {\tt NILC} and {\tt MILCA} $y$-maps. We use the  {\tt IFCAMEX} \citep[{\tt MHW2},][]{GonzalezNuevo:2006p2545,LopezCaniego:2006p2546} and the single frequency matched filter ({\tt MF}, \citealt{melin2006}) methods. The detected sources with signal-to-noise ratio $>4$ have been matched to the Planck
cluster catalogue \citep[PSZ2,][]{planck2014-a36}. We have associated the detected sources to PSZ2 objects if the distance between their positions is smaller than 10$^{\prime}$ (the resolution of the SZ all-sky maps). 
%388 PSZ2 clusters (with a mean signal-to-noise ratio of about 10) are detected by the {\tt MHW2} algorithm on both {\tt NILC} and {\tt MILCA}.
The {\tt MHW2} algorithm finds 1018 and 1522 candidates for {\tt NILC} and {\tt MILCA} respectively, out of which 500 and 457 correspond to objects present within the PSZ2 catalogue. For NILC (MILCA) we have 41 (25) positions that are not masked by the PSMASK and that do not correspond to PSZ2 clusters. However 14 (5) are located in regions excluded by the mask used to build the PSZ2 catalogue. 
With the {\tt MF} method we have 1472 and 1502 candidates for {\tt NILC} and {\tt MILCA}, out of which 1107 and 1096 correspond to objects present within the PSZ2 catalogue (867 and 835 corresponding to already validated objects, respectively). This implies a good agreement between the cluster sample and the detected sources in the $y$\/-maps.

\begin{figure}[t]
\begin{center}
\setlength{\unitlength}{\columnwidth}
\begin{picture}(1,1.5)
\put(0,1){\includegraphics[width=0.65\columnwidth]{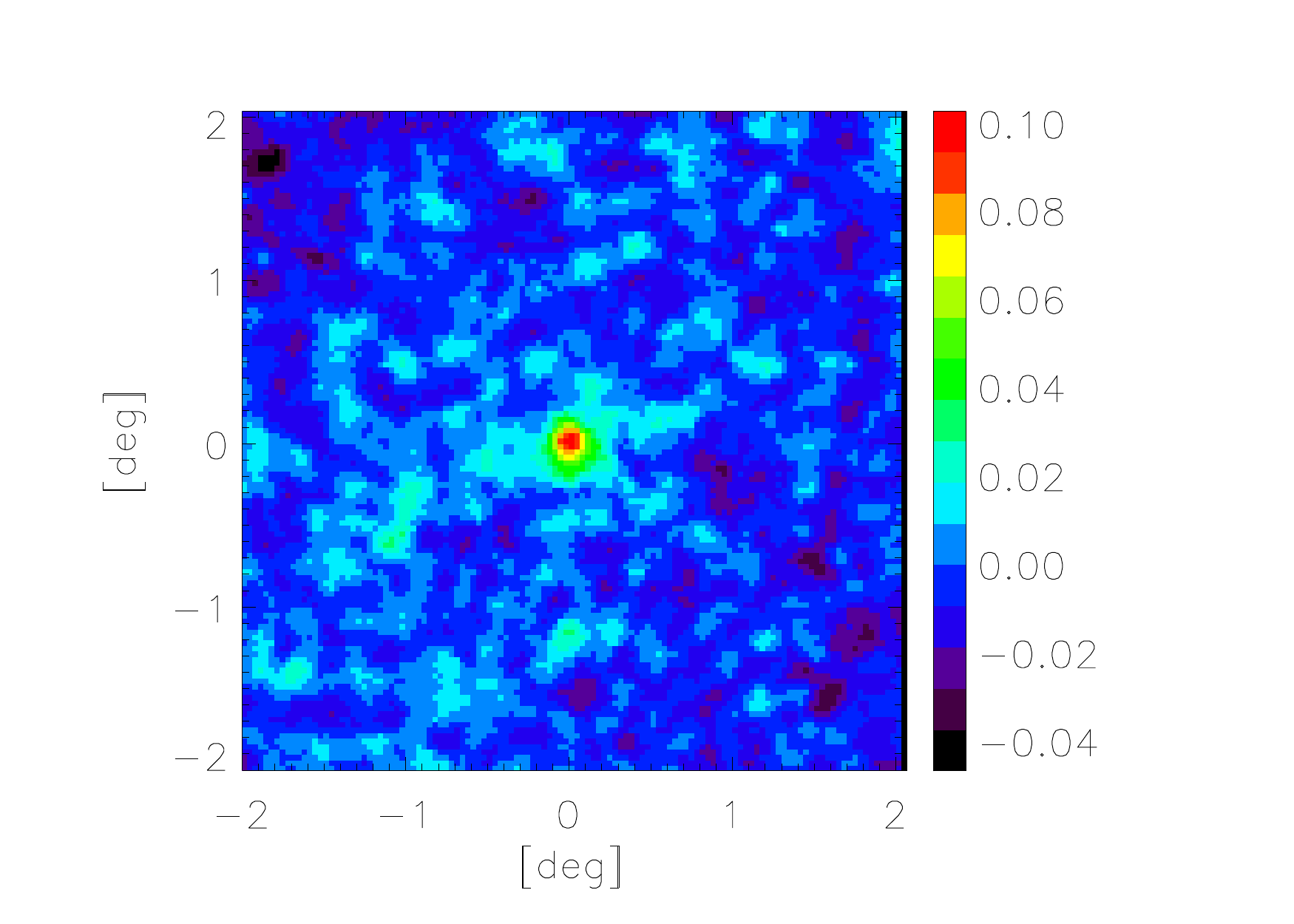}}
\put(0,0.5){\includegraphics[width=0.65\columnwidth]{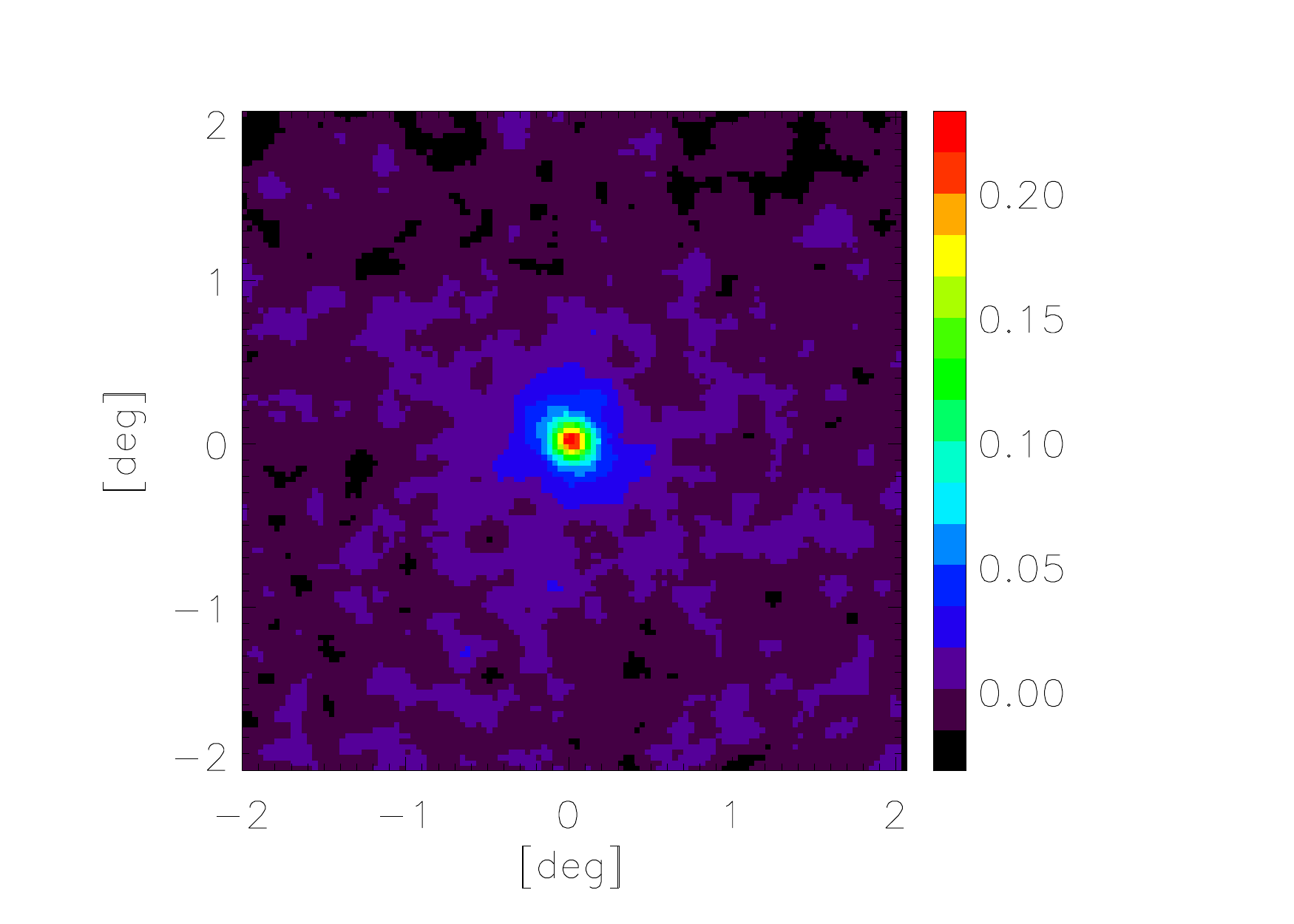}}
\put(0,0){\includegraphics[width=0.65\columnwidth]{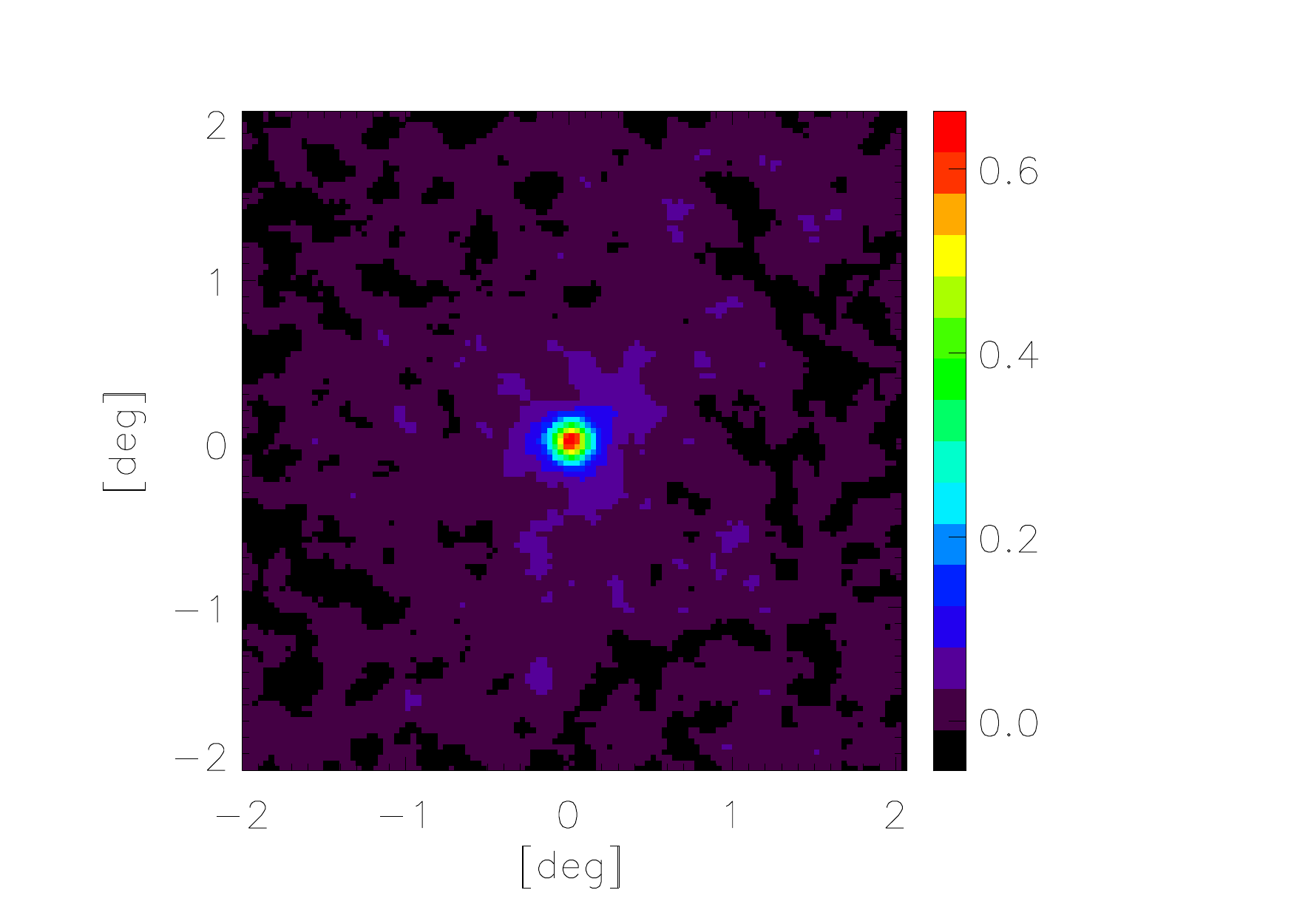}}
\put(0.5,1){\includegraphics[width=0.65\columnwidth]{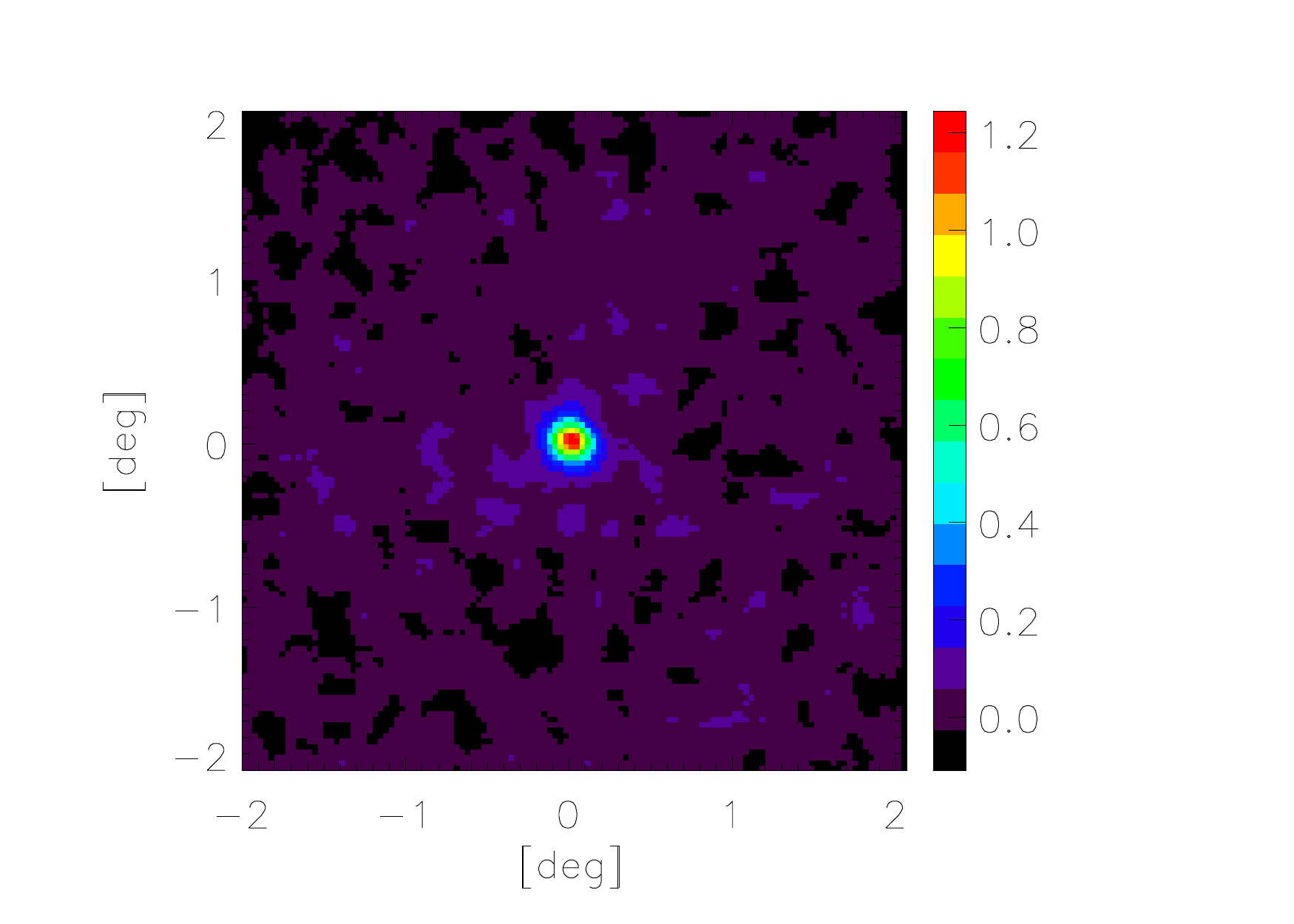}}
\put(0.5,0.5){\includegraphics[width=0.65\columnwidth]{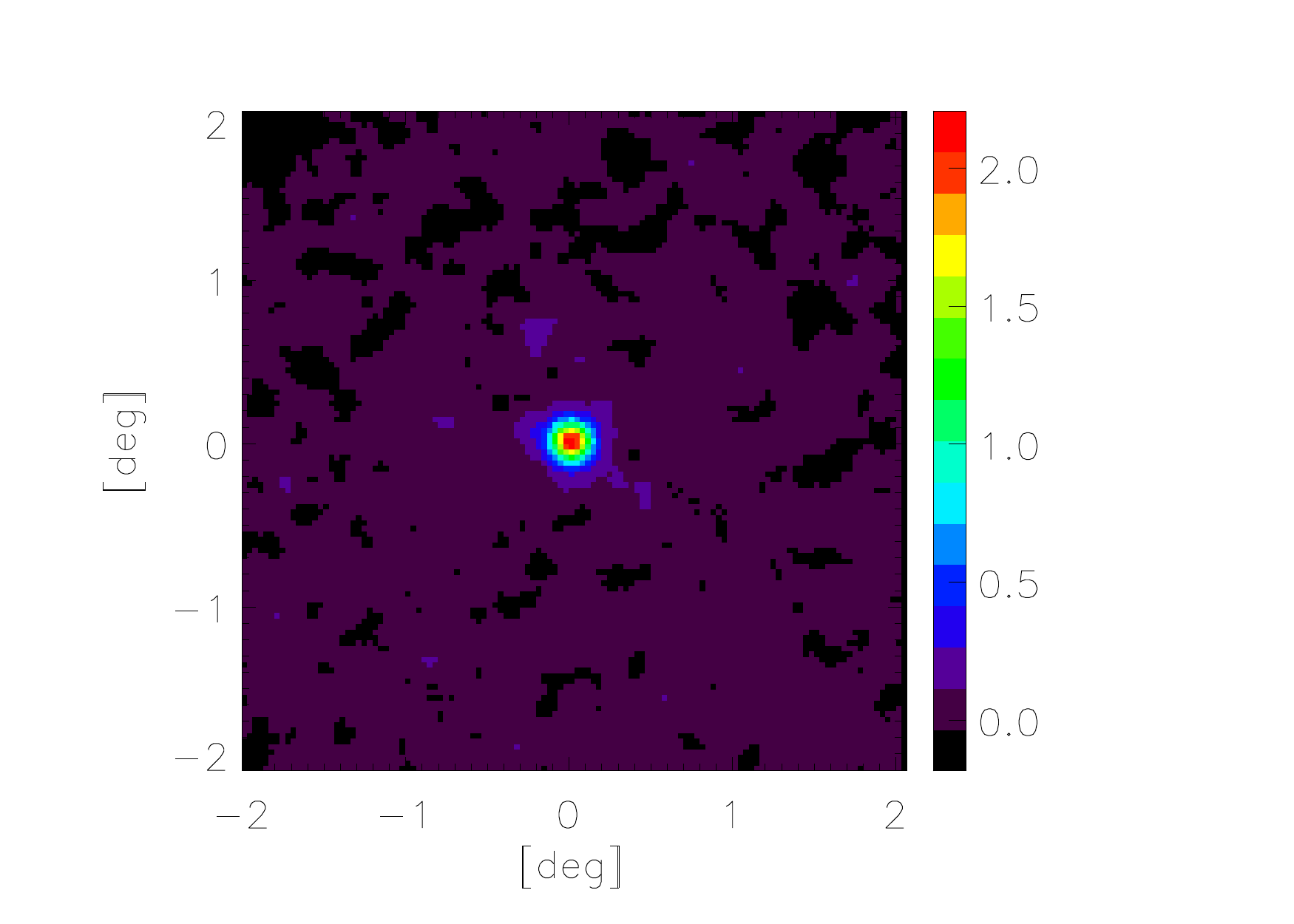}}
\put(0.5,0){\includegraphics[width=0.65\columnwidth]{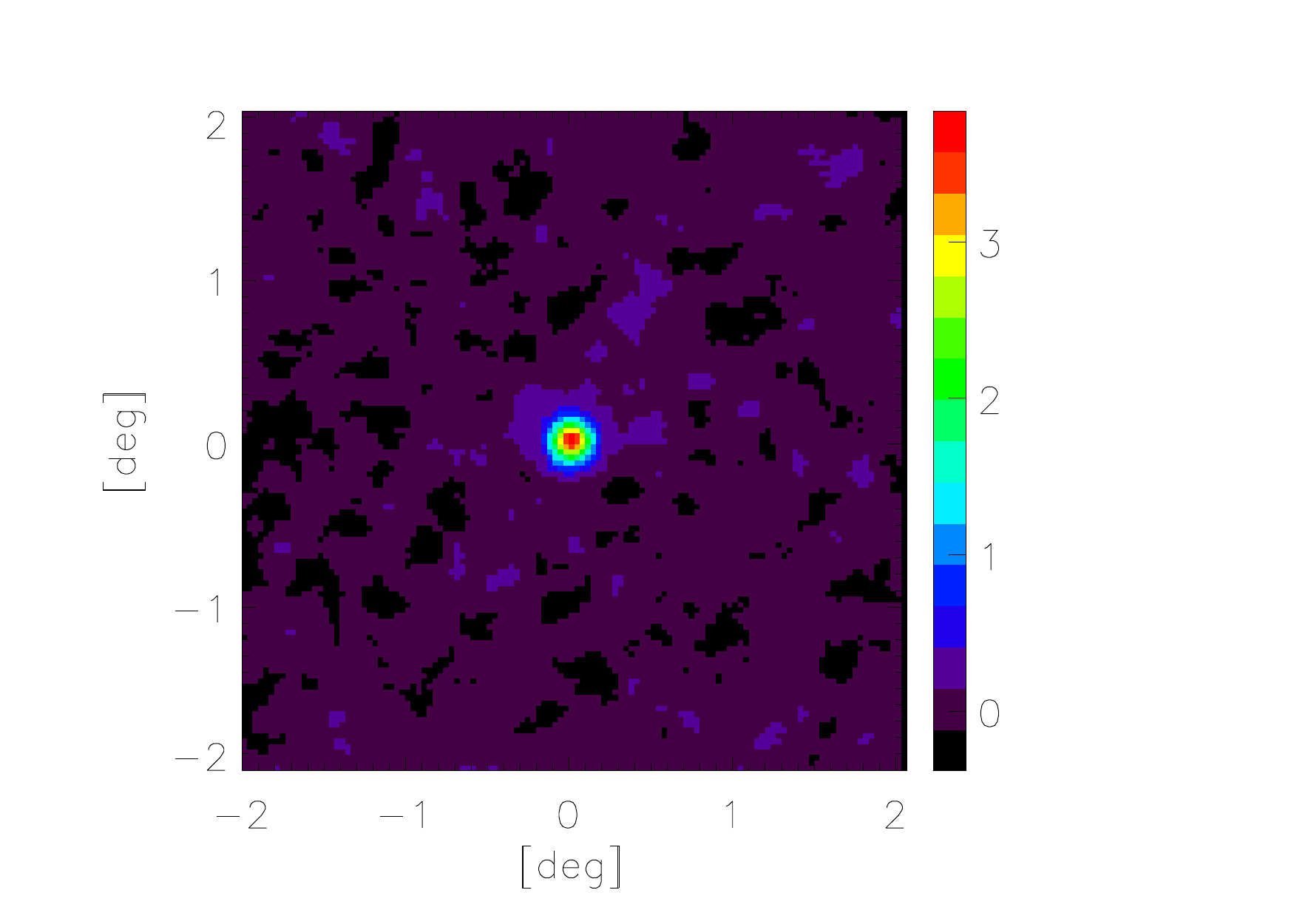}}
\put(0.05,1.45){(a) 8 $<  N_{200} \leq$ 10}
\put(0.05,0.95){(b) 10 $< N_{200} \leq$ 20}
\put(0.05,0.45){(e) 20 $< N_{200} \leq$ 30}
\put(0.55,1.45){(d) 30 $< N_{200} \leq$ 40}
\put(0.55,0.95){(e) 40 $< N_{200} \leq$ 60}
\put(0.55,0.45){(f) 60 $< N_{200} \leq$ 100}
\end{picture}
\end{center}
\caption{4$^{\circ}$ x 4$^{\circ}$ average maps for different ranges in $N_{200}$ from 8 to 100. The color scale is in unit of 10$^{-6}$ $y$. \label{SDSS_stack}} 
 
\end{figure} 

\begin{figure}
 \begin{center}
 \includegraphics [width=\columnwidth,height=\columnwidth]{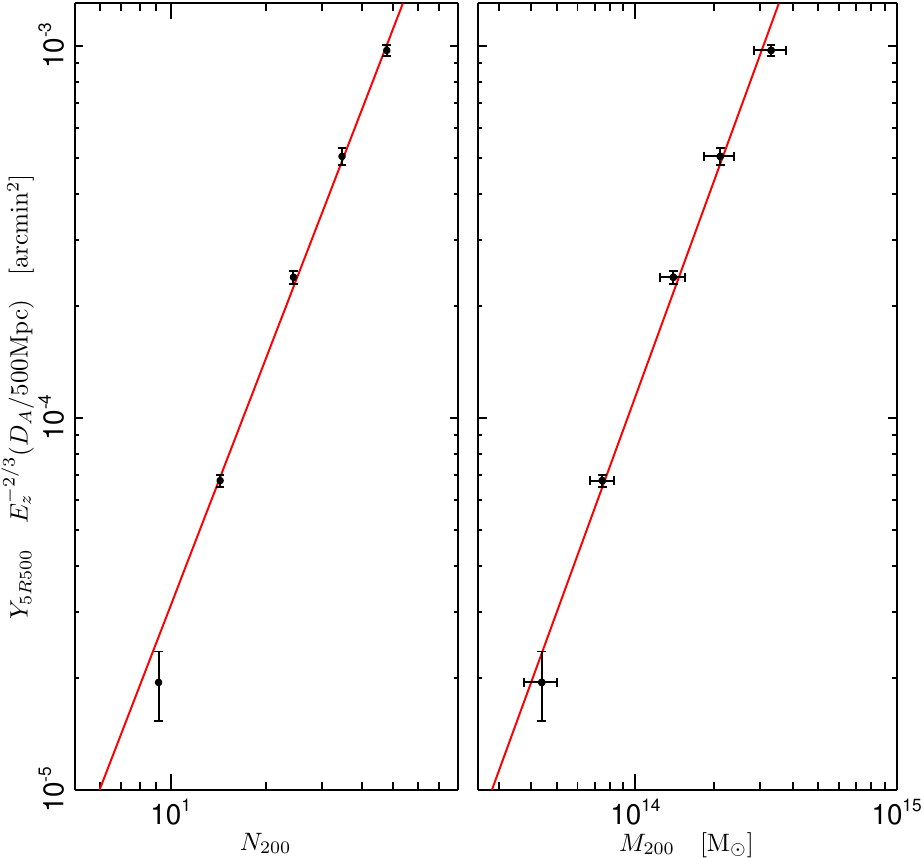}
\caption{Integrated tSZ signal as a function of cluster richness (left) and total mass (right). The black points correspond to the average signal obtained for the richness bins considered in Figure \ref{SDSS_stack}. The red line represents the corresponding best-fit power law (Eqs. \ref{sz_rich_rel} and \ref{sz_mass_rel} respectively). Considering $z~\le~$0.42, there are 13814 objects for 8 $<  N_{200} \leq$ 10, 37250 for 10 $< N_{200} \leq$ 20, 7458 for 20 $< N_{200} \leq$ 30, 2069 for 30 $< N_{200} \leq$ 40, and 1133 for 40 $< N_{200} \leq$ 60. \label{fig:scaling}}
\end{center}

\end{figure}

This agreement is improved by taking the union of the  {\tt MHW2} and the {\tt MF} catalogues. In this case, and considering only validated sources in the PSZ2 catalogue (1070 sources), we find 907 and 870 matches with blind detected cluster candidates for {\tt NILC} and {\tt MILCA}, respectively. We have also performed a visual inspection of the {\tt NILC} and {\tt MILCA} $y$\/-maps at the position of validated PSZ2 sources for which we find no counterpart and we find evidence of an excess of signal. For most of these sources low signal-to-noise ratio and/or foreground contamination can explain why they are not detected blindly in the $y$\/-map. This is consistent with the fact that we expect blind detection methods in the $y$\/-map to be less sensitive than multifrequency ones \citep{Melin:2012p2056}. In some cases these non-detected sources are extended but with a relatively high signal-to-noise (between 6 and 12). We notice that the {\tt MHW2} and {\tt MF} methods are tuned to detect point-like and compact sources primarily, and thus they may fail for extended ones.

We present in the top panel of Figure~\ref{fig:photopsz2} a comparison of the flux measured for these common sources in the {\tt NILC} $y$\/-map with that derived from the PSZ2. The $Y_{5r_{500}}$ correspond to the integrated signal within a radius equal to 5 $\times$ $r_{500}$\footnote{$r_{500}$ is the cluster-centric distance at which the mean cluster density is equal to 500 times the critical density of the Universe}.
We observe a good agreement between the two. Most of the observed outliers (low $y$\/-map flux with respect to the PSZ2 one) correspond to cluster candidates detected close to masked point sources, or in regions with strong Galactic contamination.
We also show in the bottom panel a comparison of the flux measured with the {\tt NILC} and {\tt MILCA} $y$\/-maps. We find good agreement between the two with three outliers
corresponding to sources lying either in a region badly affected by radio and/or infrared point sources or for which we observe large uncertainties in the estimated characteristic radius of the cluster candidate. For both maps ({\tt NILC} and {\tt MILCA}), the average signal-to-noise ratio of the recovered clusters with the {\tt MHW2} algorithm is $\sim$ 10, while the signal-to-noise ratio of PSZ2 clusters only found with the matched filter technique is, as expected, lower ($\sim$ 6).

\subsubsection{Maps of selected clusters}
We have performed a more detailed analysis for some of the cluster candidates in the PSZ2 sample. In Figure~\ref{fig:lowsnsources} we present Compton parameter maps centred at the position of three of the newly discovered \Planck\ clusters with signal-to-noise ratio of 9.3 (left), 6.2 (middle) and 4.6 (right) both for MILCA (top row) and NILC (middle row). In the bottom row of the figure we present the radial profiles of these clusters as obtained from the {\tt MILCA} (black lines) and {\tt NILC} (blue) maps. We find that the profiles are consistent between {\tt MILCA} and {\tt NILC}. For comparison we also show as a blue dashed line the radial profile of a 10$^{'}$ gaussian beam. We observe that even for low-signal-to-noise and compact clusters we are able to detect extended emission. Thus, these maps could be used to extend the pressure profile analysis presented in~\citet{planck2012-V,} to fainter and higher redshift clusters. \\

One of the major outcomes of the {\tt NILC} and {\tt MILCA} $y$\/-maps is the possibility to study diffuse faint emission between clusters as well
as the emission in the cluster outskirts. We show in Figure~\ref{fig:mergers} some well known merging systems including the {\tt Shapley} supercluster, and
the {\tt A3395-A3391}  and the {\tt A339-A401} interacting systems \citep[see][for a detailed description]{2013A&A...550A.134P}. 
We observe that the clusters themselves, their outskirts and the inter cluster emission are well reconstructed in the {\tt NILC} and {\tt MILCA} $y$\/-maps, which show consistent results. The quality of these maps will permit analyses similar to the one presented in \citet{2013A&A...550A.134P} with a significant increase of the signal-to-noise ratio. 

\subsection{SZ signal below the noise level}
\label{sz_bkgd}

We use the catalog of 132684 clusters of galaxies identified from the Sloan Digital Sky Survey III \citep{Wen2012} in order to quantify the tSZ signal below the noise
level in the $y$-map. This catalogue provides estimates of cluster redshift, richness ($N_{200}$) and characteristic radius ($r_{200}$).

We focus on unresolved groups and clusters for which the richness is $8 \leq N_{200} \leq 100$. These objects are
expected to be at signal-to-noise ratio well below 1 in the $y$\/-map and so their direct detection is not possible. However, their average tSZ signal can be detected using a stacking approach on the $y$-maps. Figure~\ref{SDSS_stack} shows the stacked maps (obtained with the {\it IAS stacking library}, \citealt{bavouzet_thesis, Bethermin2010}) for 6 richness intervals with $N_{200}$ going from 8 to 100, on patches of 4$^{\circ}$ x 4$^{\circ}$, for the full sky {\tt NILC $y$\/-map}. For the stacking we exclude all the positions for which point sources are found within a radius of 10 arcmin from the center of the object. 
The noise of the stacked maps scales as expected with $1/\sqrt{N}$ and the average signal increases for increasing richness. Then, when considering a sufficiently large number of objects, we are able to significantly detect the average SZ signal, even for small groups (Fig. \ref{SDSS_stack}).  We obtain consistent results for the {\tt MILCA}
and {\tt NILC} $y$\/-maps. 

For the $N_{200}$ intervals reported in Fig.~\ref{SDSS_stack} we have estimated the total stacked fluxes ($Y_{5r_{500}}$) as the integrated signal within a radius $5 \times r_{500}$, with $r_{500} = 0.7  \langle r_{200} \rangle$, $\langle r_{200} \rangle$ being the mean of the $r_{200}$ reported by \citet{Wen2012} for clusters belonging to each considered subsample. The average values and associated errors have been obtained with a bootstrap approach. For this, we have constructed and stacked cluster samples obtained by randomly replacing sources from the original sample, so that each of them contains a number of clusters equal to the initial one. By repeating the process several times, we have determined the statistical properties of the population being stacked. In Figure \ref{fig:scaling} (left panel) we report $Y_{5r_{500}}$ as a function of richness ($N_{200}$) for all the objects with $z \leq$ 0.42. In fact higher redshift clusters present in this catalogue may have a biased lower richness because of incompleteness of member galaxies, as detailed in \cite{Wen2012}. 
%In the left panel of Figure \ref{scaling} we report results for two different binning in richness, in black the one adopted in in \cite{Ford2014}, in cyan the one considered by \citep[][ 10-13, 14-17, 18-24, 25-32, 33-43, 44-58, 59-77, 78-104]{PEP2011}. 
Following \cite{PEP2011}, we have fitted a power law of the form
\begin{equation}
	Y_{5r_{500}} E^{-2/3}_z \left(\frac{D_A(z)}{500 {\mathrm~Mpc}}\right)^2=Y_{20}\left(\frac{N_{200}}{20}\right)^{\alpha},
	\label{sz_rich_rel}
\end{equation}
with $E(z)^{2} = \left(H(z)/H_0\right)^{2}$ and $D_A$ the angular diameter distance of the cluster. 

By considering only $N_{200} \le 60$ (for which the number of clusters in each richness bin is $>$ 1000), we find a slope $\alpha=2.21\pm0.10$, which is consistent with the scaling obtained by \cite{PEP2011}. 
We have verified that the scaling obtained is insensitive to wether we limit our cluster sample to high Galactic latitude objects or not, and that a change in the integration radius only affects the normalization $Y_{20}$. With $Y_{500}~=~Y_{5r_{500}}/1.79$ \citep{planck2013-p05a}, we obtain $Y_{20} = \left(8.07 \pm 0.41\right)\times 10^{-5} $arcmin$^2$.  However, for independent cluster datasets, different choices are made for the fainter magnitude of the member galaxies contributing to $N_{200}$. Then this may affect the richness associated to an object of given mass and complicate the comparison of the constant term $Y_{20}$ for analysis that are based on different cluster datasets. 

The mass-richness scaling relation,
\begin{equation}
	M_{200}=M_{0}\left(\frac{N_{200}}{20}\right)^{\beta},
	\label{mass_rich}
\end{equation}
 can then be used to also derive the $Y_{500}$-$M_{\mathrm{tot}}$ scaling (see Figure~\ref{fig:scaling}, right panel), by fitting a power low of the form
 \begin{equation}
 Y_{5r_{500}} E^{-2/3}_z \left(\frac{D_A(z)}{500 \mathrm{Mpc}}\right)^2=Y_0 \left(M_{200}\right)^B.
 \label{sz_mass_rel}
\end{equation}
\cite{Ford2014} compares different results for the scaling of the mass with richness, in the form of Eq. \ref{mass_rich}. In particular they discuss those obtained by \cite{Wen2012}, for a subsample of clusters with already known masses, and by \cite{Covone2014}, from weak lensing mass measurements of 1,176 clusters of the catalogue used here. 
Assuming $\beta=1.2\pm0.1$ and $M_0=(1.1\pm1.1) \times 10^{14} M_{\odot}$ \citep[][ and a ten percent error]{Wen2012, Covone2014} in Eq. \ref{mass_rich}, we compute via a Monte Carlo approach the average $M_{200}$ and associated uncertainties for each $Y_{5r_{500}}$ bin. From this we find $B=1.92\pm0.42$ for the $Y_{5r_{500}}$-$M_{200}$ scaling (Eq.~\ref{sz_mass_rel}). This is consistent with the self-similar value (5/3) and with what is expected for this mass range according to simulations \citep{MUSIC2014}. Through a weak lensing analysis of $\gtrsim 18000$ clusters in the CFHTLenS (Canada-France-Hawaii Telescope Lensing Survey), \cite{Ford2014} obtain a steeper calibration for the mass-richness relation, with $\beta=1.4\pm0.1$. For an higher $\beta$, the $Y_{5r_{500}}$-$M_{200}$ slope $B$ gets lower ($B=1.66\pm0.37$), but still consistent with the value obtained before. Again comparison of $M_0$ and $Y_0$ is complicated by the different choices of the fainter limit on galaxies contributing to $N_{200}$. By exploring the range $0.2~\lesssim~z~\lesssim~0.9$, \cite{Ford2014} finds no evolution with redshift. Thus, we do not expect our result to be biased by the $z \le$~0.42 cut.

\section{Angular power spectrum}
\label{sec:powerspec}

We now consider the validation of the Compton parameter maps at the power spectrum level.
 
\subsection{Power spectrum estimation}

To estimate the power spectrum of the tSZ signal we use the {\tt XSPECT} method \citep{Tristram2005} initially developed for the
cross-correlation of independent detector maps.  {\tt XSPECT} uses
standard {\tt MASTER}-like techniques \citep{2002ApJ...567....2H} to
correct for the beam convolution and the pixelization, as well as the
mode-coupling induced by masking foreground contaminated sky
regions.

In the following, all the spectra will use a common multipole binning scheme, which was defined in
order to minimize the correlation between adjacent bins at low
multipoles and to increase the signal-to-noise at high multipole
values. We will also only consider cross angular power spectra between the $y$\/-maps
reconstructed from the first  (F) and second (L) halves of the data.
This allows us to avoid the bias induced by the noise in the auto angular power spectrum, the correction of which would require a large number of noise simulations.
The cross angular spectra are named in the following {\tt NILC} F/L and {\tt MILCA} F/L.
Error bars in the spectrum are computed analytically from the auto-power and cross-power spectra of the pairs of maps, as
described in \citet{Tristram2005}. We do not consider here higher order terms to the power spectrum variance.
All of our Compton parameter maps assume a circular Gaussian beam of 10$^{\prime}$ FWHM.  
The additional filtering at large angular scales in the {\tt MILCA} Compton parameter maps is also accounted for and deconvolved.

 \begin{figure*}
 \centering
 \includegraphics [width=0.98\columnwidth]{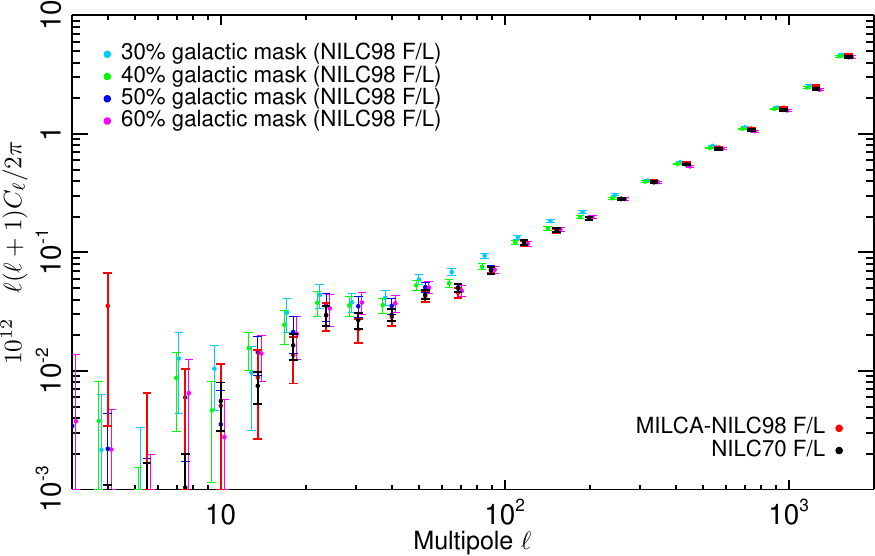}
\includegraphics [width=0.98\columnwidth]{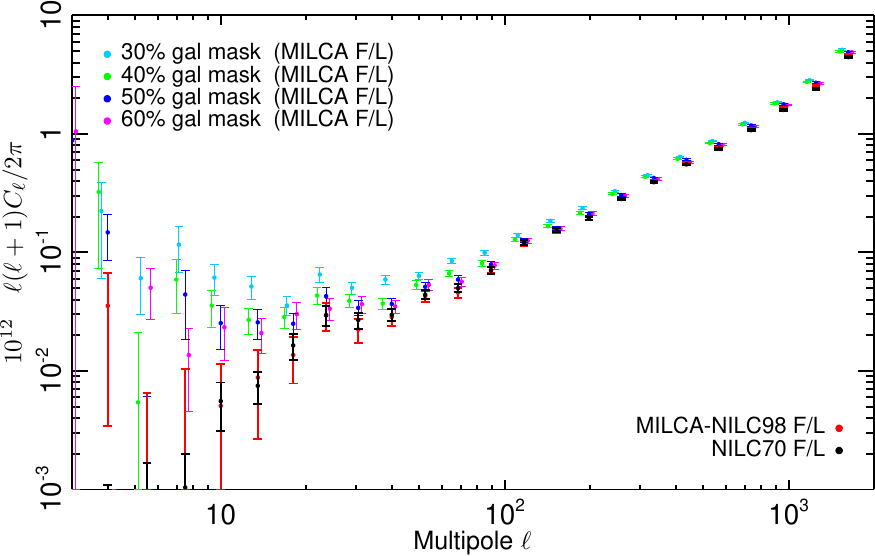}
\caption{Angular cross-power spectra of the \Planck\ {\tt NILC} F/L (left) and {\tt MILCA}  F/L (right)
reconstructed Compton parameter maps for different Galactic masks corresponding to 30 \% (cyan), 40 \% (green), 50 \% (blue)
and 60 \% (pink) of the sky. For comparison we also show {\tt MILCA}-{\tt NILC} F/L (red) and {\tt NILC70} F/L (black)
on 50 \% of the sky. See text for details. \label{fig:milca_nilc_cross_galcut}}
\end{figure*}

\subsection{Foreground contamination}
\label{subsec:forecont}

In \citet{planck2013-p05b} we have identified the dominant foreground contributions to the  angular power spectrum of the reconstructed $y$\/-maps
using the FFP6 simulated data set. We have repeated here the same analysis on updated simulated data sets (see Sect.~\ref{subsec:simulations}) for which
the description of foreground components has been improved according to our current knowledge \citep{planck2014-a14}. We find no major difference between the two analyses
and therefore we do not repeat the discussion here. At large angular scales the dominant foreground contribution is the Galactic thermal
dust emission as we have already discussed in Section~\ref{subsec:galacticemission}. At intermediate and small angular scales the major
contribution comes from the clustered CIB emission. Equally important at small angular scales are the residual contribution from radio and IR sources.

 \begin{figure*}
 \centering
 \includegraphics [width=\columnwidth]{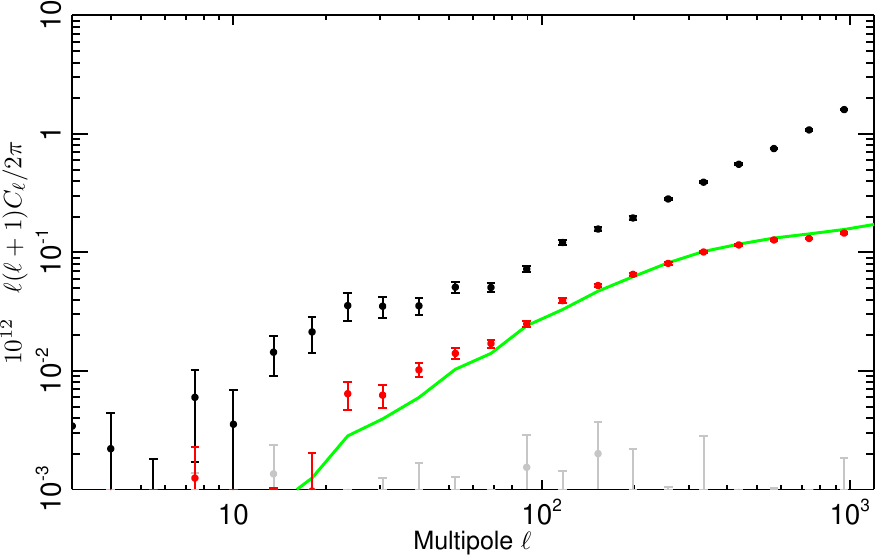}
 \includegraphics [width=\columnwidth]{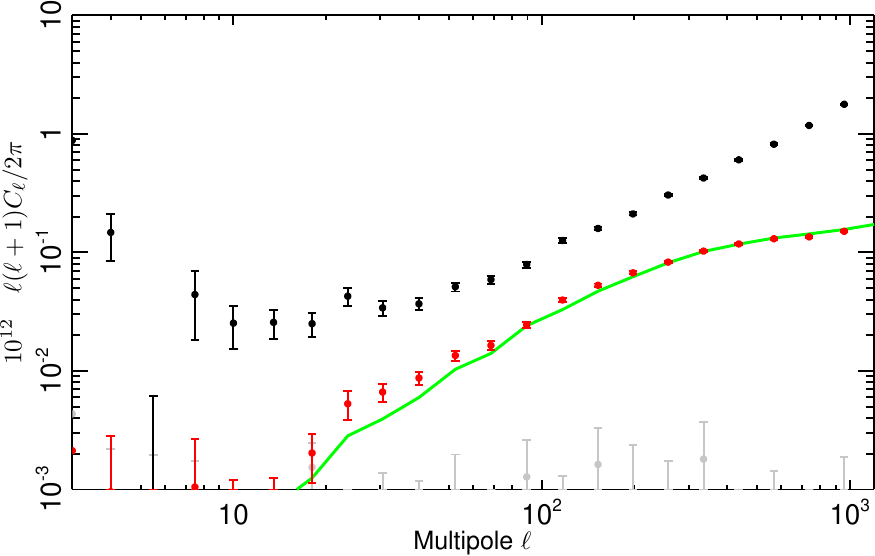}
\caption{Comparison of the tSZ angular power spectrum estimated from 
  the cross-power-spectrum of the {\tt NILC} (left) and  {\tt MILCA} (right) F/L maps (black) with the
  expected angular power spectrum of the confirmed clusters in the
  \Planck\ Cluster Sample (green line).  The angular cross-power spectrum between the {\tt NILCA} and {\tt MILCA}
  Compton parameter maps and the simulated detected cluster map is shown in red. Tthe correlation between the reconstructed $y$-map and the simulated detected cluster map, to which an arbitrary rotation has been applied, is plotted in grey. \label{fig:cross_cat}}
 \end{figure*}

\subsubsection{Low-multipole contribution \label{sec:lowell}}
%
%The diffuse Galactic foreground contribution can be significantly
%reduced by choosing a more aggressive Galactic mask.  
Assuming that at large angular scales the Compton parameter maps are mainly affected by
diffuse Galactic dust emission, we have tested several Galactic masks
by imposing flux cuts on the \Planck\ 857\,GHz channel intensity map.
In particular we investigated masking out 30\%, 40\%, 50\% and 60\% of
the sky.  The edges of these masks have been apodized to limit ringing
effects on the reconstruction of the angular power spectrum.
Figure~\ref{fig:milca_nilc_cross_galcut} shows the power spectra for {\tt NILC} (left) and
{\tt MILCA} (right). We observe that the {\tt MILCA} F/L large angular scales power
decreases when imposing a more severe masking (larger fraction of the sky is masked out).
A similar effect, but significantly smaller, is also observed for the {\tt MILCA} F/L cross power spectrum. The {\tt MILCA}  $y$\/-map is significantly more contaminated
by thermal dust emission at multipoles below 30. However, we also observe for {\tt NILC} F/L an extra noise contribution
at large angular scales as discussed in Section~\ref{subsec:noise}. These two problems limit the reliability of the
results at multipoles below 30.

To extend the measurement of the angular power spectrum of the tSZ emission to multipoles below 30 we have
considered two options: 1) as in \citet{planck2013-p05b} we apply a more severe galactic mask (30 \% of the sky  is masked) before the computation of the {\tt NILC} weights  to produce the $y$\/-map ({\tt NILC70}), and 2) compute the cross-correlation of {\tt NILC} and {\tt MILCA} $y$\/-maps.
Considering 50\%  of the sky we show in Figure~\ref{fig:milca_nilc_cross_galcut} the angular power spectrum of the {\tt NILC70} $y$\/-map (black) and
the cross spectrum of the {\tt NILC} first half and {\tt MILCA} second half ( {\tt NILC}-{\tt MILCA} F/L) $y$\/-maps. We observe that the two are compatible within error bars in the multipole range $10 < \ell <1500$. As expected the cross spectrum {\tt NILC}-{\tt MILCA} F/L shows larger error bars. 

Although the {\tt NILC70} $y$\/-map seems to be the best choice in terms of power spectrum estimation, it results in a significant reduction of the
available sky area for other kind of studies as those presented in Section~\ref{sec:pixelanalysis}. Furthermore, it is difficult to accurately estimate the ultimate residual
foreground contribution at very large angular scales. Because of this and to preserve the coherence of the delivered products and the analysis presented in this paper,
we have chosen the cross angular power spectrum of {\tt NILC}-{\tt MILCA} F/L as a baseline. This is obviously a more conservative choice in terms of noise induced uncertainties.
The {\tt NILC}-{\tt MILCA} F/L  cross angular power spectrum bandpowers and uncertainties are further discussed in Sect. \ref{sec:cosmo}.
Using the in-scan and cross-scan $y$\/-maps presented in Section~\ref{subsec:systematics} we find that stripe contamination accounts for less than 10\% of the total signal in 
the {\tt NILC}-{\tt MILCA} F/L  cross angular power spectrum. We have enlarged the error bars to account for this systematic effect.

\subsubsection{High-multipole contribution}
\label{refinedforegroundmodel}

At small angular scales the measured tSZ power spectrum is affected
by residual foreground contamination coming from clustered CIB emission
as well as radio and IR point sources. They show up in the {\tt MILCA}-{\tt NILC} F/L
cross power spectrum (see Figure~\ref{fig:milca_nilc_cross_galcut})
% (see Figures~\ref{fig:milca_nilc_cross_galcut} and \ref{fig:powerspectcosmo}) \
as an excess of power at large multipoles. 

To deal with those we adopt the same strategy as in \citet{planck2013-p05b}.
We define physically motivated models of the angular power spectrum
of the foreground components for each observation channel, including 
cross correlations between channels. In contrast to \citet{planck2013-p05b} we
also account for the cross correlation between the clustered CIB and the tSZ emission. 
A detailed description of this cross correlation as well as of the clustered  CIB model is presented
in \citet{planck2014-a29}. For the radio and IR point source models we refer to \citet{planck2013-p05b}.

Using the models described above we compute Gaussian realizations of the foreground contribution 
for each HFI frequency channel between 100 and 857~GHz.  Note that the LFI channels are only used at large angular scales.
We apply the {\tt MILCA} or {\tt  NILC} weights to these simulated maps. 
%In order to compute the foreground contribution to the {\tt NILC}-{\tt MILCA} F/L cross spectrum a set
%of 50 simulations is performed and averaged out.
From these simulations we find that the cross correlation between the CIB and tSZ
contribution can be neglected to first order with respect to the others and it will therefore not be considered hereafter.
Uncertainties on the parameters describing the foreground models were also propagated
using simulations. We find that the clustered CIB model uncertainties might be as large as 50\%
in amplitude. In addition, we notice that the amplitude of the point source models can vary significantly
with the point source mask applied. These uncertainties are taken into account hereafter.
The amplitude of the residual foreground models are jointly fitted with the cosmology-dependent tSZ model, 
as detailed in Sect. \ref{powerspectcosmo}.
%In Figure \ref{fig:powerspectcosmo} we present, for illustration purposes, the angular power spectrum of the clustered CIB (green), and radio (blue) and IR (cyan) point source contribution to the
%$y$\/-map. Notice that the amplitudes have been adjusted to fit the {\tt NILC}-{\tt MILCA} F/L cross spectrum
%over 50 \% of the sky (black data samples). The solid red line also shows the best-fit tSZ power spectrum model that is presented in Sect. 7.1.

\subsection{Contribution of resolved clusters to the tSZ power spectrum}
\label{subsec:resolvedclusters}

We simulate the expected Compton parameter map for the detected and
confirmed clusters of galaxies in the \Planck\ catalogue~\citep{planck2013-p05a} from their measured
integrated Compton parameter, $Y_{\mathrm{5r_{500}}}$ and redshift, $z$. 
We assume hydrostatic equilibrium and an \citet{Arnaud2010} pressure profile.
The green solid line in Fig.~\ref{fig:cross_cat} shows the power spectrum of this simulated map.
Figure~\ref{fig:cross_cat} also shows the cross-power spectra of the
{\tt NILC} and {\tt MILCA} F/L maps (in black). We also compute the
cross-power spectrum of the simulated cluster map and the
\Planck\ reconstructed  {\tt NILC}  and {\tt MILCA}  Compton parameter maps. This is shown
in red in the figure. Here again, the signal is consistent with the
expected power spectrum of the confirmed \Planck\ clusters of
galaxies. These results show that the tSZ signal from clusters is preserved in the $y$\/-maps.

%These results show that a significant fraction of the
%signal in the reconstructed \Planck\ Compton parameter maps is due to
%the tSZ effect of detected and confirmed clusters of
%galaxies, verifying the SZ nature of the signal. % In addition, by

\section{Higher order statistics}
\label{sec:higorderstat}

The power spectrum analysis presented above only provides 
information on the 2-point statistics of the Compton parameter
distribution over the sky. An extended characterization of the field can be
performed by studying the higher-order moments in the 1D PDF of the
map, or by measuring 3-point statistics, i.e., the bispectrum.

\begin{figure*}
\includegraphics[width=\columnwidth]{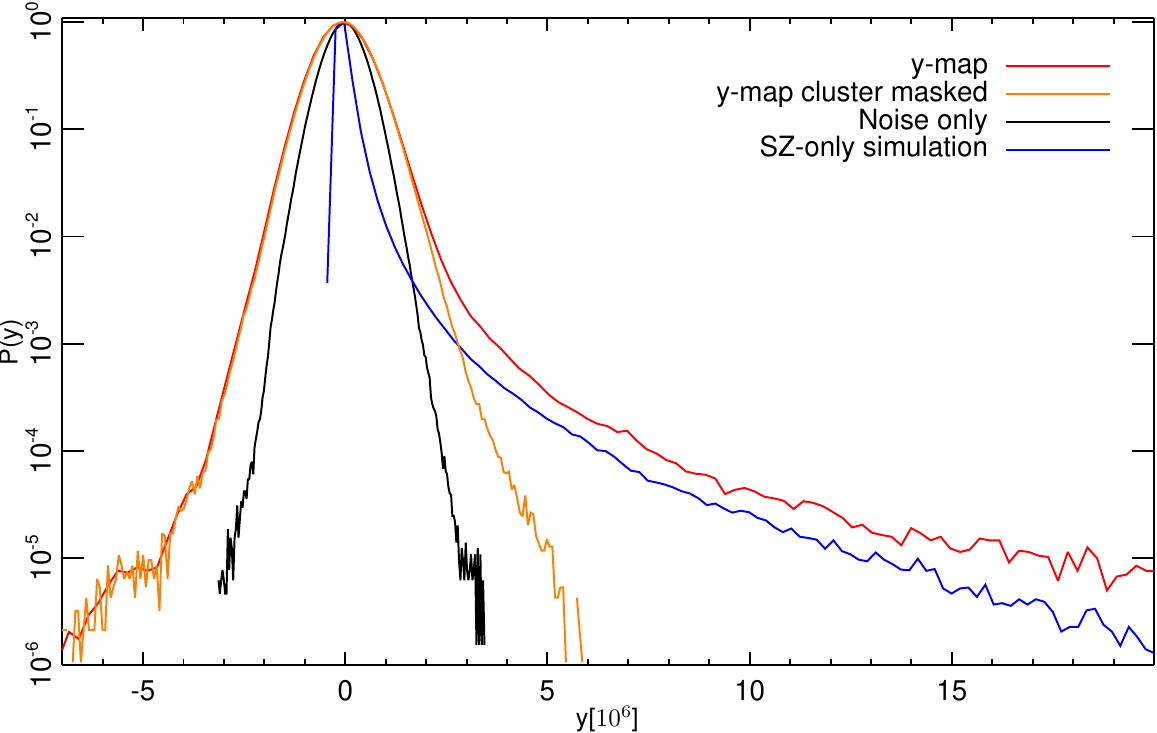}
\includegraphics[width=\columnwidth]{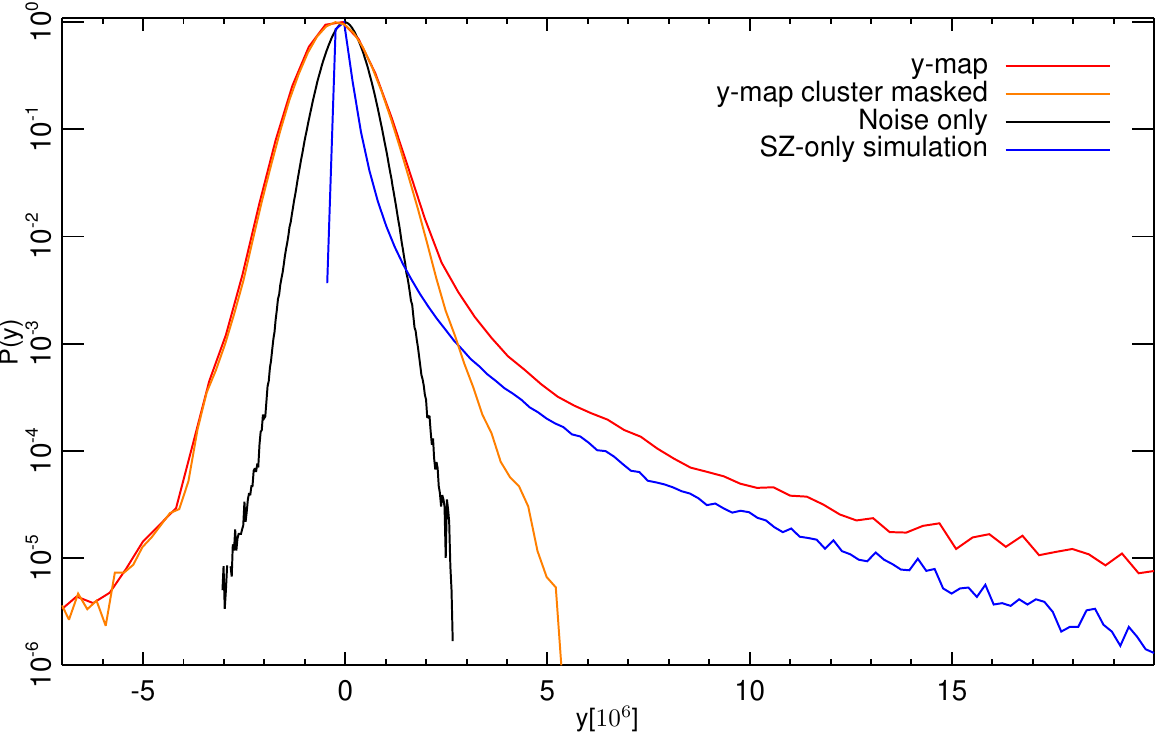}
\caption{1D PDF of the \Planck\ $y$-map before (red) and after (orange) masking the PSZ2 clusters, and of the half-difference map (black)
for the {\tt NILC} (left) and {\tt MILCA} (right) methods. We also show for comparison
the 1D PDF of the simulated PSZ2 cluster map (dark blue).
\label{fig:py_data}}
\end{figure*}

\subsection{1D PDF analysis}
\label{methods:1D PDF}

Following \cite{planck2013-p05b} we perform an analysis of the 1D PDF 
of the {\tt NILC} and {\tt MILCA} 
reconstructed Compton parameter maps. For the tSZ effect we
expect an asymmetric distribution with a significantly positive tail
\citep{RubinoMartin:2003p1790}. We thus focus on the asymmetry of the
distribution and its unnormalized skewness.  
First, we filter the maps in order to enhance the tSZ signal with
respect to foreground contamination and noise.  
We apodise the combined PSMASK and Galactic mask to
avoid residual point source ringing. 
We follow the approach of \cite{Wilson:2012p2102} and use a filter in harmonic
space. For each multipole $\ell$, the filter is constructed as the ratio between the angular power spectrum of
the expected tSZ signal ($C_{\ell}^{{\mathrm tSZ}}$, obtained from the simulations in Sect.~\ref{subsec:simulations}) and the
power spectrum of the half-difference $y$ maps ($C_{\ell}^{\mathrm N}$, see Sect.~\ref{subsec:noise}) such that 
$F_{\ell} = C_{\ell}^{{\mathrm tSZ}}/C_{\ell}^{\mathrm N}$. We smooth this filter function in multipole space using a
21-point square kernel and normalize it to one.  Notice that this filter only selects the multipole
range for which the tSZ signal is large with respect to the noise, and
thus, it does not modify significantly the non-Gaussianity properties of the $y$\/-maps.  Furthermore,
we have found that the filter used here behaves better than the more
traditionally used Wiener filter, as it is less affected by
point-source ringing. Following this procedure, the 1D PDF of the filtered
Compton parameter maps, $P(y)$, is computed from the histogram of the pixels.
As for the power spectrum, several Galactic masks have been considered in order to tests the robustness of the results.

Figure~\ref{fig:py_data} shows the 1D PDF for
the {\tt NILC} (left) and {\tt MILCA} (right) Compton parameter map in red
when masking 50\% of the sky. Each of them corresponds to the
convolution of the 1D PDF of the different components in the map: the tSZ
effect; foregrounds; and the noise. The 1D PDF of the {\tt NILC} and {\tt MILCA} $y$\/-maps clearly show three
distinct contributions: a Gaussian central part that is
slightly wider than the noise contribution, as expected from the half-difference map 1D
PDF (black curve); a small negative tail, corresponding most likely to
residual radio sources; and a positive tail corresponding mainly to
the tSZ signal as observed for the PSZ2 cluster simulated map (dark blue).
In comparison to the \citet{planck2013-p05b} we find now
a better agreement between the {\tt NILC} and {\tt MILCA} results.
This is mainly due to the improved masking of radio sources presented in
Section~\ref{subsec:pointsourcecont}. Finally we also show the 1D PDF of the
reconstructed $y$\/-maps after masking the PSZ2 clusters (orange). We observe that most of
the tSZ is removed, indicating that the 1D PDF of the $y$\/-map is dominated by
detected clusters.

A simple analysis of the measured 1D PDF can be performed by considering
the asymmetry of the distribution:
\begin{equation}
A\equiv \int_{y_{\mathrm{p}}}^{+\infty}P(y)dy - \int^{y_{\mathrm{p}}}_{-\infty}P(y)dy,
\end{equation}
where $y_{\mathrm{p}}$ is the peak value of the normalized
PDF $P(y)$.  In addition, the non-Gaussianity of
the positive tail can be quantified by
\begin{equation}
\Delta =  \int_{y_{\mathrm{p}}}^{+\infty}\left[P(y)-G(y)\right]dy,
\end{equation}
with $G(y)$ the expected PDF if fluctuations were only due to
noise. The latter can be obtained from the half-difference $y$\/-maps.

Masking 60 \% of the sky, we find $A=0.218$ and
$\Delta=0.10$ for the {\tt NILC} Compton parameter map.  
Equivalently, for the {\tt MILCA} Compton parameter
map we find $A=0.223$ and $\Delta=0.11$.  These results are 
in agreement and consistent with a positive tail in the 1D PDF, confirming the tSZ nature of the signal.
The agreement between the {\tt NILC} and {\tt MILCA} results degrade when reducing the
masked area as a consequence of a larger foreground contribution in the {\tt MILCA} $y$\/-map.
See \cite{Hill_py_2014} for a similar analysis conducted on ACT data.

\begin{figure}
\begin{center}
\setlength{\unitlength}{\columnwidth}
\begin{picture}(1,1.0)
\put(0.55,0.0){\includegraphics[width=0.45\columnwidth]{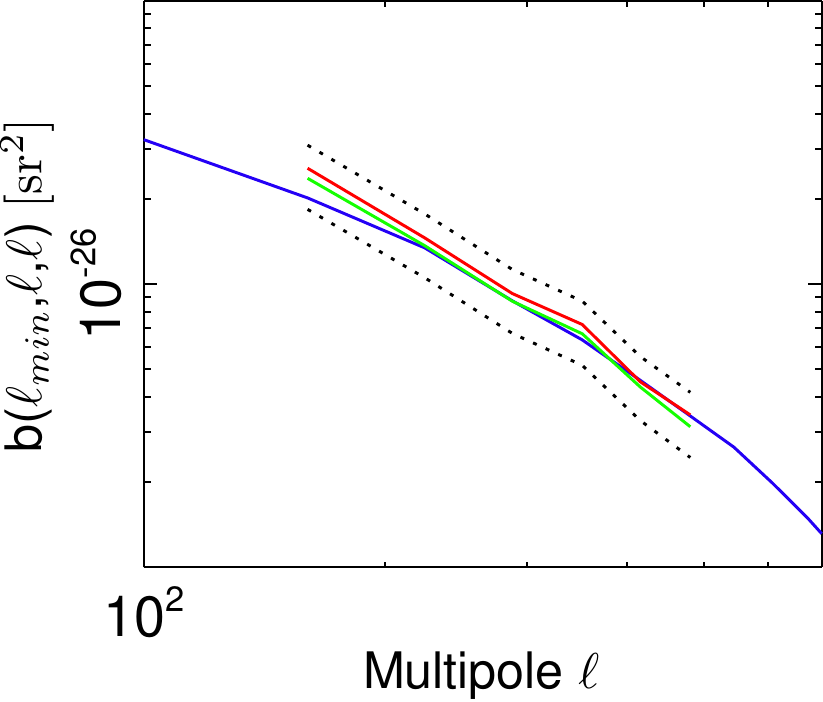}}
\put(0.55,0.50){\includegraphics[width=0.45\columnwidth]{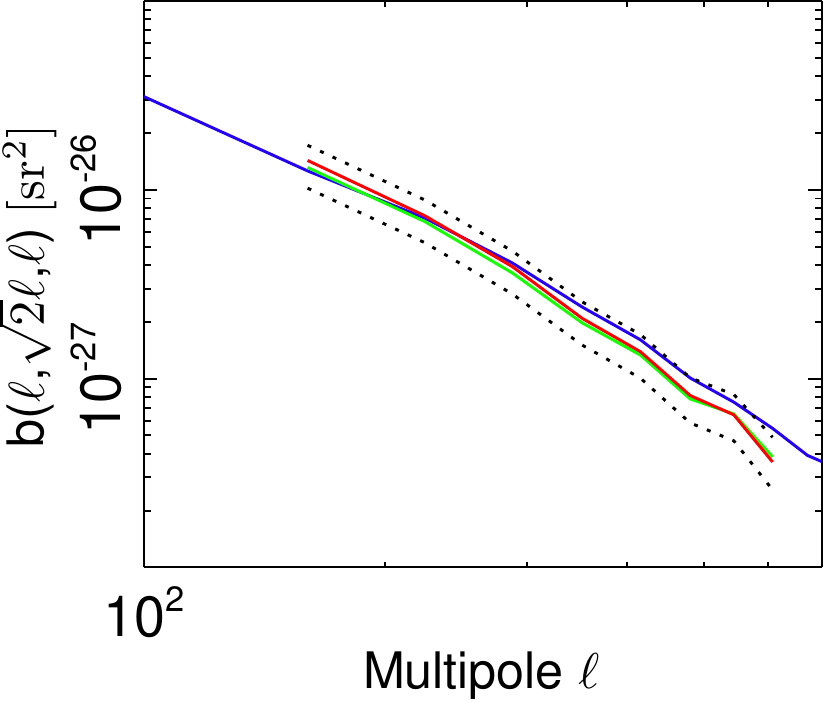}}
\put(0.0,0){\includegraphics[width=0.45\columnwidth]{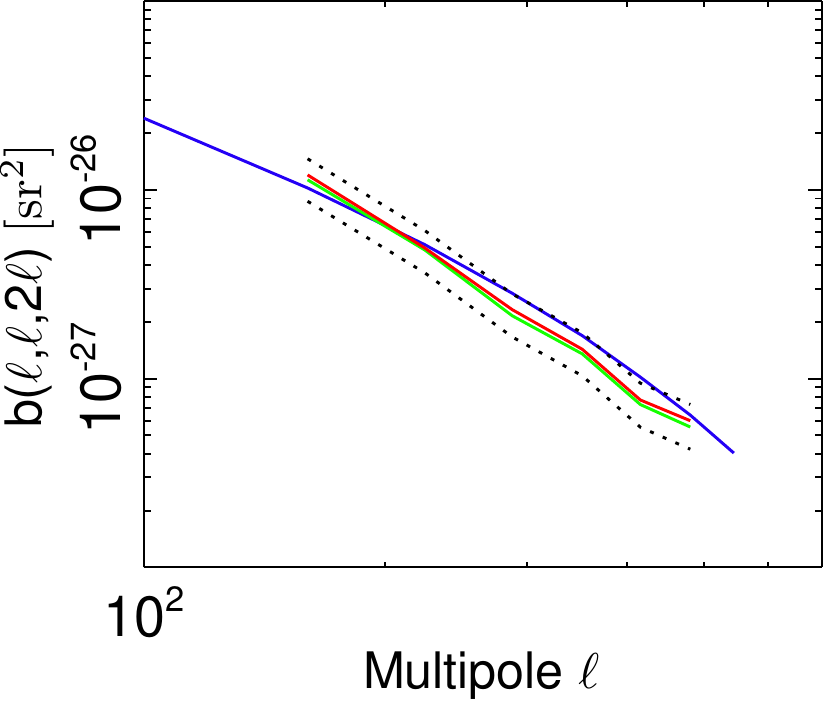}}
\put(0.0,0.50){\includegraphics[width=0.45\columnwidth]{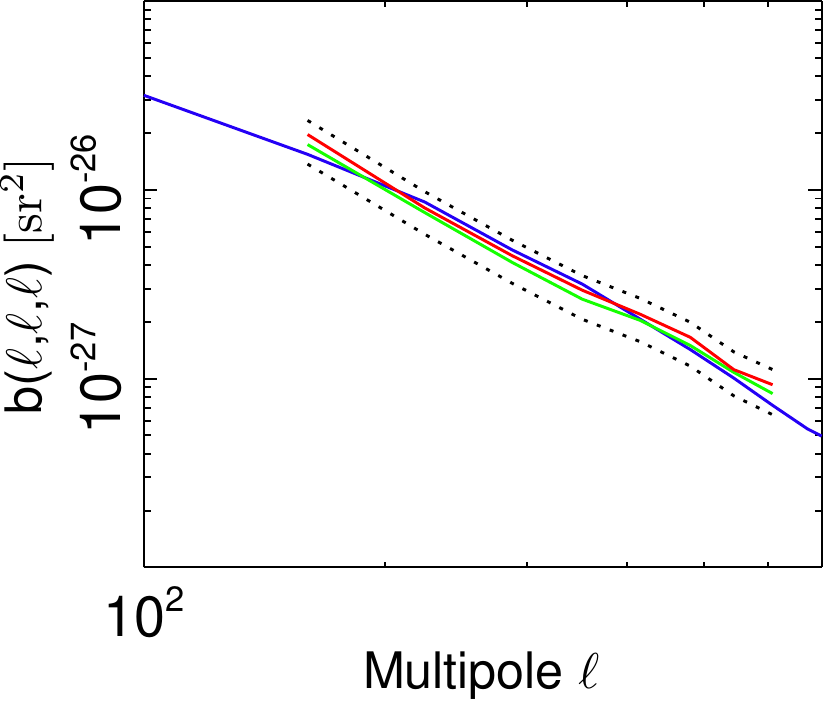}}
\put(0.05,0.95){(a) Equilateral}
\put(0.05,0.45){(c) Flat}
\put(0.6,0.95){(b) Orthogonal}
\put(0.6,0.45){(d) Squeezed}
\end{picture}
\caption{Bispectra of the {\tt NILC} (green) and  {\tt MILCA} (red) $y$\/-maps for four different configurations (equilateral, orthogonal, flat and squeezed), compared with the bispectrum of the projected map of the PSZ2 clusters (blue). $\pm 1\sigma$ uncertainties are indicated as black dotted lines.  \label{Fig:bispectre}}
\end{center}
\end{figure}
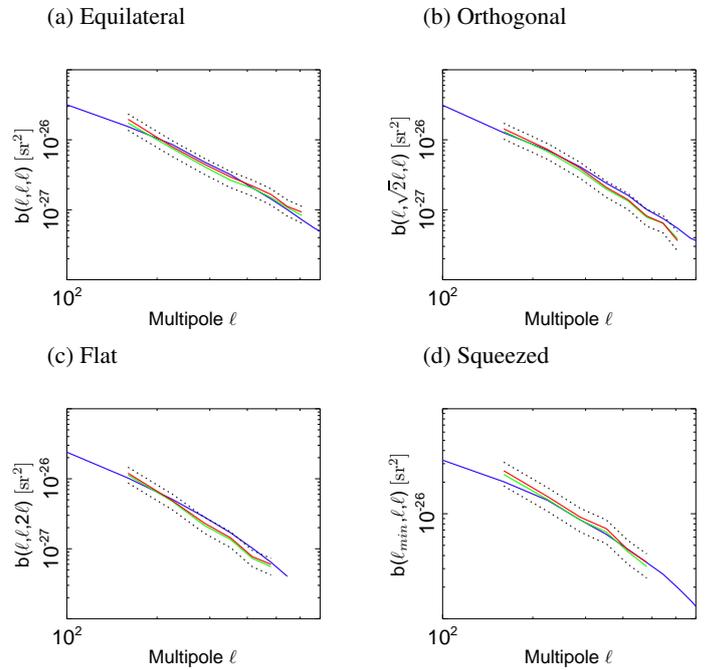

\begin{figure}[t]
 \centering
\includegraphics [width=\columnwidth]{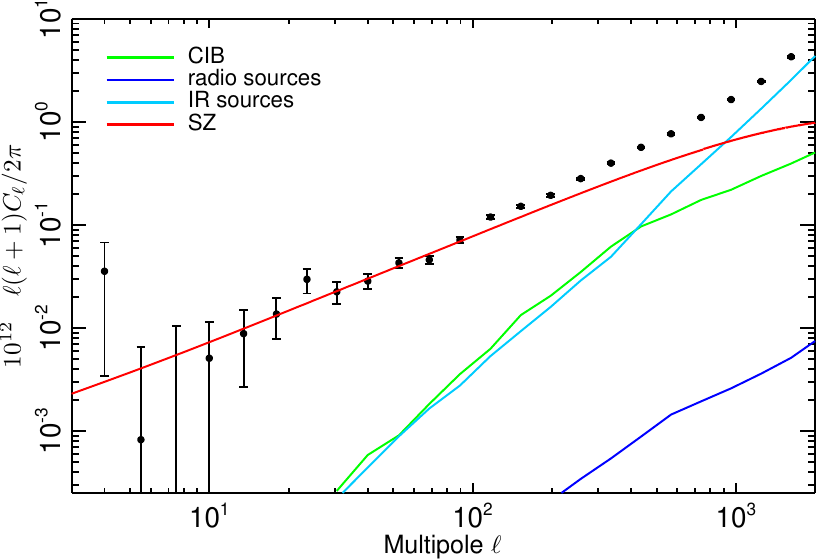}
\caption{{\tt NILC} - {\tt MILCA} F/L cross-power spectrum (black) compared to the power spectra of the
physically motivated foreground models.  The considered foregrounds are: clustered CIB
(green line); infrared sources (cyan line); and radio sources (blue line). 
 Additionally, the best-fit tSZ power spectrum model presented in
Sect.~\ref{powerspectcosmo} is also plotted as a solid red line.
 \label{fig:powerspectcosmo}}
 \end{figure}

\subsection{Bispectrum}
\label{bispectrumm}

Since the SZ signal is non-Gaussian, significant statistical information is  contained in the bispectrum, complementary to the power spectrum \citep{RubinoMartin:2003p1790,Bhattacharya:2012p2458}. Results on SPT data have been obtained by \citet{Crawford_2014} and \citet{George_2014}. The bispectrum also provides an additional statistic to assess the compatibility of the {\tt NILC} and {\tt MILCA} reconstructed Compton parameter maps, as well as their reliability in terms of foreground contamination.

We therefore analyze the bispectra of the {\tt NILC} and {\tt MILCA} maps. The estimation method is essentially the same as for the 2013 results \citep{planck2013-p15}, briefly recapped here. We use the binned
bispectrum estimator described in \citet{Bucher2010} and \citet{Lacasa2012}, which is also used for the \Planck\ primordial non-Gaussianity analysis \citep{planck2013-p09a, planck2014-a19}. We use a multipole bin size $\Delta \ell = 64$ and a maximum multipole $\ell_\mathrm{max} = 2048$ for the analysis, working at a resolution $N_{\mathrm{side}}=1024$ to reduce computing time. The {\tt NILC} and {\tt MILCA} maps are masked with the combination of PSMASK, described in Sect.~\ref{subsec:pointsourcecont}, and a Galactic mask at 50, 60 or 70\%, described in Sect.~\ref{sec:lowell} (in the rest of this section we will simply denote as X\% mask the combination of PSMASK and the Galactic mask at X\%). The best-fit monopole and dipole outside the mask are finally removed before estimation.

An important part of the pipeline is then to correct for the bias introduced by masking. To this end, we compute the ratio of the full-sky and masked sky bispectra, on highly non-Gaussian simulations with a tSZ-like bispectrum and a 10$^{\prime}$ resolution. This ratio is used to correct the measured bispectra and flag unreliable $(\ell_1,\ell_2,\ell_3)$ configurations. Specifically we flag configurations where the ratio is different by $>25\%$ from the naive expectation $f_\mathrm{SKY} \; B(\ell_1) \, B(\ell_2) \, B(\ell_3)$, where $B(\ell)$ is the Gaussian 10$^{\prime}$ beam.

For both {\tt NILC} and {\tt MILCA}, we find that the bispectra computed on the 50, 60 and 70\% masks are consistent. This indicates that there is no detectable residual galactic contamination in these bispectra. However we did find Galactic contamination on less aggressive Galactic masks, specifically positive Galactic dust. As Galactic dust is highly non-Gaussian, we warn the use against the measurement of higher order statistics using Galactic masks smaller than 50\%. In the following we adopt the 50\% mask as baseline, as it leaves the most sky available for estimation and minimizes masking effects in the measurement.

Figure~\ref{Fig:bispectre} shows the obtained bispectra as a function of multipole for the  {\tt NILC} (green) and  {\tt MILCA} (red) Compton parameter maps.  We observe a good agreement between the bispectra of the two maps, and the bispectral behaviour is consistent with that expected from a tSZ signal \citep[see e.g.][chapter 5]{Lacasa2014thesis}. We furthermore compare these measurements with the bispectrum of the simulated map for the PSZ2 clusters, which is presented in blue in Fig.~\ref{Fig:bispectre}. We observe a good agreement between the bispectra of the {\tt NILC} and  {\tt MILCA} and that of the PSZ2 clusters. We therefore conclude that the observed bispectrum in the $y$\/-map is dominated by detected clusters.

Finally, in Fig.~\ref{Fig:bispectre} are shown the $\pm 1\sigma$ uncertainties of the measurements, in black dotted lines. The error bars were computed in a similar manner to that of the 2013 results \citep{planck2013-p15}, see Appendix~\ref{bispectrumth} for a more detailed discussion.

With a detection per configuration at an average significance of 3.5$\sigma$, and a total significance of $\sim60\sigma$, the \Planck\ data thus provide a high quality measurement of the non-Gaussianity of the thermal Sunyaev-Zel'dovich signal, with undetectable contamination from foregrounds.

\begin{figure*}[]
\begin{center}
\includegraphics[trim=0cm 0cm 0cm 0cm, clip=true,width=2\columnwidth]{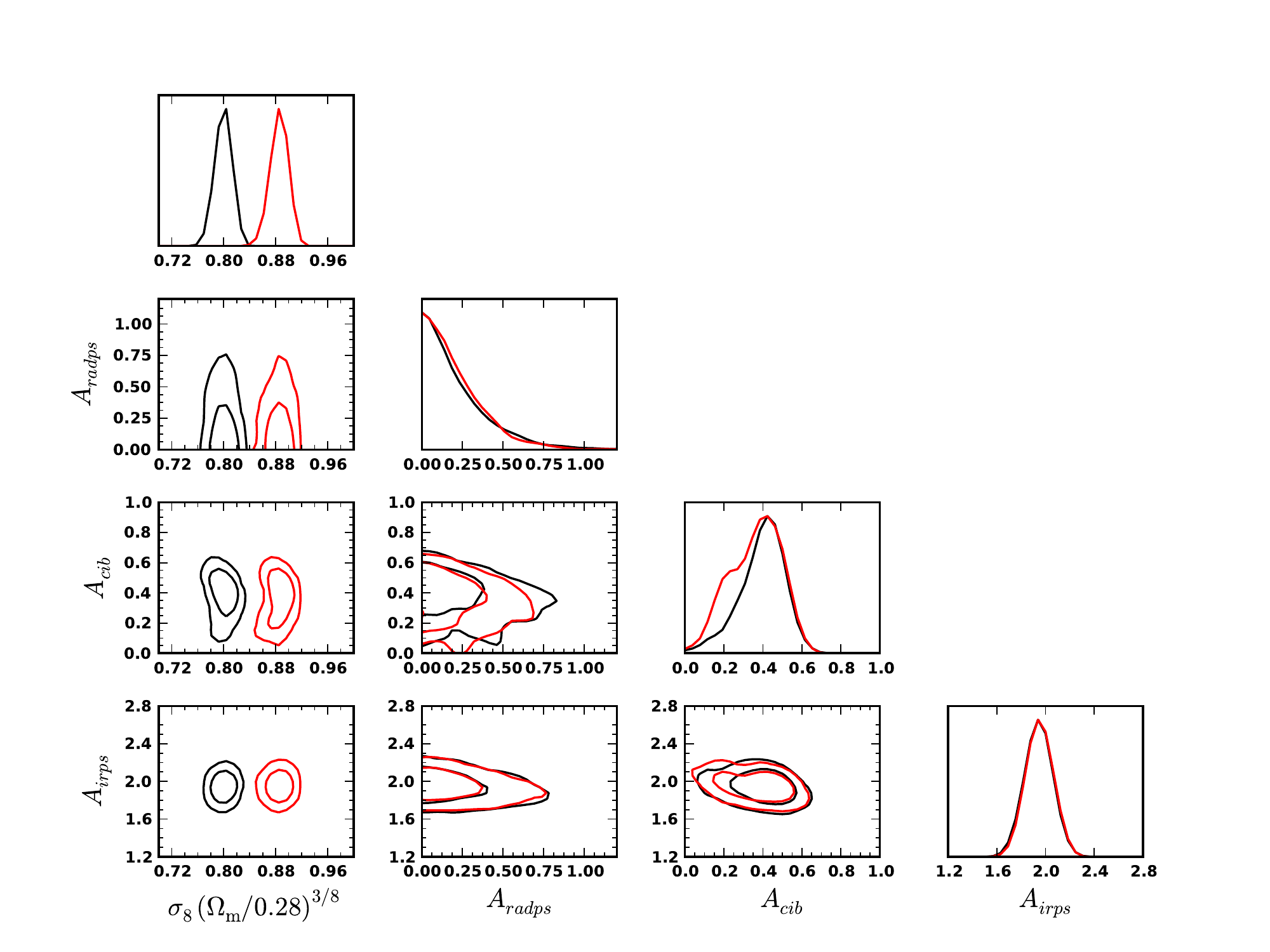}
\end{center}
\caption{2D and 1D likelihood distributions for the combination of
 cosmological parameters $\sigma_{8} (\Omega_{\mathrm{m}}/0.28)^{3/8}$,
 and for the foreground parameters $A_{\mathrm{Rad. PS}}$, $A_{\mathrm{CIB}}$ 
 and $A_{\mathrm{IR. PS}}$.
 We show the 68.3\% and 95.4\% C.L. contours. The red and black contours
 correspond to a fixed mass bias of 0.2 and 0.4 respectively.
\label{fig:compfitall}}
\end{figure*}

\section{Cluster physics and cosmology}
\label{sec:cosmo}

\begin{table}[tmb]
\begingroup
\newdimen\tblskip \tblskip=5pt
\caption{Marginalized bandpowers of the angular power spectrum of the \Planck\
tSZ Compton parameter map (in dimensionless $(\Delta T/T)^2$ units),
statistical and foreground errors, and
best-fit tSZ power spectrum and number counts models
(also dimensionless).\label{tab:bandpowers}}                          
\nointerlineskip
\vskip -1mm
\footnotesize
\setbox\tablebox=\vbox{
   \newdimen\digitwidth 
   \setbox0=\hbox{\rm 0} 
   \digitwidth=\wd0 
   \catcode`*=\active 
   \def*{\kern\digitwidth}

   \newdimen\dpwidth 
   \setbox0=\hbox{.} 
   \dpwidth=\wd0 
   \catcode`!=\active 
   \def!{\kern\dpwidth}
   
\halign{\tabskip 0em\hfil#\hfil\tabskip 1em&
     \hfil#\hfil \tabskip 1em&
     \hfil#\hfil \tabskip 1em&
     \hfil#\hfil \tabskip 1em&
     \hfil#\hfil \tabskip 1em&
    \hfil#\hfil \tabskip 1em&
     \hfil#\hfil \tabskip 0em \cr
\noalign{\doubleline}
*$\ell_{\mathrm{min}}$& *$\ell_{\mathrm{max}}$& *$\ell_{\mathrm{eff}}$&
 ${\ell(\ell+1)C_{\ell}/2\pi}$& $\sigma_{\mathrm{stat}}$&
 $\sigma_{\mathrm{fg}}$&  Best-fit\cr
\noalign{\vskip 5pt}
\omit& & & $[10^{12} {y}^2]$& $[10^{12} {y}^2]$& $[10^{12} {y}^2]$&
 $[10^{12} {y}^2]$\cr
\noalign{\vskip 3pt\hrule\vskip 5pt}
***9  &  **12  &  **10.0  &  0.00506  &  0.00629  &  0.00002  &  0.00726 \cr
\noalign{\vskip 2pt}
**12  &  **16  &  **13.5  &  0.00876  &  0.00615  &  0.00007  &  0.00984 \cr
\noalign{\vskip 2pt}
**16  &  **21  &  **18.0  &  0.01353  &  0.00579  &  0.00015  &  0.01320 \cr
\noalign{\vskip 2pt}
**21  &  **27  &  **23.5  &  0.02946  &  0.00805  &  0.00021  &  0.01737 \cr
\noalign{\vskip 2pt}
**27  &  **35  &  **30.5  &  0.02191  &  0.00522  &  0.00053  &  0.02274 \cr
\noalign{\vskip 2pt}
**35  &  **46  &  **40.0  &  0.02744  &  0.00464  &  0.00109  &  0.03008 \cr
\noalign{\vskip 2pt}
**46  &  **60  &  **52.5  &  0.04093  &  0.00468  &  0.00172  &  0.03981 \cr
\noalign{\vskip 2pt}
**60  &  **78  &  **68.5  &  0.04227  &  0.00429  &  0.00320  &  0.05236 \cr
\noalign{\vskip 2pt}
**78  &  *102  &  **89.5  &  0.06463  &  0.00454  &  0.00567  &  0.06901 \cr 
\noalign{\vskip 2pt}
*102  &  *133  &  *117.0  &  0.10738  &  0.00562  &  0.00969  &  0.09102 \cr
\noalign{\vskip 2pt}
*133  &  *173  &  *152.5  &  0.12858  &  0.00594  &  0.01889  &  0.11956 \cr
\noalign{\vskip 2pt}
*173  &  *224  &  *198.0  &  0.15696  &  0.00611  &  0.02895  &  0.15598 \cr
\noalign{\vskip 2pt}
*224  &  *292  &  *257.5  &  0.21738  &  0.00687  &  0.04879  &  0.20306 \cr
\noalign{\vskip 2pt}
*292  &  *380  &  *335.5  &  0.28652  &  0.00824  &  0.08374  &  0.26347 \cr
\noalign{\vskip 2pt}
*380  &  *494  &  *436.5  &  0.36682  &  0.00958  &  0.13524  &  0.33848 \cr
\noalign{\vskip 2pt}
*494  &  *642  &  *567.5  &  0.42666  &  0.01242  &  0.19500  &  0.42930 \cr
\noalign{\vskip 2pt}
*642  &  *835  &  *738.0  &  0.53891  &  0.01645  &  0.27718  &  0.53577 \cr
\noalign{\vskip 2pt}
*835  &  1085  &  *959.5  &  0.71103  &  0.02402  &  0.37576  &  0.65454 \cr
\noalign{\vskip 2pt}
1085  &  1411  &  1247.5  &  0.82294  &  0.04172  &  0.55162  &  0.77885\cr
% 2013 RESULTS
%**21& **27& $**23.5$&  $<0.045$& $ \dots$& $ \dots$& $0.014$\cr
%\noalign{\vskip 2pt}
%**27& **35& $**30.5$&  $<0.052$& $ \dots$& $ \dots$& $0.019$\cr
%\noalign{\vskip 2pt}
%**35& **46& $**40!*$&     $<0.053$& $ \dots$& $ \dots$& $0.025$\cr
%\noalign{\vskip 2pt}
%**46& **60& $**52!*$&    $**0.046$& $0.007$& $^{+0.014}_{-0.011}$& $0.032$\cr
%\noalign{\vskip 2pt}
%**60& **78& $**68!*$&    $**0.047$& $0.007$& $^{+0.015}_{-0.012}$& $0.042$\cr
%\noalign{\vskip 2pt}
%**78& *102& $**89!*$&    $**0.056$& $0.007$& $^{+0.015}_{-0.013}$& $0.055$\cr
%\noalign{\vskip 2pt}
%*102& *133& $*117!*$&    $**0.077$& $0.008$& $^{+0.020}_{-0.016}$& $0.072$\cr
%\noalign{\vskip 2pt}
%*133& *173& $*152!*$&    $**0.084$& $0.008$& $^{+0.029}_{-0.025}$& $0.094$\cr
%\noalign{\vskip 2pt}
%*173& *224& $*198!*$&    $**0.092$& $0.009$& $^{+0.040}_{-0.033}$& $0.121$\cr
%\noalign{\vskip 2pt}
%*224& *292& $*257!*$&    $**0.158$& $0.009$& $^{+0.046}_{-0.040}$& $0.157$\cr
%\noalign{\vskip 2pt}
%*292& *380& $*335!*$&    $**0.232$& $0.012$& $^{+0.056}_{-0.050}$& $0.203$\cr
%\noalign{\vskip 2pt}
%*380& *494& $*436!*$&    $**0.264$& $0.013$& $^{+0.069}_{-0.064}$& $0.261$\cr
%\noalign{\vskip 2pt}
%*494& *642& $*567!*$&    $**0.341$& $0.017$& $^{+0.080}_{-0.081}$& $0.332$\cr
%\noalign{\vskip 2pt}
%*642& *835& $*738!*$&    $**0.340$& $0.024$& $^{+0.102}_{-0.110}$& $0.417$\cr
%\noalign{\vskip 2pt}
%*835& 1085& $*959!*$&    $**0.436$& $0.035$& $^{+0.149}_{-0.171}$& $0.515$\cr
%\noalign{\vskip 2pt}
%1085& 1411& $1247!*$&    $**0.681$& $0.059$& $^{+0.222}_{-0.272}$& $0.623$\cr
\noalign{\vskip 3pt\hrule\vskip 5pt}
}
}
\endPlancktable 
\endgroup
\end{table}

\subsection{Power spectrum analysis}
\label{powerspectcosmo}

\subsubsection{tSZ power spectrum modelling}
As a measure of structure growth, the tSZ power spectrum can provide
independent constraints on cosmological parameters. As shown by \citet{2002MNRAS.336.1256K}, the power spectrum
of the tSZ effect is highly sensitive to the normalization of the
matter power spectrum, commonly parameterized by the rms of the $z =
0$ mass distribution on $8\, h^{-1}\,\mathrm{Mpc}$ scales, $\sigma_{8}$,
and to the total amount of matter $\Omega_\mathrm{m}$.  We expect the tSZ power spectrum to
also be sensitive to other cosmological parameters, e.g., the baryon density parameter $\Omega_\mathrm{b}$,
the Hubble contant $H_{0}$, and the primordial special index $n_{\mathrm{s}}$. For reasonable external priors
on those parameters, however, the variations are
expected to be negligible with respect to those introduced by changes in
$\Omega_\mathrm{m}$ and $\sigma_{8}$ and so they are not considered here. \\

Following \cite{planck2013-p05b} we consider here a two-halo model for the tSZ power spectrum, which is fully described in Appendix~\ref{poowerspecth}.
This model accounts for both intra-halo (1-halo term) and inter-halo (2-halos) correlations.
Following Eq.~(\ref{twohalomodel}), the tSZ spectrum is computed using
the 2-halo model, the \citet{Tinker:2008p1782} mass function, and the
\citet{Arnaud2010} pressure profile. In particular, we use
the numerical implementation presented in~\citet{Taburet2009,Taburet2010a,Taburet11}, and integrating in
redshift from 0 to 3 and in mass from $10^{13}\,\mathrm{M}_{\odot}$ to
$5 \times 10^{15}\,\mathrm{M}_{\odot}$. Our model allows us to compute the
tSZ power spectrum at the largest angular scales. It is consistent with the tSZ
spectrum presented in~\citealt{2012MNRAS.423.2492E} (EM12), which was used as a
template in the CMB cosmological analysis in \citet{planck2013-p11} and \citet{planck2014-a13}.
We also include the mass bias parameter, $b$, which accounts for bias between the 
observationally deduced ($M^{obs}$) and true ($M^{true}$) mass 
of the cluster \citep[see][for details]{planck2013-p15} such that $M^{obs}  = (1-b) M^{true}$.

Cosmological parameter results are very sensitive to the mass bias and in particular we expect 
$\sigma_8$ and $\Omega_{\mathrm{m}}$ to be strongly degenerated with $b$ \citep{planck2013-p05b}. 
By contrast to \cite{planck2014-a30}, for which external priors in the mass bias have been used, 
we consider here only two distinct values: $b=0.2$ and $b=0.4$. The former 
corresponds to the average value that numerical simulations seem to indicate 
\citep[Fig. A.2 in][]{planck2013-p15}. The latter value for the bias alleviates 
the inconsistency with the constraints derived from the analysis based on primary CMB anisotropies \citep[see][]{planck2013-p15}. 

This value is however larger than that obtained by mass comparison of
%\cite{WtG_Planck_2014} ($1-b$~=~0.688~$\pm$~0.072) 
%when considering 22 
clusters present in both the Planck cosmology sample \citep{planck2013-p15} and 
the {\it Weighing the Giants} \citep[WtG][]{WtG_2014} project. Even if
studies based on lensing mass measurements still provide different and inconsistent results for the cluster 
mass calibration, their number and their accuracy has incredibly improved in the very recent 
past ({\it i.e.} \citealp{Mahdavi_2013} for the {\it CCCP} project, \citealp{Umetsu_clash_2014} for {\it CLASH}, 
\citealp{Israel_2014} for {\it 400d WL}, \cite{Ford2014} for {\it CFHTLenS}, \citealp{Gruen_2014} for {\it WISCy}) and they
are expected to provide useful information.

\subsubsection{Maximum likelihood analysis}

As in \cite{planck2013-p05b}, cosmological constraints are obtained from a fit of the
tSZ power spectrum. As discussed in Section~\ref{subsec:forecont} we take the 
{\tt NILC}-{\tt MILCA} F/L cross-power spectrum (black dots in Figure~\ref{fig:powerspectcosmo}) as a reference and limit the analysis
to  50\% of the sky to minimize foreground residuals. In terms of astrophysical signal we consider a four-component model:
tSZ, clustered CIB, radio point sources, and infrared point sources. We restrict the analysis to multipoles
$ \ell > 10$ so that we can neglect the residual thermal dust contamination (see Section~\ref{sec:lowell}).
For $\ell > 2000$ the total signal in the tSZ map is dominated by correlated noise which is also accounted for in the fit. 
Because of this correlated noise and the expected high value of foreground contamination we limit the fit to multipoles $\ell < 1411$. 
Finally, the observed $y$\/-map power spectrum, $C_{\ell}^{\mathrm{m}}$, is modelled as:
\begin{eqnarray}
C_{\ell}^{\mathrm{m}}  & = & C_{\ell}^{\mathrm{tSZ}} (\Omega_{\mathrm{m}}, \sigma_{8}) + A_{\mathrm{CIB}} \ C_{\ell}^{\mathrm{CIB}} +  \\
& & \nonumber A_{\mathrm{IR }} \  C_{\ell}^{\mathrm{IR}} + A_{\mathrm{Rad}} \  C_{\ell}^{\mathrm{rad}} + A_{\mathrm{CN}} \ C_{\ell}^{\mathrm{CN}}. 
\end{eqnarray}

Here $C_{\ell}^{\mathrm{tSZ}} (\Omega_{\mathrm{m}},\sigma_{8})$ is the
tSZ power spectrum (in red in Fig. \ref{fig:powerspectcosmo}), $C_{\ell}^{\mathrm{CIB}}$ is the clustered CIB
power spectrum (in green), and $C_{\ell}^{\mathrm{IR}}$ and
$C_{\ell}^{\mathrm{rad}}$ are the infrared and radio source power
spectra, respectively (in cyan and in blue). $C_{\ell}^{\mathrm{CN}}$ is an empirical model for the high multipole correlated noise.

Foreground contamination is modelled following Sect.~\ref{refinedforegroundmodel}. 
As discussed there, the main uncertainties in the residual power spectrum translate into up to 50\% uncertainty in the
amplitude of the clustered CIB. We have not considered in the analysis the CIB-tSZ cross-correlation that
was proved to be negligible in terms of power spectrum. The amplitude of the IR and radio point-source contribution will depend very
much on the exact Galactic mask used for the analysis. However, we expect the shape of the their power spectra
to remain the same. We thus allow for a variation of the normalization amplitudes for the clustered CIB,
$A_{\mathrm{CIB}}$, and for the point sources, $A_{\mathrm{IR }}$ and $A_{\mathrm{rad }}$.
%, with Gaussian priors centred on 1 and standard deviation 0.5.

We assume a Gaussian approximation for the likelihood
function. Best-fit values and uncertainties are obtained using an
adapted version of the {\tt CosmoMC} algorithm~\citep{PhysRevD.66.103511}.
Only $\sigma_{8}$ and $\Omega_{\mathrm{m}}$ are allowed to vary. 
All other cosmological parameters are fixed to their best-fit
values as obtained in Table~2 of ~\citet{planck2013-p11}. The
normalization amplitudes, $A_{\mathrm{CIB}}$, $A_{\mathrm{rad}}$ and $A_{\mathrm{IR}}$,
considered as nuisance parameters, are allowed to vary between 0 and
3. For the range of multipoles considered here, the tSZ angular power
spectrum varies like $C_\ell \propto \sigma_8^{8}
\Omega_\mathrm{m}^{3}$. The results are thus presented in terms of
this parameter combination.

\subsection{Best-fit parameters and tSZ power spectrum}
Figure~\ref{fig:compfitall} presents the 2D and 1D likelihood
distributions for the cosmological parameter combination $\sigma_{8}
{(\Omega_{\mathrm{m}}/0.28)}^{3/8}$ and for the foreground nuisance
parameters.  We present the results obtained assuming a
mass bias of $0.2$ (black) and $0.4$ (red). We obtain very similar values for the nuisance parameters in both cases.
In particular the best-fit values for a mass bias of $0.2$ are $A_{\mathrm{CIB}}=0.29^{+0.34}_{-0.20}$,  $A_{\mathrm{rad}}=0.01^{+0.70}_{-0.01}$
and $A_{\mathrm{IR}}=1.97^{+0.20}_{-0.30}$. However, there is a significant shift in the value of  $\sigma_{8} {(\Omega_{\mathrm{m}}/0.28)}^{3/8}$  
as one would expect \citep{planck2013-p15}. In the case of a mass bias of $0.2$ we have $\sigma_{8} (\Omega_{\mathrm{m}}/0.28)^{3/8}=0.80^{+0.01}_{-0.03}$, while for a mass bias of $0.4$ we have $\sigma_{8} (\Omega_{\mathrm{m}}/0.28)^{3/8}=0.90^{+0.01}_{-0.03}$.
Notice that these values are obtained in a specific framework: all other cosmological parameters being fixed and a
fiducial fixed model used for the signals. Relaxing this framework would likely weaken the constraints presented here
as discussed below.

Figure~\ref{fig:powerspectcosmo} shows the {\tt NILC}-{\tt MILCA} F/L angular cross-power spectrum before correcting (black dots) for foreground contribution. We also show the best-fit foreground models: clustered CIB (green line), and radio (blue line) and IR (cyan line) point sources.
The statistical (thick line) and total (statistical plus foreground, thin line) are also shown. The best-fit tSZ power spectrum is presented as a solid red line.
We conclude that the {\tt NILC}-{\tt MILCA} F/L angular cross-power spectrum is dominated by tSZ for multipoles $\ell < 700$, and by foreground contribution for multipoles $\ell > 1200$. We also note that for the best-fit model the radio point-sources contribution seems to be negligible with respect to the IR one. This is not a physical result and it is most probably explained by the strong degeneracy observed between the radio and IR point-source amplitude (see Fig.~~\ref{fig:compfitall}).

\begin{figure}[t]
 \centering
\includegraphics [width=\columnwidth]{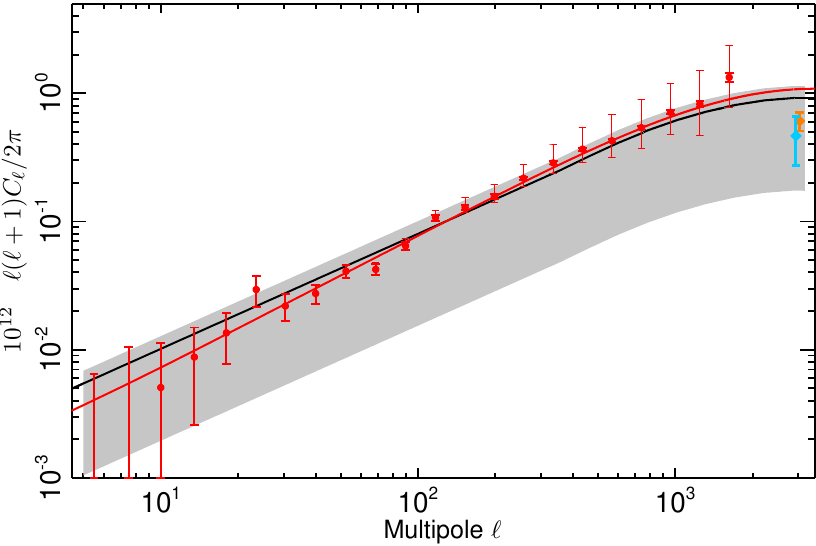}
\caption{{\tt NILC} - {\tt MILCA} F/L cross-power spectrum after
foreground subtraction (red points), compared to the Atacama Cosmology Telescope (ACT; cyan dot) and the
South Pole Telescope \citep[SPT; orange, ][]{George_2014} power spectrum estimates.
 The black line shows the tSZ power spectrum template \citep[EM12][]{2012MNRAS.423.2492E} used in the \Planck\ CMB cosmological
 analysis \citep{planck2013-p11,planck2014-a13} with its best fit amplitude $A_{tSZ}$ \citep{planck2014-a13}, the grey region allows comparison with the 2$\sigma$ interval. 
 \label{fig:powerspectcosmo_CMB}}
 \end{figure}

Finally we present in Figure~\ref{fig:powerspectcosmo_CMB} the {\tt NILC}-{\tt MILCA} F/L angular cross-power spectrum after correcting for foreground contribution (red dots).
Uncertainties account for statistical and systematic errors as well as for uncertainties in the foreground subtraction. The marginalized bandpowers and uncertainties are also
presented in  Table~\ref{tab:bandpowers}. We note that foreground induced uncertainties dominate at multipoles $\ell > 100$. Bandpowers for the best-fit model for the angular tSZ power spectrum are also given for comparison. We also show in the Figure~\ref{fig:powerspectcosmo_CMB} tSZ power spectrum estimates at high multipoles obtained in CMB oriented analyses by  the Atacama Cosmology Telescope (ACT; cyan dot) and the South Pole Telescope \citep[SPT; orange, ][]{George_2014}. The black line shows the tSZ power spectrum template \citep[EM12][]{2012MNRAS.423.2492E} used in the \Planck\ CMB cosmological analysis \citep{planck2013-p11,planck2014-a13} assuming the best-fit amplitude $A_{tSZ}$ and the grey region 2$\sigma$ uncertainties from \cite{planck2014-a13}. We observe that the amplitude of the tSZ signal found in this paper is consistent with the high multipole based measurements. 

\begin{figure}[t]
\begin{center}
\includegraphics[width=0.975\columnwidth]{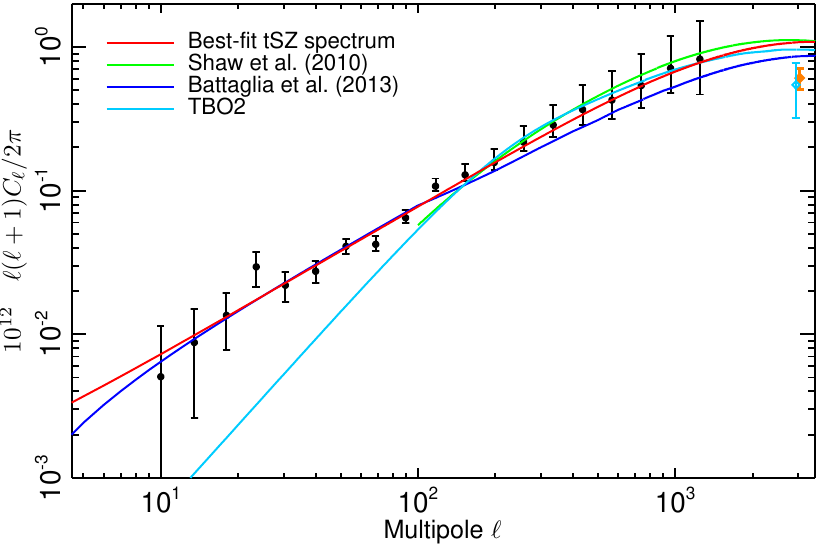}
\end{center}
\caption{tSZ power spectrum for existing models in the literature. 
{\tt NILC}-{\tt MILCA} F/L cross-power spectrum after foreground correction (black dots)
 compared to the Atacama Cosmology Telescope (ACT; cyan dot) and the
South Pole Telescope \citep[SPT; orange, ][]{George_2014} power spectrum estimates.
  We also show the tSZ power spectrum models from
 hydrodynamic simulations \citep[][blue]{Battaglia:2012p1841},
 from $N$-body simulations plus semi-analytical dust gas models
 \citep[][cyan; TBO2]{Trac:2011p1795}, and from analytical calculations
 (\citealt[][green]{Shaw2010}).
\label{fig:comp:models}}
\end{figure}

\subsubsection{Cluster physics dependence}
As discussed in \citet{planck2013-p05b}, we also expect the tSZ power spectrum amplitude to be sensitive to the
physics of clusters of galaxies. To explore this dependence we have considered a set of predicted tSZ spectra
for various physical models. In Fig.~\ref{fig:comp:models} we compare these models to the foreground cleaned
\Planck\ tSZ power spectrum derived above (grey dots), as well as to the Atacama Cosmology Telescope (ACT; cyan dot) and the
South Pole Telescope \citep[SPT; orange, ][]{George_2014} power spectrum estimates. We consider the
predictions derived from hydrodynamical simulations
\citep[][blue]{Battaglia2010,Battaglia:2012p1841}, from $N$-body
simulations plus semi-analytical models
\citep[][cyan; TBO2]{Trac:2011p1795} and from analytical calculations
(\citealt[][green]{Shaw2010}).
These models were originally computed for the set of
cosmological parameters in~\citet{Hinshaw:2012p2598} with
$\sigma_{8}=0.8$ and have been rescaled in amplitude to our best-fit value
for $\sigma_8^{8} \Omega_\mathrm{m}^{3}$. 
We note that there is some dispersion in the predicted amplitudes and shapes of the tSZ power spectrum.
These differences reflect the range of methodologies and assumptions
used both in the physical properties of clusters and in the
technical details of the computation. The latter includes
differences in the redshift ranges and also in the
mass intervals probed by the limited sizes of the simulation boxes of the
hydrodynamical simulations. Analytical predictions are also sensitive to
the model ingredients, such as the mass function, mass bias and scaling
relations adopted.

We see from Fig.~\ref{fig:comp:models} that the models presented
above (the tSZ template for CMB analyses, plus
the \citealt{Battaglia:2012p1841}, \citealt{Shaw2010} and TBO2 models)
provide reasonable fits to the data for multipoles above $200$.
For lower multipoles the \citealt{Shaw2010} and TBO2 models are not consistent
with the data. 

We have also performed a simplified likelihood analysis to evaluate the
uncertainties in cosmological parameters induced
by the uncertainties in the modelling of the cluster physics. 
We replace our own model of the tSZ power spectrum by the models discussed
above and recompute
$\sigma_8(\Omega_{\mathrm{m}}/0.28)^{3/8}$, $A_{\mathrm{CIB}}$, $A_{\mathrm{Rad}}$ and
$A_{\mathrm{IR}}$ from a simple linear fit to the {\tt NILC}-{\tt MILCA} F/L
cross-power spectrum. In the case of mass bias of $0.2$, we obtain values for $\sigma_{8} (\Omega_{\mathrm{m}}/0.28)^{3/8}$
between 0.77 and 0.80, which lie within the 1$\,\sigma$ uncertainties (0.03) presented above.

In the case of our fiducial model (see Appendix~\ref{poowerspecth}) we can also consider uncertainties in the parameters 
describing the scaling relations allowing us to relate the observed tSZ flux to the mass of the cluster for
a given redshift. Following Eq. (7) in \citet{planck2013-p05} the main parameters to be considered are the mass bias $b$, the 
overall amplitude $Y_{*}$ and the scaling slope $\beta$. As discussed above the mass bias is fully degenerate with 
$\sigma_{8}$. Similar conclusions can be drawn for $Y_{*}$, which is expected to be known at the percent level (see Table 1 in \citet{planck2013-p05})
and therefore it is subdominant with respect to the uncertainties in the mass bias. Although the uncertainties in the slope of the scaling relation
are relatively large, we have checked that they lead to negligible uncertainties on cosmological parameters.

\subsection{Higher order statistics}
\label{subsec:cosmohighorder}

\subsubsection{Skewness measurements}
The skewness of the 1D PDF distribution, $\int y^{3} P(y)dy/\left(\int y^{2} P(y)dy
\right)^{3/2}$ can also be used to derive constraints on cosmological parameters.
Following \cite{Wilson:2012p2102,planck2013-p05b} we have chosen
a hybrid approach, by computing the skewness of the
filtered Compton parameter maps outside the 50\% sky mask. In particular, we
have computed the skewness of the \Planck\ data Compton parameter maps
$\langle y^{3}\rangle$, and of the half-difference maps $\langle y_{\mathrm{N}}^{3}\rangle$. 

Using the models presented in Sect.~\ref{sec:theory} we can show that the unnormalized
skewness of the tSZ fluctuation, $\langle T^{3}(\mathbf{n})\rangle$
scales approximately as $\sigma_{8}^{11}$, whereas the
amplitude of the bispectrum scales as $\sigma_8^\alpha$ with
$\alpha=11$--$12$, as shown by \cite{Bhattacharya:2012p2458}.
In the following we do not consider the dependency of
the bispectrum and the unnormalized skewness on other cosmological
parameters, since such dependencies are expected to be significantly lower
than for $\sigma_{8}$ \citep{Bhattacharya:2012p2458}.

We derive constraints on $\sigma_{8}$ by comparing the measured
unnormalized skewness and bispectrum amplitudes with those obtained from
simulations of the tSZ effect. The tSZ contribution was obtained from a hybrid
simulation including a hydrodynamic component for $z<0.3$ plus extra
individual clusters at $z>0.3$, and with $\sigma_{8} = 0.789$. 
This approach is strongly limited by systematic uncertainties and the details 
of the theoretical modelling \cite[see][]{Hill:2013p2067}.
Uncertainties due to foreground contamination are computed using the 
simulations and are accounted for in the final error bars.  

We obtain $\sigma_{8} = 0.77$ for {\tt NILC} and
$\sigma_{8} = 0.78$ for {\tt MILCA}. Combining the two results and
considering model and foreground uncertainties we obtain $\sigma_{8} =
0.78 \pm 0.02 \ (68\% \ \mathrm{C.L.})$. Notice that
the reported uncertainties are mainly dominated by foreground contamination.
However the model uncertainties only account for the expected dependence of the
unnormalized skewness upon $\sigma_{8}$, as shown in
Appendix~\ref{sec:theory}. We have neglected, as was also the case
in~\citet{Wilson:2012p2102}, the dependence on other cosmological
parameters. We have also not considered any uncertainties coming from the
combination of the hydrodynamical and individual cluster
simulations. Because of these limitations, our error bars might be
underestimated.
 
\subsubsection{Fit of the 1D PDF distribution}

We also derived constraints on $\sigma_{8}$ by fitting the 1D PDF obtained in
Sect.~\ref{methods:1D PDF}. Here, we follow the formalism described in
\cite{Hill_py_2014} that evaluates the tSZ 1D PDF theoretically integrating across
individual cluster contributions.   
We use the \citet{Tinker:2008p1782} mass function and the \citet{Arnaud2010} pressure profile. The later is normalised following
the $Y$--$M$ scaling relations described in \cite{planck2013-p15}, and
considering a mass bias parameter of $b=0.2$. All cosmological parameters are fixed to 
the \Planck\ 2015 CMB analysis best-fit values, and we fit only for $\sigma_{8}$.

We note that the \citet{Hill_py_2014} formalism explicitly neglects any effects
due to overlapping clusters along the line of sight. For this reason, and given
the uncertainties in the modeling of the foreground residuals, the best-fit
solution to the observed 1D PDF is fitted in the region which is dominated by
non-overlapping cluster and where the noise and foregrounds contributions are
minimal. In our case, we use $y > 4.5 \times 10^{-6}$.

The confidence limits on $\sigma_{8}$ are obtained from a maximum likelihood
approach, in which the likelihood has a multivariate Gaussian shape with a
covariance matrix which only depends on $\sigma_{8}$. This covariance matrix is
evaluated numerically, accounting only for Poisson terms (both for pixel to
pixel and due to the correlations introduced by the cluster's $y$-profile).
We obtain $\sigma_{8} = 0.77 \pm 0.02$ (68\,\% C.L.) both for {\tt NILC} and {\tt MILCA} maps including
statistical and systematic uncertainties.

 \begin{figure}
  \begin{center}
   \includegraphics[width=\columnwidth]{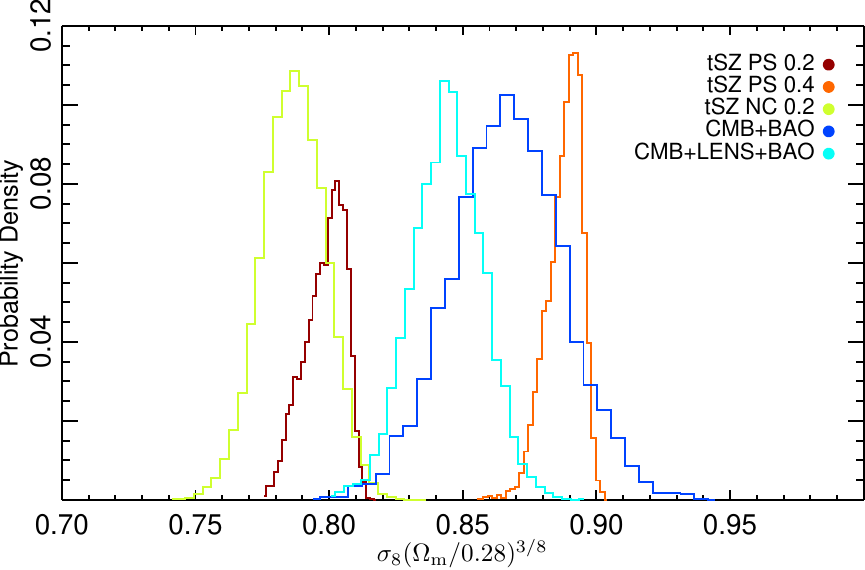}
  \end{center}
\caption{Marginalised likelihood distribution for $\sigma_{8} (\Omega_{\mathrm{m}}/0.28)^{3/8}$ for tSZ and CMB based analyses. We represent the tSZ power spectrum analysis results assuming a mass bias, $b$, of 0.2 (red) and 0.4 (orange), the cluster number count analysis results \citep[green;][]{planck2014-a30}, and the combined \Planck\ CMB and BAO analysis \citep{planck2014-a15} with (cyan) and without (blue) extra lensing constraints.
\label{fig:likecmbszcomparison}}
\end{figure}
 
\subsection{Comparison to other \Planck\ cosmological probes}
\label{subsec:compacosmo}

We have shown in Figure~\ref{fig:powerspectcosmo_CMB} that the amplitude of the tSZ power spectrum measured in this paper and from the \Planck\ CMB  analysis are in good agreement. However, the \Planck\ 2013 results \citep{planck2013-p15} have shown tension between CMB and tSZ derived constraints on $\sigma_8$ for wide range of
experiments including \Planck\ . Figure~\ref{fig:likecmbszcomparison} shows the marginalised likelihood distribution for $\sigma_{8} (\Omega_{\mathrm{m}}/0.28)^{3/8}$ as obtained from the combined  \Planck\ CMB and BAO analysis \citep{planck2014-a15}  with (cyan) or without (blue) lensing constraints. We also presents the results obtained for the \Planck\ 2015 cluster number count analysis assuming a mass bias of 0.2 \citep[green;][]{planck2014-a30}, and for the tSZ power spectrum analysis in this paper assuming a mass bias of 0.2 (red) and 0.4 (orange). We observe that assuming a mass bias of 0.2 the two tSZ analyses are in good agreement but the tension with the CMB measurements remains. This tension can be alleviated by assuming a larger mass bias that increases the value of $\sigma_{8} (\Omega_{\mathrm{m}}/0.28)^{3/8}$. Furthermore, we find that including lensing constraints leads to smaller values of $\sigma_{8} (\Omega_{\mathrm{m}}/0.28)^{3/8}$, which as a consequence are in better agreement with the tSZ results. From this we can conclude that the LSS data would have a marginal preference for small values of the mass bias.

\section{Summary and conclusions}
\label{conclusions}

Because of its wide frequency coverage from 30 to 857\,GHz, the
\Planck\ satellite mission is particularly well suited for the
measurement of the thermal Sunyaev-Zeldovich effect.
Working with the \Planck\ frequency channel
maps from 30 to 857\,GHz, we have reconstructed the tSZ signal over the full
sky using tailored component separation methods.  \\

We tested and validated the \Planck\ $y$\/-maps extensively and characterized them in terms of noise and foreground contamination. As expected the
noise in the $y$\/-map is inhomogeneous and can be characterized by pixel dependent variance
and a homogeneous correlated Gaussian noise. Foreground contamination by thermal
dust emission is found to be important at large angular scales. Additional foreground contamination is due to radio and IR 
point sources for which a mask is provided. In terms of tSZ signal we find good agreement
between the flux of blindly detected clusters in the $y$\/-map and that measured for clusters in the Planck cluster sample.
Furthermore, we find that the sensitivity of the $y$\/-map is sufficient to detect faint and diffuse structures 
such as bridges between merging clusters. Moreover, we have proved via a stacking analysis that the
very low signal-to-noise regions in the $y$\/-map preserve the tSZ signal even for small galaxy groups
(tens of galaxies). 

After accounting for foreground contribution, mainly thermal dust emission at large angular scales,
and clustered CIB and point sources at small angular scales, we have derived from the $y$\/-map  the tSZ angular power
spectrum in the multipole range from $9 < \ell <1411$. This extends significantly the range with respect to
previous measurements \citep{planck2013-p05b} giving for the first time access to the 2-halo term contribution.
The cosmological analysis of the tSZ power spectrum allows us to set constraints on 
cosmological parameters representing matter content in the Universe, mainly $\sigma_8$ and
$\Omega_m$. These constraints are consistent with those obtained from cluster number counts \citep{planck2014-a30}
and in soft tension with those derived from CMB analysis \citep{planck2014-a15}.  

The analysis of the non-Gaussian properties of the $y$\/-map using the 1D PDF, the unnormalized skewness
and the bispectrum of the map have confirmed the tSZ nature of the signal.  \\

The \Planck\ $y$\/-maps and additional ancillary data (noise variance maps, foreground masks and ILC weights) are made available to the public for the  \Planck\ 2015 release (see Appendix \ref{products} for details). These $y$\/-maps are expected to be useful in a wide range of astrophysical and cosmological analyses with clusters. For any of these analyses,
and depending on the scientific goals, the inhomogeneous properties of the noise, and the systematics and foreground contamination should be taken 
into account in different ways as described in this paper. Regions masked by the point source mask should never be used. In the case of pixel based analyses quality flags can be defined by combining the information from the variance map and the various foreground masks. For power spectrum, cross correlation and higher order statistic analyses we remind the fact that the $y$\/-maps present significant foreground contamination that needs to be taken into account both by masking highly contaminated regions (namely the Galactic plane region) and by using adequate foreground models to which the ILC weights are applied. Taking these necessary precautions, the \Planck\ $y$\/-maps will prove a very useful tool for the community.

\begin{acknowledgements}
The Planck Collaboration acknowledges the support of: ESA; CNES and CNRS/INSU-IN2P3-INP (France); ASI, CNR, and INAF (Italy); NASA and DoE (USA); STFC and UKSA (UK); CSIC, MINECO, JA, and RES (Spain); Tekes, AoF, and CSC (Finland); DLR and MPG (Germany); CSA (Canada); DTU Space (Denmark); SER/SSO (Switzerland); RCN (Norway); SFI (Ireland); FCT/MCTES (Portugal); ERC and PRACE (EU). A description of the Planck Collaboration and a list of its members, indicating which technical or scientific activities they have been involved in, can be found at \url{http://www.cosmos.esa.int/web/planck/planck-collaboration}.
\end{acknowledgements}

\appendix

\section{Modelling the expected tSZ signal }
\label{sec:theory}

\subsection{tSZ power spectrum}
\label{poowerspecth}
The representation of the $y$\/-map in spherical harmonics, $Y_{\ell m}$,  reads
\begin{equation}
y(\vec{n}) = \sum_{\ell m} \ y_{\ell m} \ Y_{\ell m} (\vec{n}).
\end{equation}
Thus, its angular power spectrum is given by
\begin{equation}
C^{\mathrm{tSZ}}_{\ell} = \frac{1}{2\ \ell +1}
 \sum_{m} y_{\ell m}  y^{*}_{\ell m}.
\end{equation}
Note that $C^{\mathrm{tSZ}}_{\ell}$ is a dimensionless quantity here, like $y$.

As in \citep{planck2013-p15} the tSZ power spectrum is modelled using a 2-halo model to
account both for intra-halo and inter-halo correlations:
\begin{equation}x
C_\ell^{\mathrm{SZ}} = C_\ell^{\mathrm{1halo}}+ C_\ell^{\mathrm{2halos}}.
\end{equation}
Following \citep{Komatsu:2002p1799} the 1-halo term reads:
\begin{equation}
C_\ell^{\mathrm{1halo}}
 = \int_0^{z_\mathrm{max}}dz\frac{dV_\mathrm{c}}
 {dzd\Omega}\int_{M_{\mathrm{min}}}^{M_{\mathrm{max}}}dM
 \frac{dn(M,z)}{dM}\left|\tilde{y_\ell}(M,z)\right|^2,
\end{equation}
where $dV_{\mathrm{c}}/(dz  d\Omega)$ is the comoving volume per
unit redshift and solid angle and
$n(M,z)dM \ dV_{\mathrm{c}}/(dz d\Omega)$ is the probability of having
a galaxy cluster of mass $M$ at a redshift $z$ in the direction
$d\Omega$. The quantity
$\tilde{y}_\ell=\tilde{y}_\ell(M,z)$ is the 2D Fourier
transform on the sphere of the 3D radial profile of the Compton
$y$-parameter of individual clusters,
\begin{equation}
\tilde{y}_\ell(M,z) = \frac{4 \pi r_{\mathrm{s}}}{l_{\mathrm{s}}^2} \left( \frac{\sigma_{\mathrm{T}}}{m_{\mathrm{e}}c^{2}}\right) \int_{0}^{\infty} \ dx \ x^{2} P_{\mathrm{e}} (M,z,x) \frac{\sin(\ell_{x}/\ell_{\mathrm{s}})}{\ell_{x}/\ell_{\mathrm{s}}} 
\end{equation}
where $x=r/r_{\mathrm{s}}$, $\ell_{\mathrm{s}} =
D_{\mathrm{A}}(z)/r_{\mathrm{s}}$, $r_{\mathrm{s}}$ is the scale
radius of the 3D pressure profile, $D_{\mathrm{A}}(z)$ is the angular diameter
distance to redshift $z$ and $P_{\mathrm{e}}$ is the electron pressure
profile.
 
The 2-halos term \citep{Komatsu:1999p2519,Diego2004,Taburet11}  is given by:
\begin{eqnarray}
\nonumber
C_\ell^{\mathrm{2halos}}&=&\int_0^{z_\mathrm{max}}dz
 \frac{dV_\mathrm{c}}{dzd\Omega}  \times \\
 & & \left[ \int_{M_{\mathrm{min}}}^{M_{\mathrm{max}}}dM
 \frac{dn(M,z)}{dM}\left|\tilde{y}_\ell(M,z)\right| \ B(M,z) \right]^2 \! P(k,z),
\label{twohalomodel}
\end{eqnarray}
where $P(k,z)$ is the 3D matter power spectrum at redshift $z$. 
$B(M,z)$ is the time-dependent linear bias factor that relates the
matter power spectrum, $P(k,z)$, to the power spectrum of the cluster
correlation function. Following
\citet[][see also \citealt{Mo:1996p2742}]{Komatsu:1999p2519} we adopt
$B(M,z) = 1 + (\nu^{2}(M,z)-1)/\delta_{\mathrm{c}}(z) $, where $\nu(M,z) =
\delta_{\mathrm{c}}(M)/D(z)\sigma(M)$, $\sigma(M)$ is the present-day rms mass
fluctuation, $D(z)$ is the linear growth factor, and
$\delta_{\mathrm{c}}(z)$ is the threshold over-density of spherical
collapse.

Finally, we use the \cite{Tinker:2008p1782} mass function, $dn(M,z)/dM$, including an observed-to-true mass bias 
$b$, as discussed in detail in \citet{planck2013-p15} , and we model the
SZ Compton parameter using the pressure profile of \cite{Arnaud2010}.
%This approach is adopted in order to be consistent with the ingredients
%of the cluster number count analysis in \citet{planck2013-p15}.

\subsection{$N\mathrm{th}$ moment of the tSZ field}
\label{nthmoment}
Assuming a Poisson distribution (1-halo term) of the clusters on the sky and neglecting clustering between clusters
the $N\mathrm{th}$ moment of the tSZ signal \citep{Komatsu:1999p2519,Wilson:2012p2102,planck2013-p15} reads
\begin{equation}
 \int_0^{z_\mathrm{max}}dz\frac{dV_\mathrm{c}}{dzd\Omega}
 \int_{M_{\mathrm{min}}}^{M_{\mathrm{max}}}dM\frac{dn(M,z)}{dM}
 \int d^{2}\mathbf{\theta} \ {y(\mathbf{\theta},M,z)}^{N},
\end{equation}
where $y(\mathbf{\theta},M,z)$ is the integrated Compton parameter
along the line of sight for a cluster of mass $M$ at redshift $z$.

\subsection{Bispectrum}
\label{bispectrumth}
The angular bispectrum is given by
\begin{equation}
B^{m_{1} m_{2} m_{3}}_{\ell_{1} \ell_{2} \ell_{3}}
 = \left< y_{\ell_{1} m_{1}}  y_{\ell_{2} m_{2}}  y_{\ell_{3} m_{3}} \right>,
\end{equation}
where the angle-averaged quantity in the full-sky limit can be written as
\begin{equation}
b(\ell_{1},\ell_{2},\ell_{3}) = \sum_{m_{1}m_{2}m_{3}}
 \left( \begin{array}{ccc}  \ell_{1} & \ell_{2} & \ell_{3} \\
 m_{1} & m_{2} & m_{3} \end{array} \right)
 B^{m_{1} m_{2} m_{3}}_{\ell_{1} \ell_{2} \ell_{3}},
\end{equation}
and satisfies the conditions $m_{1}+m_{2}+m_{3}=0$,
 $\ell_{1}+\ell_{2}+\ell_{3}=\mathrm{even}$, and
 $\left|\ell_{i}-\ell_{j}\right|\leq\ell_{k}\leq\ell_{i}+\ell_{j}$,
for the Wigner $3j$ function in brackets. Assuming a Poissonian spatial distribution
of the clusters as above, the bispectrum reads \citep{Bhattacharya:2012p2458}
\begin{eqnarray}
\nonumber
b(\ell_{1},\ell_{2},\ell_{3}) \approx
 \sqrt{\frac{(2\ell_{1}+1)(2\ell_{2}+1)(2\ell_{3}+1)}{4 \pi}}
 \left( \begin{array}{ccc}  \ell_{1} & \ell_{2} & \ell_{3} \\
 0 & 0 & 0 \end{array} \right) \\
\nonumber \times \int_0^{z_\mathrm{max}}\!\!dz\frac{dV_\mathrm{c}}{dzd\Omega} \int_{M_{\mathrm{min}}}^{M_{\mathrm{max}}}\!\!dM\frac{dn(M,z)}{dM}  \tilde{y}_{\ell_{1}}(M,z) \tilde{y}_{\ell_{2}}(M,z) \tilde{y}_{\ell_{3}}(M,z).
\end{eqnarray}

\section{Bispectrum cosmic variance}
\label{sec:bispcosmicvariance}

Following \citep[][chapter 2]{Lacasa2014thesis}, the bispectrum cosmic variance is composed of a Gaussian term, a bispectrum$\times$bispectrum term, a spectrum$\times$trispectrum term and a connected 6-point term. Due to the lack of model or measurement of the trispectrum and 6-point function, we neglected the last two terms. Note that they are however expected to yield a subdominant contribution. Thus we have in full-sky~:
\begin{itemize}
\item Gaussian cosmic variance~:
\begin{equation}
\mathrm{Var}_G(b_{\ell_1 \ell_2 \ell_3}) = \frac{C_{\ell_1} \, C_{\ell_2} \, C_{\ell_3}}{N_{\ell_1 \ell_2 \ell_3}} \times \left\{\begin{array}{ll} 6 &  \mathrm{equilateral} \\ 2 & \mathrm{isosceles} \\ 1 & \mathrm{general} \\ \end{array}\right.
\end{equation}
where
\begin{equation}
\quad  N_{\ell_1 \ell_2 \ell_3} = \frac{(2\ell_1+1)(2\ell_2+1)(2\ell_3+1)}{4\pi} \ 
\left(\begin{array}{ccc} \ell_1 & \ell_2 & \ell_3 \\ 0 & 0 & 0 \end{array}\right)^2
\end{equation}
and where $C_\ell$ is the auto power spectrum of the Compton parameter map, thus containing the noise contribution.
\item Bispectrum$\times$bispectrum cosmic variance
\begin{eqnarray}
\nonumber \mathrm{Cov}_{3\times 3}(b_{\ell_1 \ell_2 \ell_3}, b_{\ell'_1 \ell'_2 \ell'_3}) = \\
\nonumber  b_{\ell_1 \ell_2 \ell_3} \, b_{\ell'_1 \ell'_2 \ell'_3} \times \left( 
\frac{\delta_{\ell_1 \ell'_1}}{2\ell_1+1} + \frac{\delta_{\ell_1 \ell'_2}}{2\ell_1+1} + \frac{\delta_{\ell_1 \ell'_3}}{2\ell_1+1} \right. \\
\nonumber  + \frac{\delta_{\ell_2 \ell'_1}}{2\ell_2+1} + \frac{\delta_{\ell_2 \ell'_2}}{2\ell_2+1} + \frac{\delta_{\ell_2 \ell'_3}}{2\ell_2+1} \\
 + \left.\frac{\delta_{\ell_3 \ell'_1}}{2\ell_3+1} + \frac{\delta_{\ell_3 \ell'_2}}{2\ell_3+1} + \frac{\delta_{\ell_3 \ell'_3}}{2\ell_3+1}
\right)
\end{eqnarray}
This is the only term which gives off-diagonal contributions to the covariance matrix.
\end{itemize}
For our purpose, this cosmic variance is multiplied by $f_\mathrm{SKY}$ and binned to the appropriate binning scheme.

We also consider systematic errors induced by foreground residuals or masking effects.
We estimate systematic errors due to component separation uncertainties from the half difference of the {\tt NILC} and  {\tt MILCA} bispectra.
Masking effects are normally corrected for using simulations, which may under- or overestimate leakage from large to small scales. We thus take a conservative $\pm 25\%$ error on the debiasing ratio, consistently with the fact that the selected configurations have a ratio within $\pm 25\%$ of $f_\mathrm{SKY} \; B(\ell_1) \, B(\ell_2) \, B(\ell_3)$. This error is most likely a conservative overestimate. 

%The last term dominates the error budget, while the three others are comparable. The signal-to-noise ratio of the detection, using the full covariance matrix, is 59.8 for {\tt NILC} and 63.5 for {\tt MILCA}. If instead we neglect the bispectrum$\times$bispectrum term, the signal-to-noise rises respectively to 63.5 for {\tt NILC} and 67.4 for {\tt MILCA}. The bispectrum$\times$bispectrum term thus has a small (although not completely negligible) effect on the covariance. We therefore expect the other non-Gaussian terms (spectrum$\times$trispectrum and 6-point function) to have a small impact on the covariance.\\

\section{Products}\label{products}
In the following we list the $y$\/-map related products delivered in the \Planck\ 2015 data release\footnote{A more detailed description is given in the \Planck\ explanatory supplement.}:

\begin{itemize}
	\item [-] Full-sky {\tt MILCA} and {\tt NILC} $y$\/-maps for the full mission and for the first (F) and second (L) halves of \Planck\ stable pointing period in Compton parameter units (see Section~\ref{subsec:ymaps}).
	\item [-] ILC weights per filter and per frequency used for the reconstruction of the {\tt MILCA} and {\tt NILC} full mission $y$\/-maps, as described in Section~\ref{subsec:ymaps}.
	\item [-] Variance map accounting for the non homogeneous coverage and power spectrum of the correlated homogeneous counterpart, $C^{N}_{\ell}$, for the {\tt MILCA} and {\tt NILC}  full mission $y$\/-maps in Compton parameter units  (see Sect.~\ref{subsec:noise}).
	\item [-] Point source masks including known radio and IR sources as described in Section~\ref{subsec:pointsourcecont}.
	\item [-] Galactic masks used in the analyses presented and in Sections~\ref{sec:powerspec} and \ref{sec:higorderstat}.
\end{itemize}

\bibliographystyle{aat}

\bibliography{szCls.bib,Planck_bib.bib}

\raggedright

\end{document}